\documentclass[twocolumn,secnumarabic,amssymb,nofootinbib,aps, prd]{revtex4-2}
\usepackage{graphicx}
\usepackage{comment}
\usepackage{setspace}
\usepackage{amsmath}
\usepackage{amssymb}
\usepackage{color}
\usepackage{indentfirst}
\usepackage{xspace}
\usepackage{hyperref}
\usepackage{verbatim}
\usepackage{epstopdf}

\usepackage{lineno}
{\large }

\usepackage{slashed,verbatim,subfigure}
\usepackage[sort&compress]{natbib}
\usepackage{appendix}
\usepackage{dcolumn}
\usepackage{bm}
\usepackage[a4paper]{geometry}
\newcommand{\be}{\begin{eqnarray}}

\begin{document}
\title{Study of identified  particle production as a function of transverse event activity classifier, $S_{T}$  in p$-$p collisions} 

\author{Rahul Verma}%
\email{rahul.verma@iitb.ac.in}
\author{Vishu Saini}%
\email{vishusaini220301@gmail.com}
\author{Basanta Nandi}%
\email{basanta@iitb.ac.in}
\author{Sadhana Dash}
\email{sadhana@phy.iitb.ac.in}

\affiliation{ $^1$Department of  Physics\\ Indian Institute of Technology Bombay, Mumbai 400076, India}

\begin{abstract}
A new observable, $S_{T}$, is introduced in terms of the sum of the transverse momentum of charged particles ($\sum_{i} p_{T_{i}}$ ) produced in proton proton (p$-$p) collisions at LHC energies to probe the 
underlying events (UE). The UE are defined as those aspects of proton-proton collisions that are not attributed to the primary hard scattering process, but rather to the accompanying interactions of the rest of the proton. 
The conventional approach of studying underlying events is usually carried out by defining topological regions with respect to the leading particle in an event. The transverse region is generally sensitive to UE and various 
classifiers have been used to discriminate the extent of UE activity regions. The  production of  identified particles like $\pi^{\pm}$, $K^{\pm}$, p , $K_{S}^{0}$, and $\Lambda^{0}$ are studied in different ranges of transverse activity classifier in p$-$p collisions at $\sqrt{s} = 13 $ TeV  using pQCD inspired PYTHIA 8 event generator. A comparative analysis of the identified particle spectra, mean multiplicity  and mean transverse momentum has been carried out with respect to $S_{T}$ and the performance of this new observable is gauged by comparing the results with previously defined  $R_{T}$ observable.  
\end{abstract}
\maketitle

\section{Introduction}
Recently, several effects typical of heavy-ion phenomenology have been observed in high-multiplicity proton-proton (p$-$p) collisions and p$-$Pb collisions. The strangeness enhancement \cite{alicestrange}, collectivity \cite{rhic2part, lhc2part} and anisotropic flows \cite{aliceflow} etc. which were well accepted signatures of creation of a hot and dense state composed of 
deconfined partons, called the Quark Gluon Plasma, have been reported in heavy-ion collisions. However, these signatures have also been reported in small systems like p$-$p and p$-$Pb \cite{cms1, cms2, alicenature, atlas}. This has intrigued the heavy-ion community to understand the underlying dynamics in  small collision systems. In order to understand the origin of such effects in small collision systems where the effects of  hot and dense medium is not expected to manifest,  it becomes imperative to study and comprehend the production of identified particles as a function of the underlying event activity \cite{atlasue, pskands} observables. 
The “underlying event” can be defined as those aspects of proton-proton collision that are not attributed to the hard scattering process, but rather to the accompanying interactions of the rest of the proton \cite{atlasue,reviewue}. Underlying event is an essential element of the hadron-hadron interaction environment. Its contributions to the final state of a given hadron-hadron collision are vast and thus to perform precise standard model measurements (and also search for physics beyond the standard model at hadron colliders like LHC), it is necessary to have a good understanding  of these events.  However, such events have contributions from both hard (perturbative QCD) and soft (non-perturbative QCD) processes. Therefore, one resorts to various phenomenological models (usually implemented in Monte Carlo event generators), to constrain parameters tuned to relevant experimental data. Various UE-sensitive variables have been proposed \cite{atlasue, pskands,sdashue} to disentangle hard process-dominated events from soft ones.
Moreover, UE activity of an event is also found to be related to geometry of energy momentum flow in the event \cite{atlasue,reviewue}. The conventional way to disentangle particle production associated with soft and hard processes is usually carried out by defining three kinematic regions depending on the angular distance with respect to the leading charged particle. These regions have different sensitivities to UE activity. Various event shape variables like transverse spherocity , thrust, sphericity etc. also provide an alternative way to classify events as jetty (dominated by 
hard processes) and isotropic (dominated by soft processes) events. A transverse activity classifier, $R_{T}$ \cite{pskands} was also  proposed to quantify UE activity of an event.\\
In this work, a new transverse activity classifier, $S_{T}$ is introduced and the production of identified particles like $\pi^{\pm}$, $K^{\pm}$, $p$, $K_{S}^{0}$, and $\Lambda^{0}$ as a function of the proposed UE activity classifier has been investigated in p$-$p collisions at $\sqrt{s} = 13 $ TeV. One of the primary aims of the present study is to explore the $p_{T}$ spectra, multiplicity distribution and strangeness  and baryon production in regions of varying UE activity quantified by $S_{T}$.  The study has been  performed  by dividing the kinematic region into {\bf toward}, {\bf away}, and {\bf transverse} regions where the {\bf transverse} region is most sensitive to UE activity. The particle production has been studied as a function of transverse activity classifiers like $R_{T}$ and $S_{T}$  in order to compare their performances. 

\section{Underlying Event Observable}

\begin{figure}[th!]
	\centering
	\includegraphics[width=0.5\textwidth,height=0.5\textwidth]{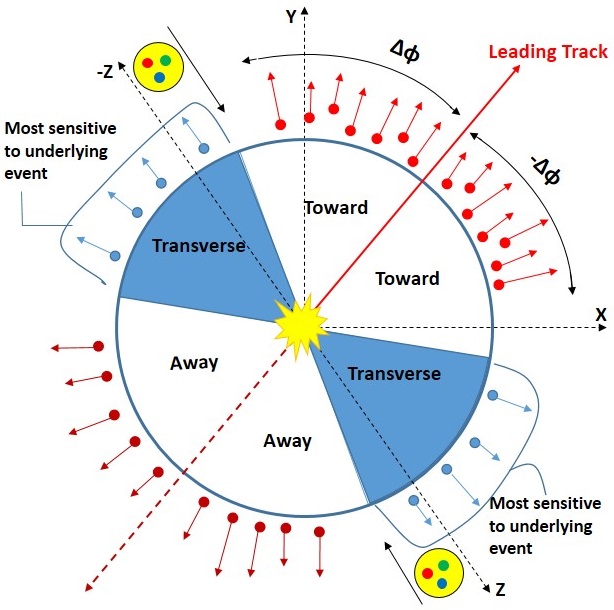}
	\caption{The three topological regions} 
	\label{Three regions}
\end{figure}

It is nearly impossible to uniquely separate the underlying event activity from the hard scattering process on an event-by-event basis in experimental studies. However, UE activity is intimately related to the topology of an event. For example, a typical hard scattering event is characterized by  a burst, of hadrons travelling approximately in the same direction as that of initial beam particles or as a  collection of hadrons (called \textit{jets}) with large momentum in transverse direction, that are approximately back to back in azimuthal angle, $\phi$. As two high $p_{T}$ jets are back to back in azimuthal angle, $\phi$, one can use the topology of collision event to study the UE. This is carried out on an event-by-event basis, by selecting a leading object in the event which is generally the charged particle having the highest transverse momentum in the event, the $p_{T}^{lead}$ . The axis of direction of the $p_{T}^{lead} $ in the event  is used to define regions in the $\eta - \phi$ plane which have different sensitivities to the UE. The axis of direction of $p_{T}^{lead}$ particle is well-defined for all events. It is also correlated with the axis of hard scattering in high $p_{T}$ events \cite{atlasue}. The azimuthal angle difference between the $p_{T}^{lead}$ and other associated particles, $|\Delta \phi| = |\phi - \phi_{lead}|$, is used to define the following three kinematic regions as shown in Table \ref{Three regions} and Figure \ref{Three regions})
\begin{table}[h]
\centering
\begin{tabular}{|l|l|}
\hline
\textbf{Azimuthal angle difference} & \textbf{Region}     \\ \hline
$|\Delta \phi|< 60^{\circ}$                   & Toward region     \\ \hline
$ 60^{\circ} < |\Delta \phi| <120^{\circ}    $              & Transverse region \\ \hline
$|\Delta \phi|> 120^{\circ}$                  & Away region       \\ \hline
\end{tabular}
\caption{ The three topological regions.}
\label{Table:Three regions}
\end{table}

\begin{figure}[th!]
	\centering
	\includegraphics[width=0.5\textwidth,height=0.3\textwidth]{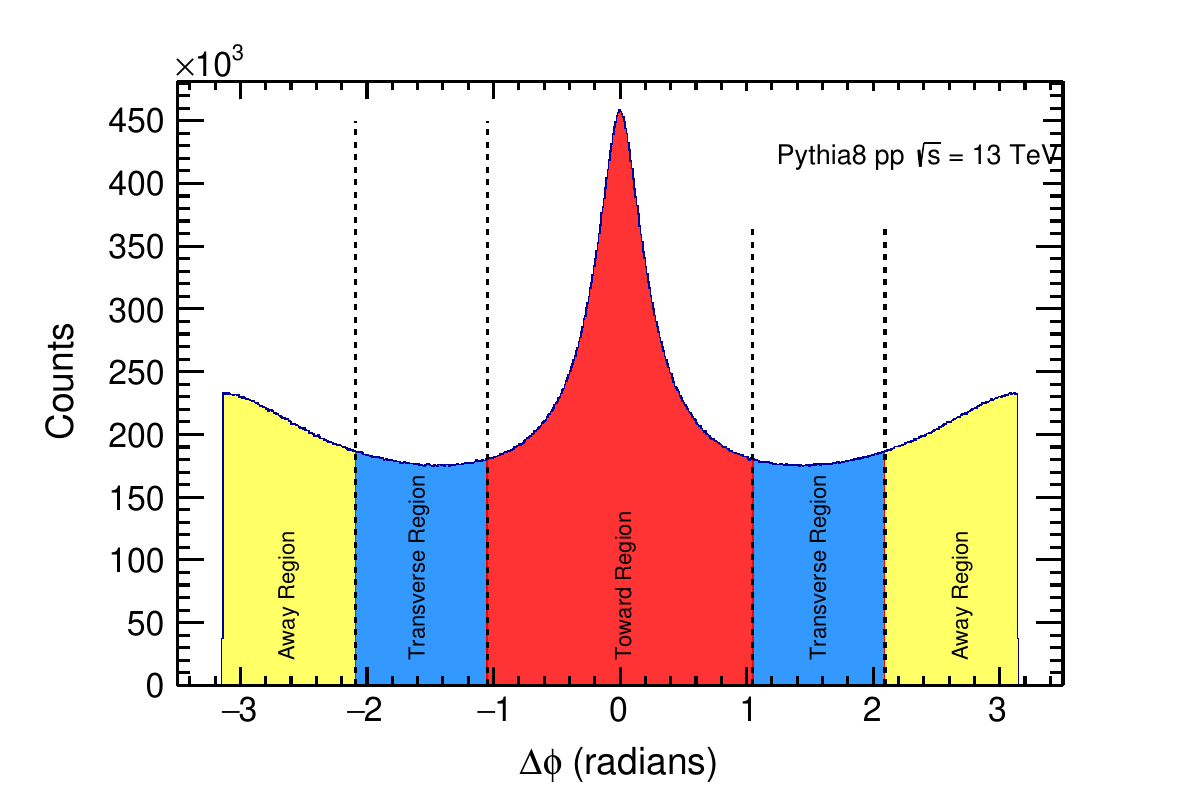}
	\caption{The $\Delta \phi$ distribution of charged particles with respect to the leading particle in the three topological regions.} 
	\label{dphi}
\end{figure}

Out of these three regions, the {\bf transverse} region is most sensitive to UE activity as it is perpendicular to the axis of hardest scattering and thus has the lowest level of activity from hard scattering. The particle production associated with hard and soft processes is explored by investigating the particles produced in {\bf toward}, {\bf away} and {\bf transverse} regions. 
The $\Delta \phi$ distribution for the three regions is shown in Figure \ref{dphi}.

\subsection{Transverse activity classifiers}
\label{subsec: RT and ST}
The UE activity of an event is directly proportional to the activity in the transverse region of that event as this region is most sensitive of the three regions to UE activities. One can use UE-sensitive variables like the charged-particle multiplicity, $N^{ch}$, and the scalar sum of $p_{T}$ of considered particles, $\sum_{i} p_{T_{i}}$ to construct transverse activity (or UE activity) classifier.
One such classifier, called the relative transverse activity classifier, $R_{T}$ was proposed \cite{pskands,aliceresult}, to quantify UE activity. $R_{T}$ of an event is defined as:
\begin{equation} 
    R_{T} = \frac{N^{ch}_{T}}{<N^{ch}_{T}>}
    \label{eq:RT}
\end{equation}
where $N^{ch}_{T}$ is the number of charged particles in the transverse region of an event and  $<N^{ch}_{T}>$ refers to the mean charged particle multiplicity. If UE activity in a particular event is high leading to the production of large number of particles in the transverse region, the corresponding $R_{T}$ of that event will be higher. Similarly, if the UE activity of an event is low, a lesser number of particles are produced in the transverse region and thus $R_{T}$ of the considered event will be low. One can observe that the choice of construction of such a variable classifies the events around $R_{T} = 1$ with ``higher-than-average" UE from the ``lower-than-average" UE. 
In similar lines, it is interesting and worth trying to explore the scalar sum of $p_{T}$,  $\sum_{i} p_{T_{i}}$, of charged particles in transverse region to quantify UE activity of an event. The proposed classifier is denoted as $S_{T}$ and is defined as :
\begin{equation}	
S_{T} =  \frac{\sum_{i} p_{T_{i}}^{T}}{<\sum_{i} p_{T_{i}}>}
\end{equation}
where $\sum p_{T}^{T}$ is the $\sum_{i} p_{T_{i}}$ of charged particles in the transverse region of an event and  $<\sum_{i} p_{T_{i}}>$ is the mean of $\sum p_{T}$. As observed for $R_{T}$, $S_{T} = 1$ also divides the events with "higher-than-average" UE activity from the ``lower-than-average" UE activity. One of the primary goals  of this study is to investigate 
the performance of  this novel classifier $S_{T}$  to probe UE activity of events as compared to the $R_{T}$ classifier.\\
It was observed previously that above a certain lower threshold of $p_{T}^{lead}$ ($p_{T}^{lead} > $  5 GeV/c), a plateau region in the transverse region was reached \cite{atlasue}. This plateau region signifies that the mean values of UE quantities (like mean charged-particle multiplicity density, $d\langle N_{ch} \rangle/d\eta$, and the  $\sum_{i} p_{T_{i}}$) calculated over a fully inclusive set of particles show little dependence on the $p_{T}^{lead}$ i.e the soft processes that contribute to UE are independent of $p_{T}^{lead}$. The slow rise of UE plateau in the transverse regions is possibly due to the additional contributions from wide-angle radiation associated with the hard scattering.
In the present study,  the particles that lie in the plateau region ($5 \ \ \text{GeV/c } < p_{T}^{lead} < 40$ GeV/c) are considered to ensure that the UE observables in the transverse region remain nearly independent of the $p_{T}$ of the leading particle. 

\section{Analysis Method}
\label{sec:Method}
Pythia 8.308 \cite{pythia8,monash}, a general-purpose Monte Carlo (MC) event generator that can simulate initial and final-state parton showers, multipartonic interactions, hadronization, and particle decays has been used to generate p$-$p collisions  at $\sqrt{s} = $13 TeV. In Pythia, multi-partonic interactions (MPI) allow the generation of a large number of final state string objects over limited transverse space. It has been observed that inclusion of MPI is necessary for a correct description of UE in event generators like PYTHIA or Herwig \cite{herwig}. From an experimental point of view, as UE consists of all activity that accompanies the primary hard scattering, it essentially consists of MPI and interactions between constituents of beam remnants. This is implemented in the Lund string model \cite{lund1,lund2}. However, this traditional model is expected to be modified in high parton density environments due to the inter-string interactions. Rope Hadronization (RH) is a model that extends the Lund string model to describe environments with many overlapping strings, such as in heavy ion collisions or high multiplicity p$-$p collisions. The idea is to let the strings overlap and act coherently forming a rope which  hadronizes with a larger, effective string tension \cite{rope1,rope2}. This model is tested and observed to improve strange to non-strange ratios and the near-side long range correlations observed in high multiplicity p$-$p collisions \cite{dash2019,dash2020}.  One essential ingredient used in the hadronization models is Color-Reconnection (CR). It is the mechanism that describes the formation of hadronizing strings between the outgoing partons. The implementation of this mechanism is based on the minimization of the total string potential energy by connecting different final state partons such that the total string length remains as short as possible \cite{color1,color2}. In Pythia 8, the QCD-based color reconnection scheme and rope hadronization has been shown to explain the enhancement of baryons \cite{dash2021}. It also explained the similarities in the multiplicity evolution of the  $p_{T}$ spectra among different colliding systems (p$-$p, p$-$Pb, and Pb$-$Pb). The present study includes the effect of color reconnection as well as rope hadronization to explore the strange particle and baryon production in varying topological regions of UE activity. The study also aims to compare the sensitivity of $R_{T}$ and $S_{T}$ . The kinematic cuts $ p_{T} \geq 0.15$  GeV/c  and $|\eta| \leq 0.8$ are applied to the particles to approximate experimental sensitivity.

\section{Results and Discussions}
\label{sec:Results}

\subsection{$R_{T}$ and $S_{T}$ distributions}

\begin{figure}[h!]
\includegraphics[width=0.5\textwidth, height = 0.35\textwidth ]{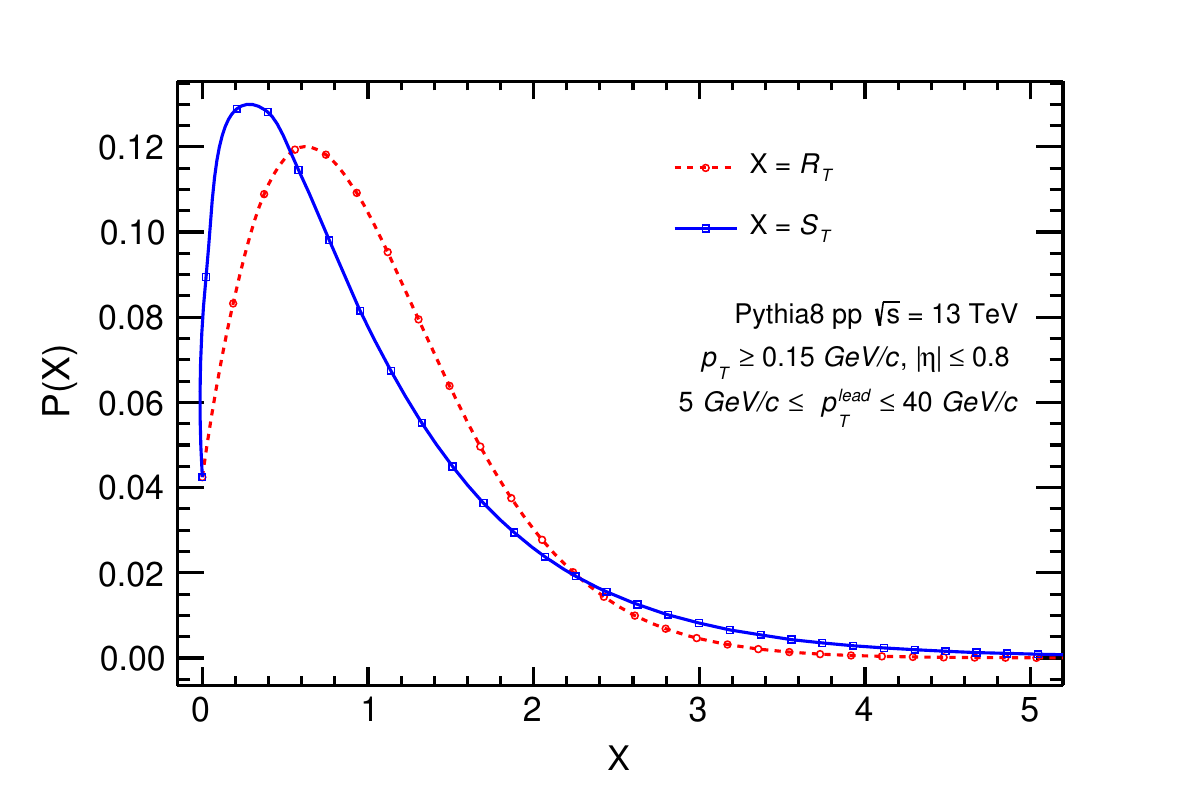}
\includegraphics[width=0.5\textwidth, height = 0.35\textwidth ]{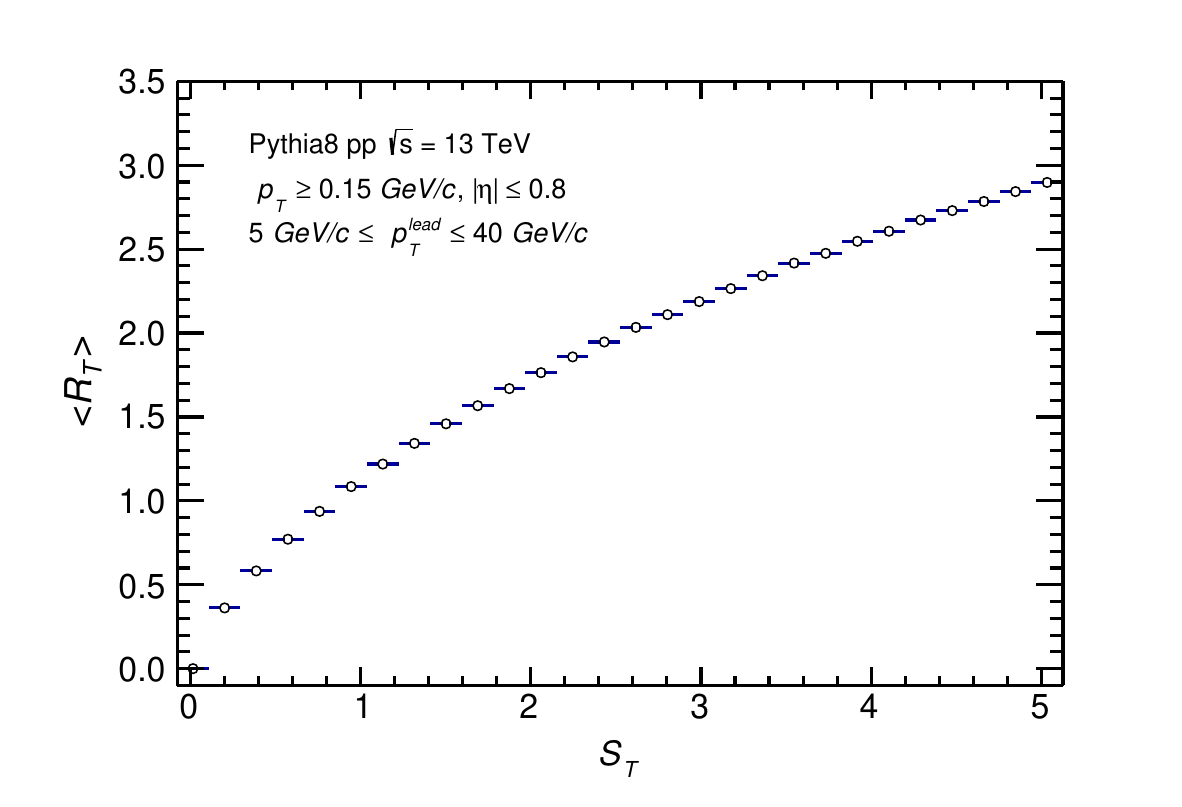}

\caption{ (Upper Panel )The $S_{T}$ and $R_{T}$ distribution for events with $p_{T}^{lead}  \geq 5$ GeV/c in p$-$p collisions at $\sqrt{s} =$ 13 TeV.  (Lower panel)The  variation of  $<R_{T}>$ with $S_{T}$.}
\label{FigRTST}
\end{figure}

Figures \ref{FigRTST} shows the $S_{T}$ and $R_{T}$ distributions obtained using charged particles in p$-$p collisions at  $\sqrt{s} =$ 13 TeV.  One can observe that higher $S_{T}$ (or $R_{T}$) events (i.e. high UE activity) are rarer than low UE activity events. In the bottom panel, where the variation of  $<R_{T}>$ is shown as a function of $S_{T}$, one notes that it deviates from linearity at values of $S_{T}$ greater than 1.5. Nevertheless, large values of $S_{T}$ corresponds to large values of  $<R_{T}>$ showing a positive correlation between the two observables.

\begin{figure}[!h]
\centering
\includegraphics[width=0.5\textwidth, height = 0.35\textwidth ]{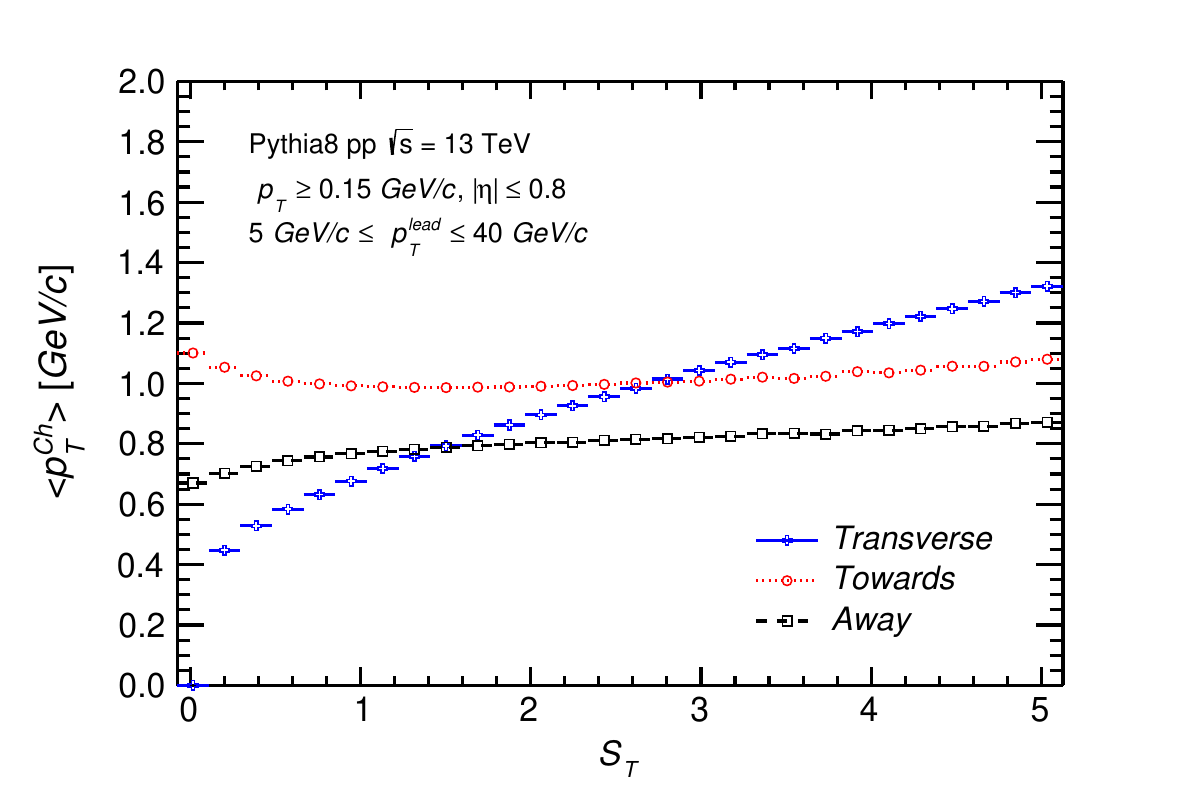}
\includegraphics[width=0.5\textwidth, height = 0.35\textwidth ]{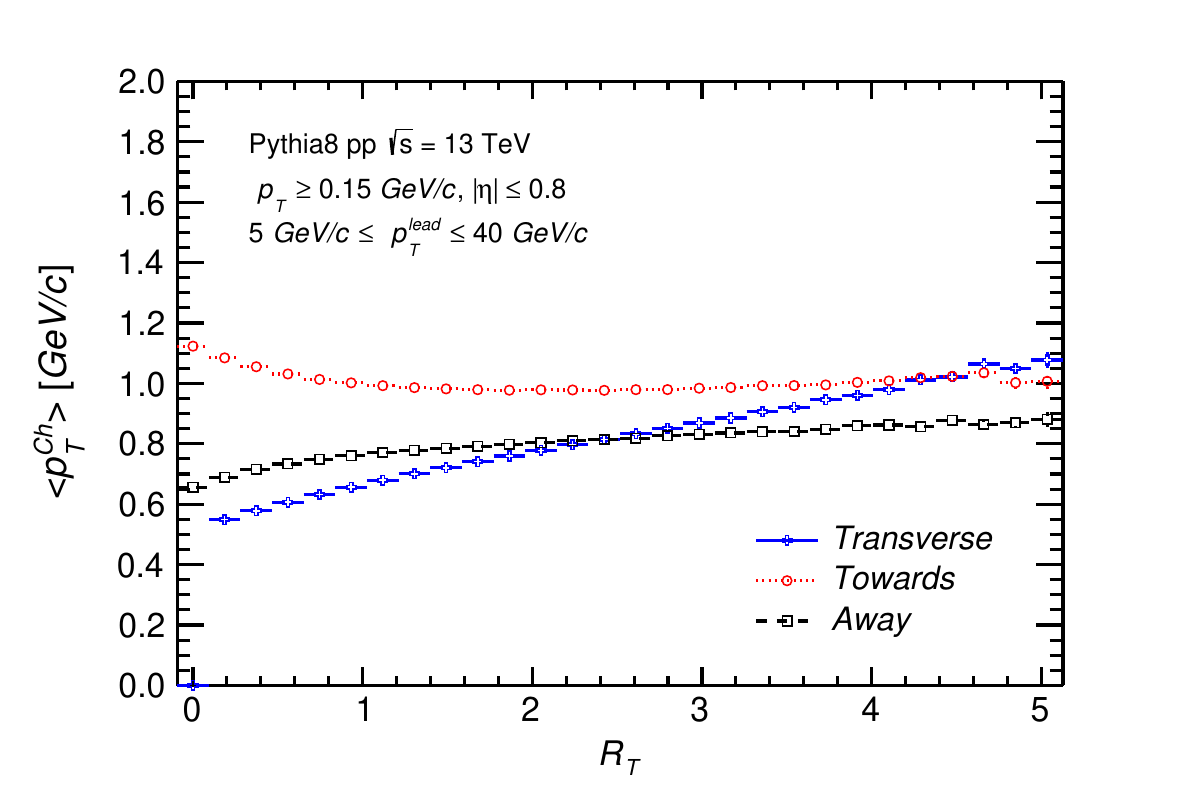}
\caption{The variation of  $<p_{T}>$ of charged particles with respect to $S_{T}$ (upper panel) and $R_{T}$ (lower panel) for events with $p_{T}^{lead}  \geq 5$ GeV/c  in p$-$p collisions at $\sqrt{s} =$ 13 TeV for the three topological regions.}
\label{meanptST}
\end{figure}

\begin{figure}[h]
\centering
\includegraphics[width=0.5\textwidth, height = 0.35\textwidth ]{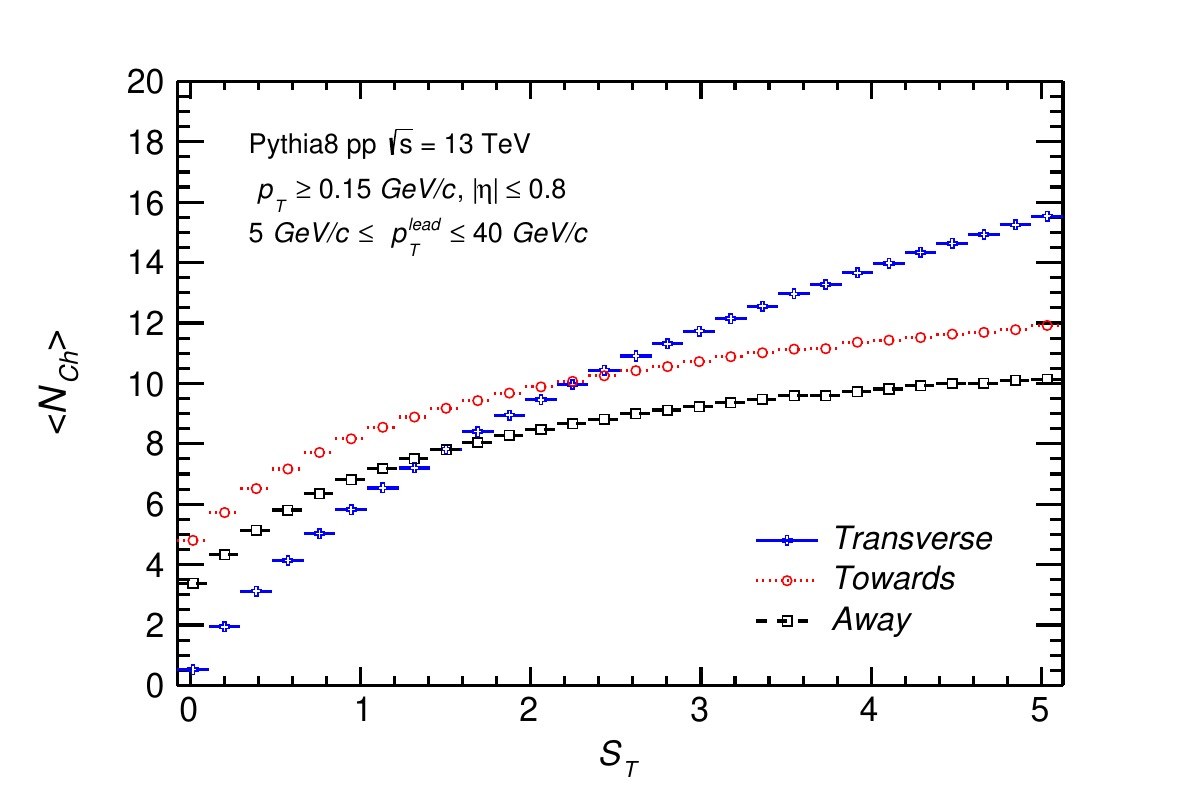}
\includegraphics[width=0.5\textwidth, height = 0.35\textwidth ]{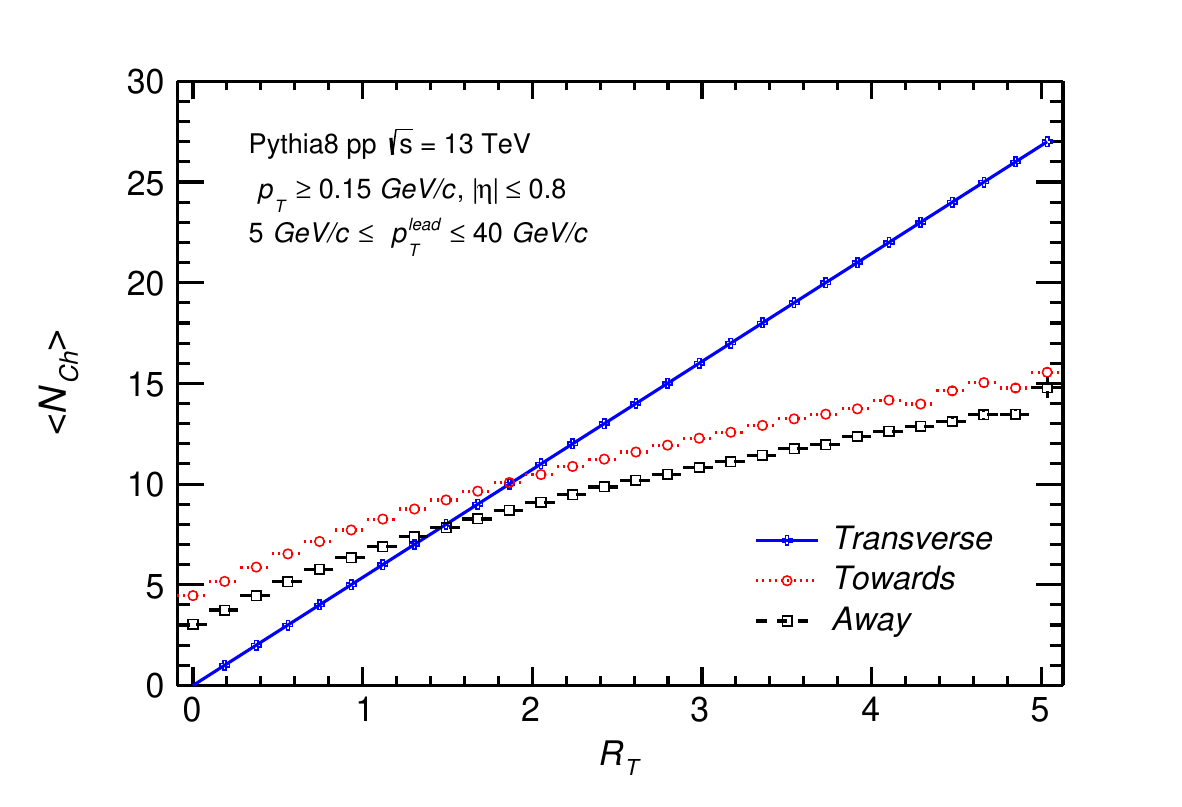}
\caption{The variation of $<N_{ch}>$ with $S_{T}$ (upper panel) and $R_{T}$ (lower panel) for events with $p_{T}^{lead}  \geq 5$ GeV/c  in p$-$p collisions at $\sqrt{s} =$ 13 TeV for the three topological regions.}
\label{meanmultST}
\end{figure}

\begin{figure}[!h]
\centering
\includegraphics[width=0.5\textwidth, height = 0.35\textwidth ]{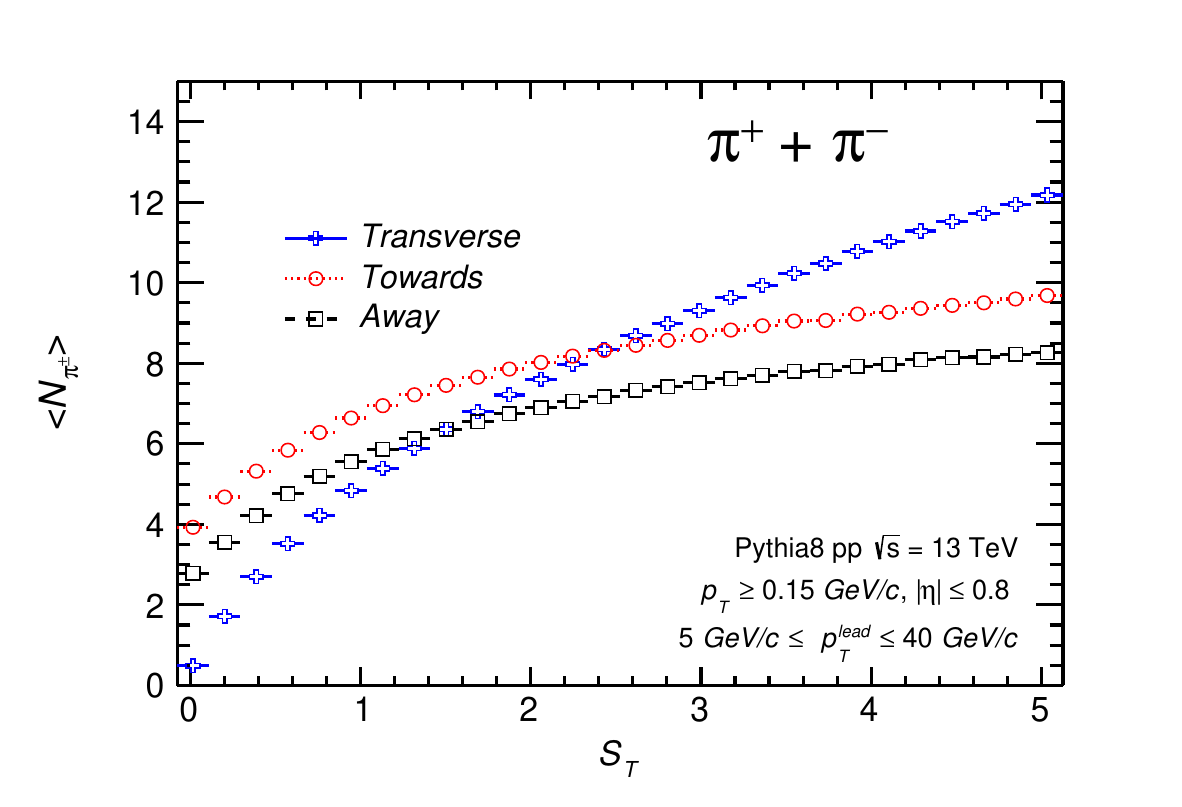}
\includegraphics[width=0.5\textwidth, height = 0.35\textwidth ]{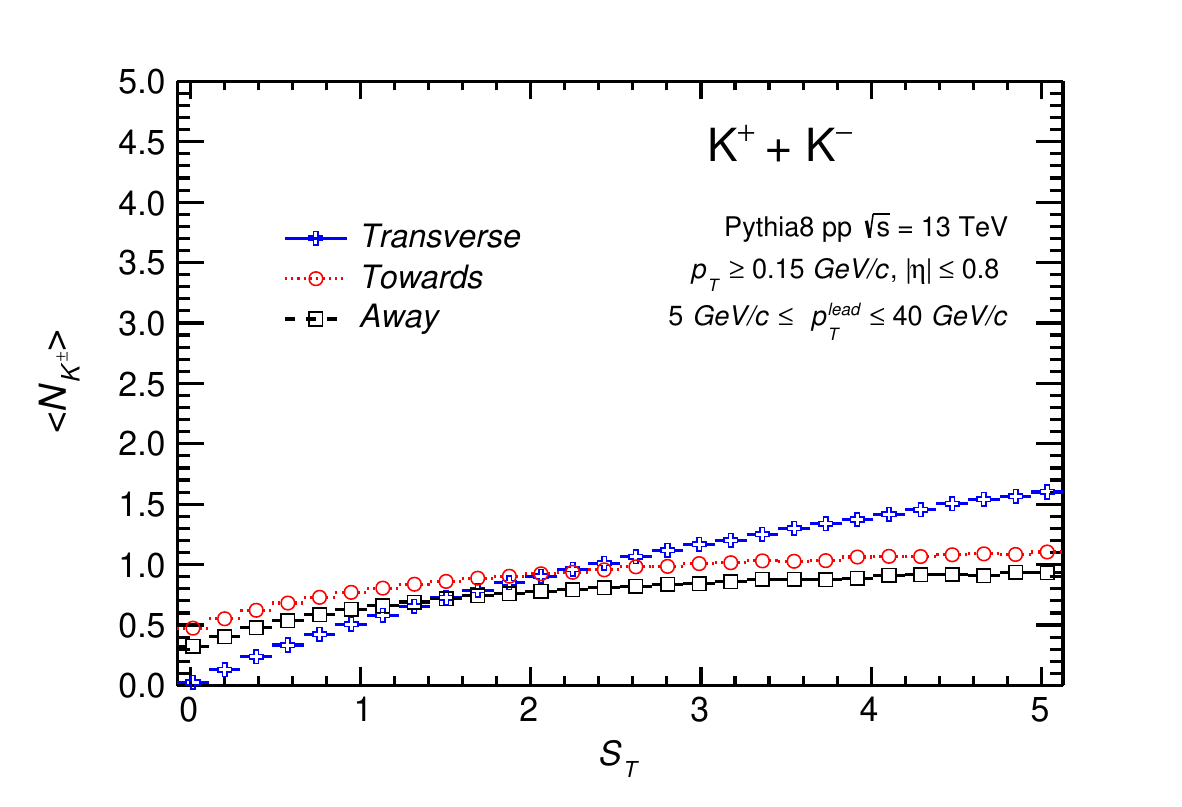}
\includegraphics[width=0.5\textwidth, height = 0.35\textwidth ]{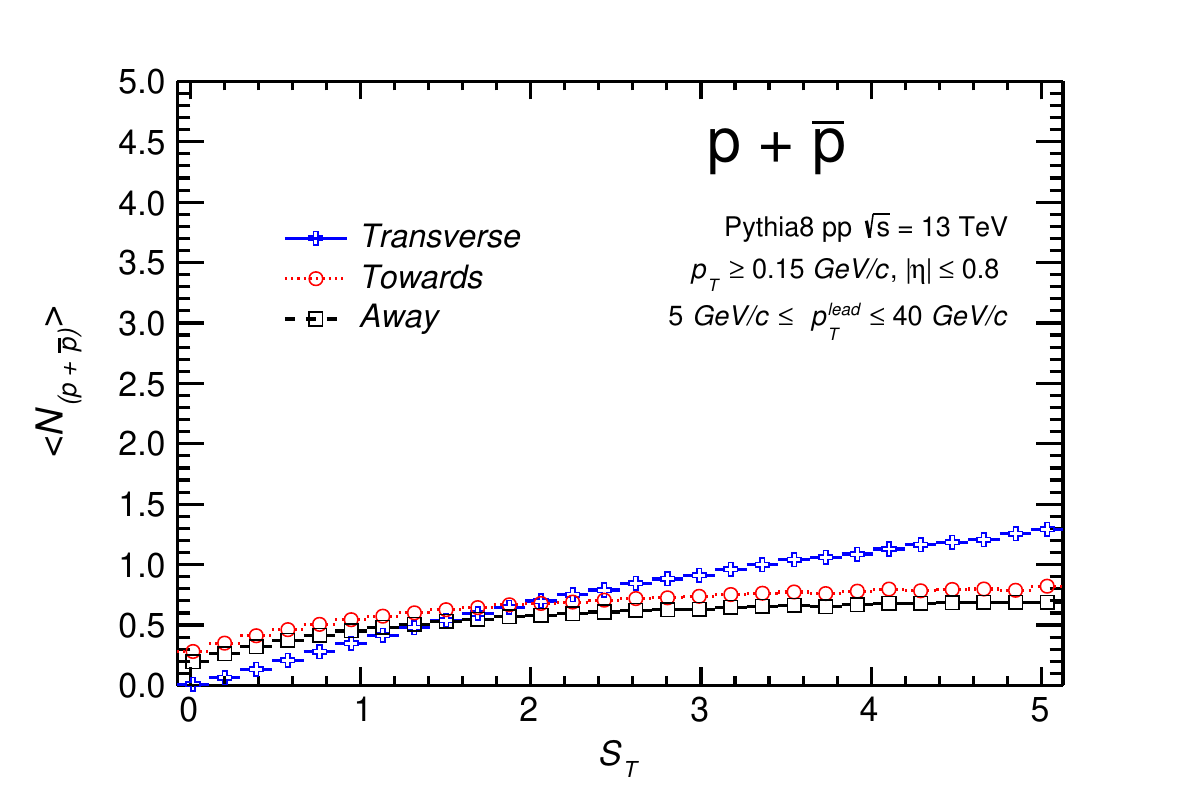}
\caption{The variation of mean multiplicity of pions(upper panel), kaons(middle panel) and protons(bottom panel) with $S_{T}$ in p$-$p collisions at $\sqrt{s} =$ 13 TeV for the three topological regions.}
\label{imeanmultST}
\end{figure}

\begin{figure}[h]
\centering
\includegraphics[width=0.5\textwidth, height = 0.35\textwidth ]{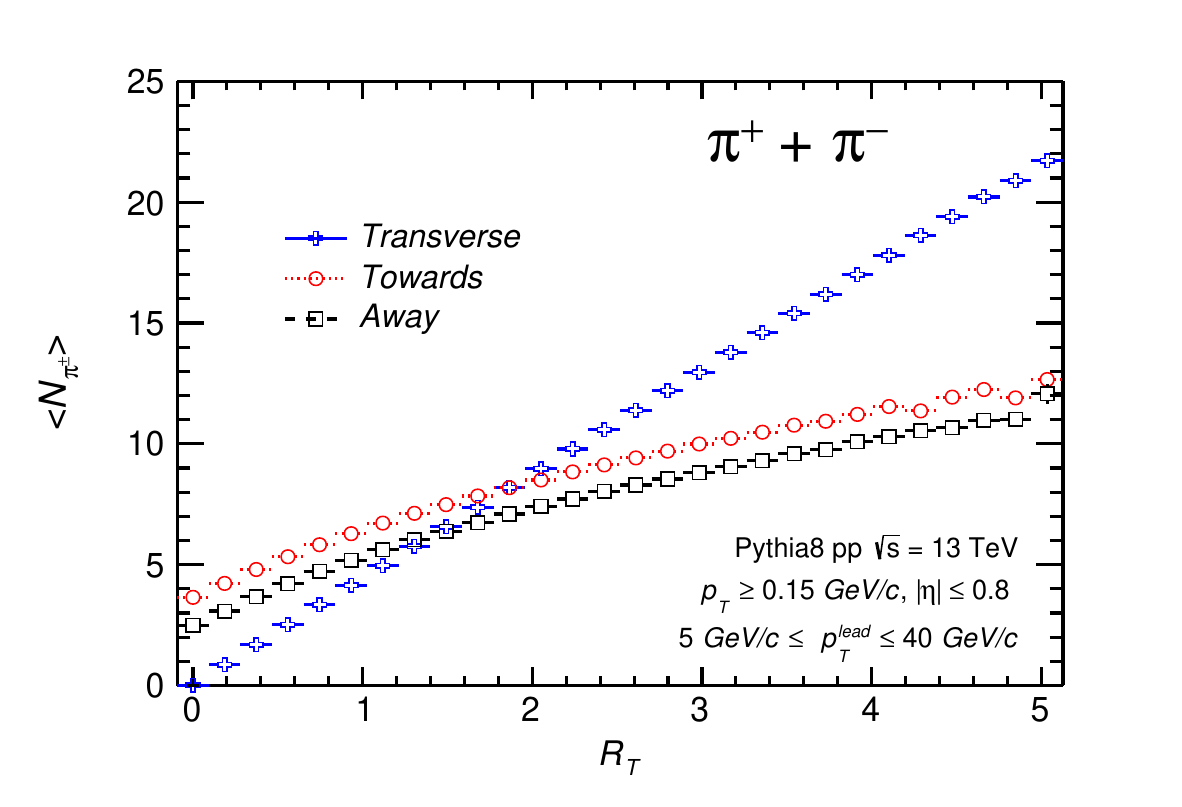}
\includegraphics[width=0.5\textwidth, height = 0.35\textwidth ]{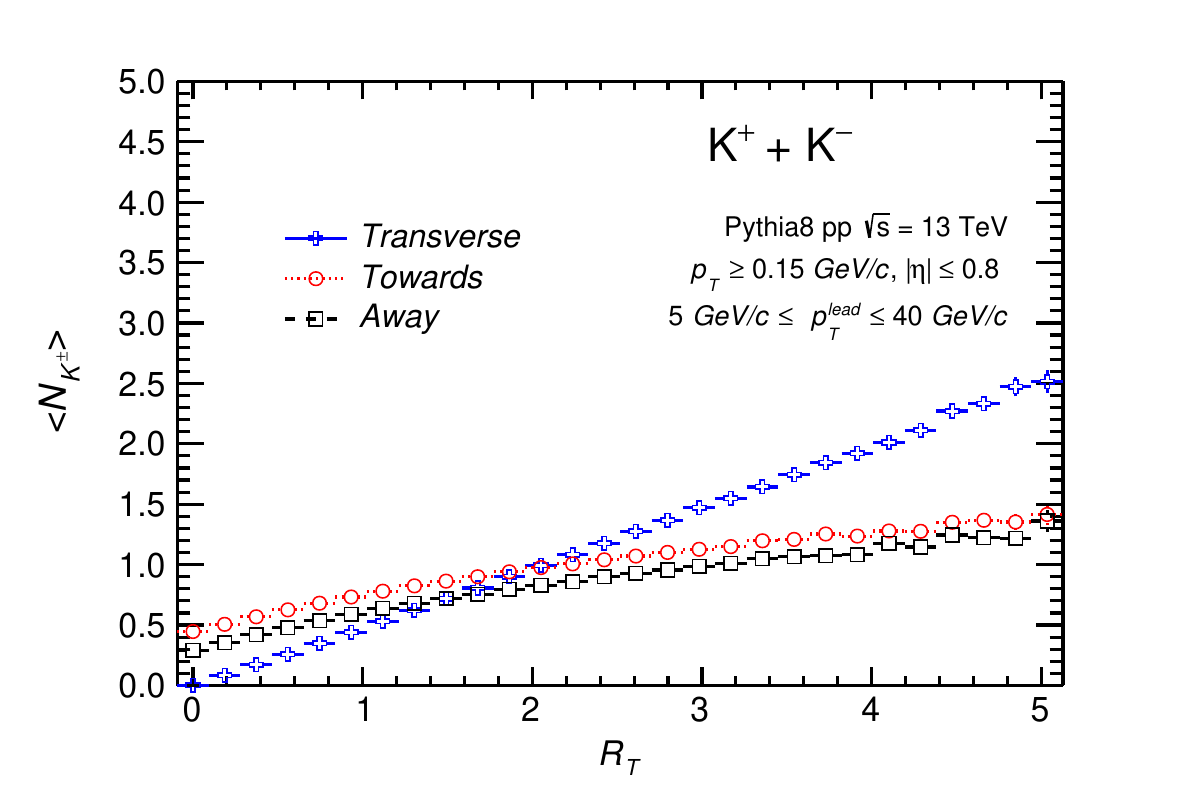}
\includegraphics[width=0.5\textwidth, height = 0.35\textwidth ]{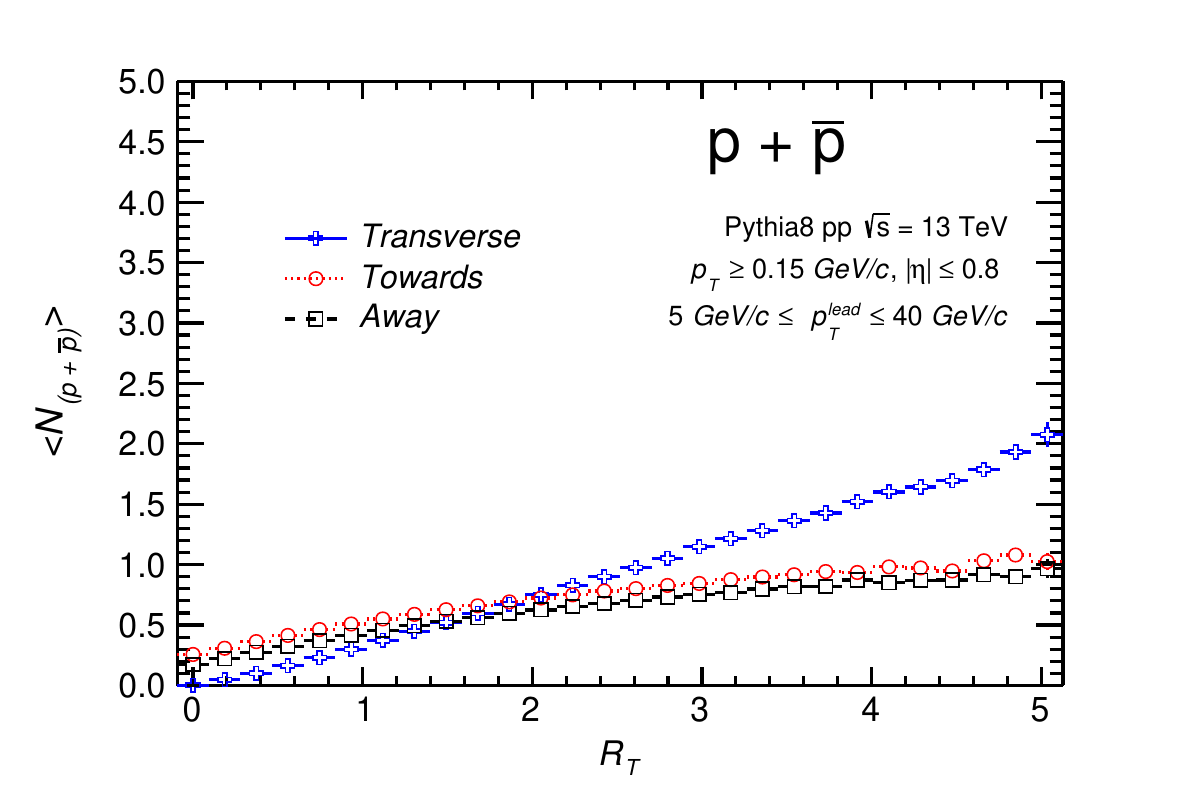}
\caption{The variation of mean multiplicity of pions(upper panel), kaons(middle panel) and protons(bottom panel) with $R_{T}$ in p$-$p collisions at $\sqrt{s} =$ 13 TeV for the three topological regions.}
\label{imeanmultRT}
\end{figure}

\begin{figure}[!h]
\centering
\includegraphics[width=0.5\textwidth, height = 0.35\textwidth ]{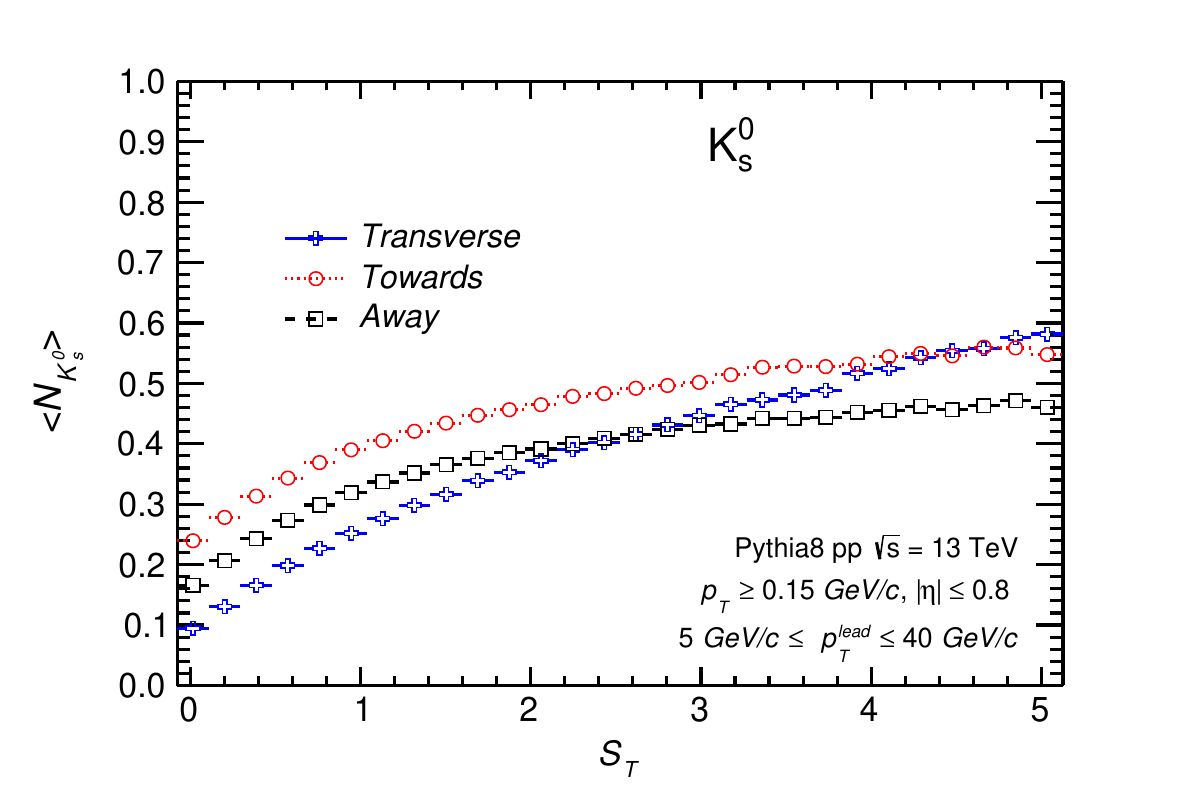}
\includegraphics[width=0.5\textwidth, height = 0.35\textwidth ]{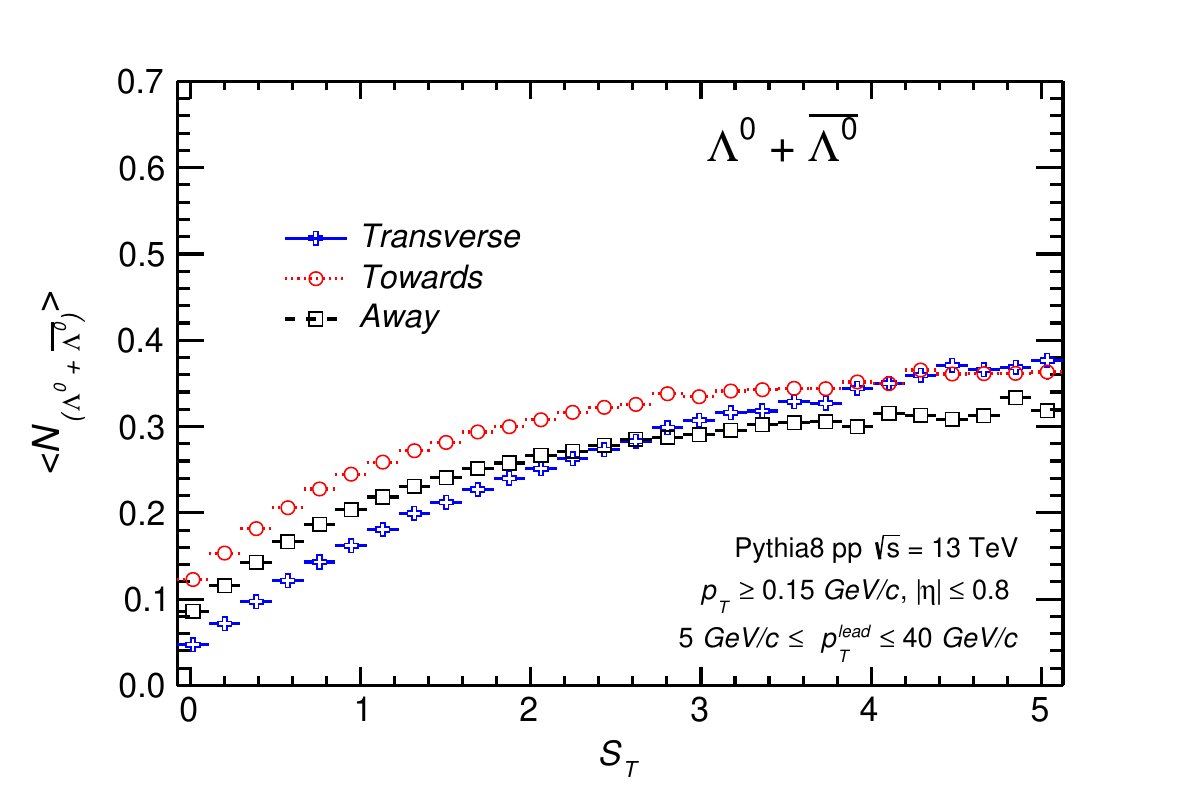}
\caption{The variation of mean number of $K^{0}_{S}$ (upper panel) and $\Lambda^{0}$ (lower panel) with  $S_{T}$ in p$-$p collisions at $\sqrt{s} =$ 13 TeV for the three topological regions.}
\label{meanmultv0ST}
\end{figure}

\begin{figure}[!h]
\centering
\includegraphics[width=0.5\textwidth, height = 0.35\textwidth ]{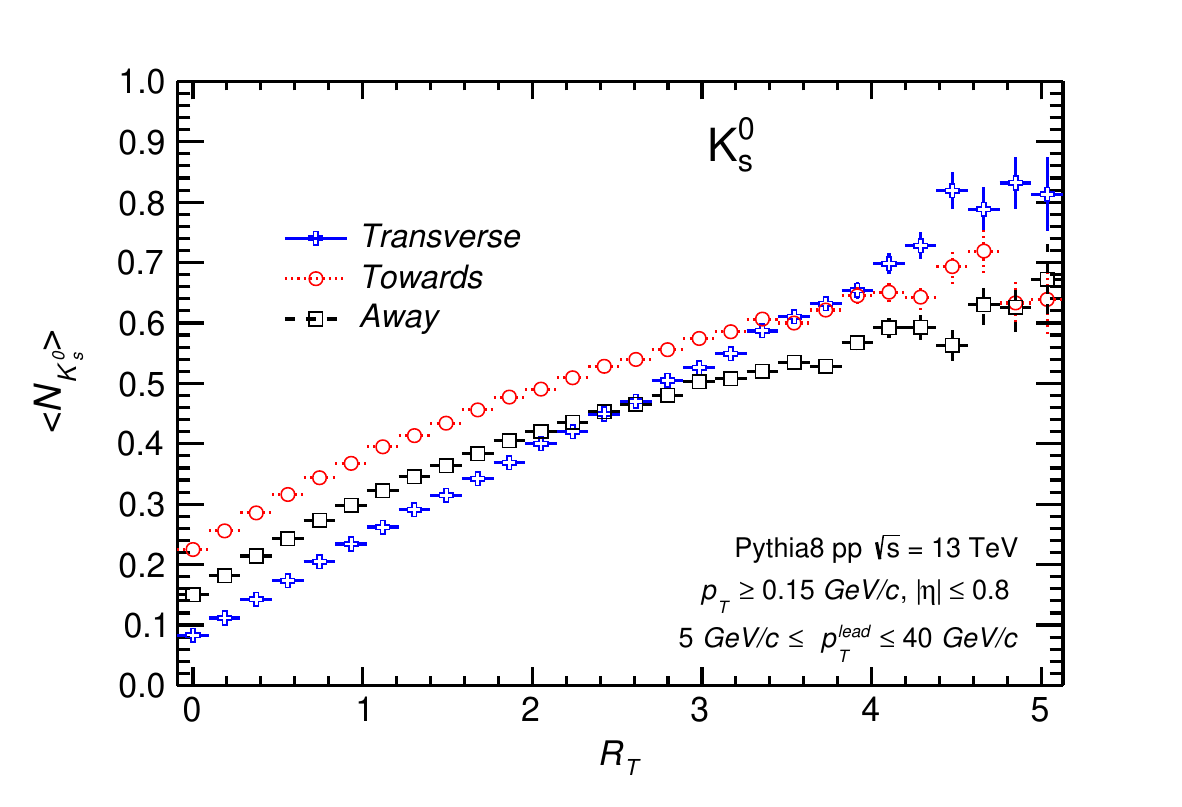}
\includegraphics[width=0.5\textwidth, height = 0.35\textwidth ]{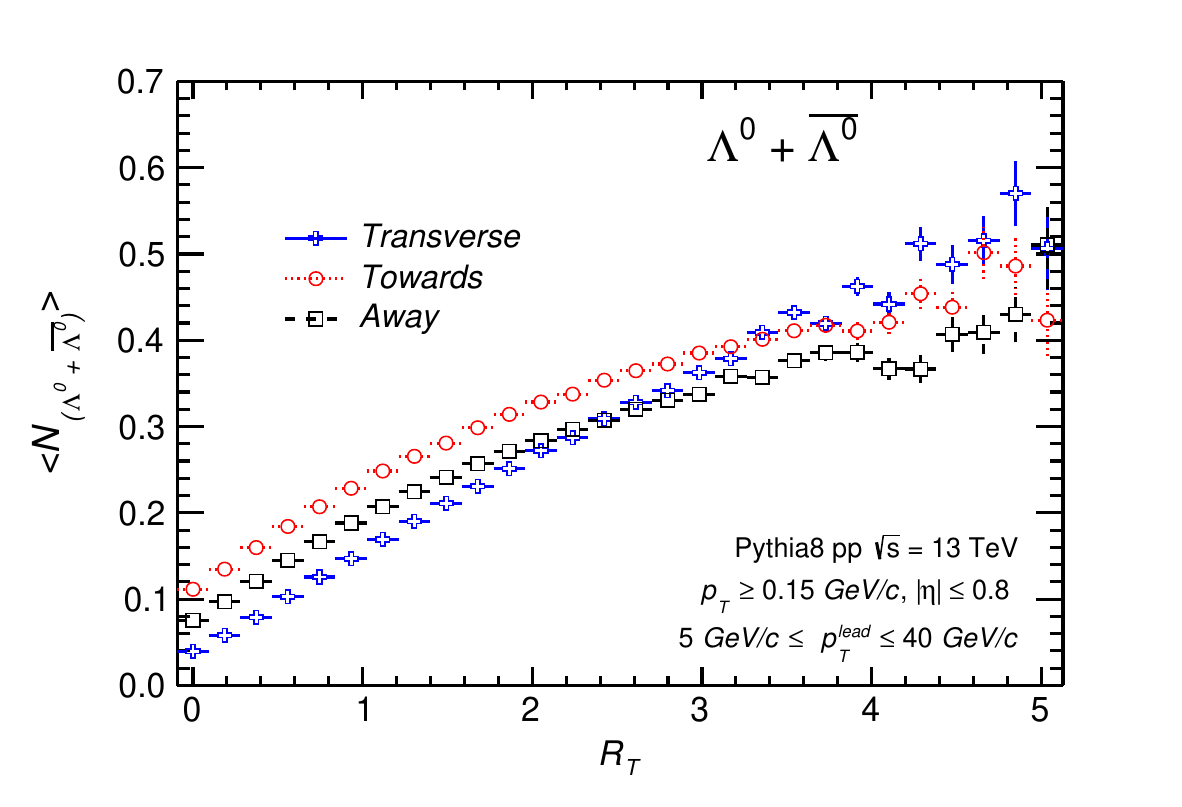}
\caption{The variation of mean number of $K^{0}_{S}$ (upper panel) and $\Lambda^{0}$ (lower panel) with  $R_{T}$ in p$-$p collisions at $\sqrt{s} =$ 13 TeV for the three topological regions.}
\label{meanmultv0RT}
\end{figure}

Figure  \ref{meanptST}  shows the mean transverse momentum ($<p_{T}>$) of the charged particles as a function of $S_{T} $( and $R_{T}$ ) in the three topological regions. In the toward region, $<p_{T}>$  of the charged particles is largest for low $R_{T}$ and $S_{T}$ regime due to dominance of jets fragmenting into numerous particles in this region. It slowly decreases and saturates for $S_{T}$ (and $R_{T}$) values $>$ 1.5.  In the away side region which is dominated by away-side jet, the $<p_{T}>$ slightly increases with $R_{T}$ and $S_{T}$. However, in the transverse region, $<p_{T}>$ values are lower than toward and away regions for lower ranges of $S_{T}$ and $R_{T}$. It increases with $R_{T}$ and $S_{T}$ and the increase can be attributed to dominance of UE activity. The increase is steeper in case of $S_{T}$ and can be seen as a consequence of the auto-correlation effect arising due the structure of the observable.  Therefore, it is worth observing the variation of mean charged particle multiplicity ($<N_{ch}>$) as a function of $S_{T}$. Figure \ref{meanmultST} shows the evolution of $<N_{ch}>$  as a function of $S_{T} $( and $R_{T}$ ) in the three topological regions. Here, one clearly observes the  $<N_{ch}>$  in toward region is consistently higher than away region for the considered range of $S_{T}$ (and $R_{T}$). However, in the transverse region, for lower values of $S_{T}$ (and $R_{T}$) there is a crossing over with the towards and away region. There is a strong rise in $<N_{ch}>$  in the transverse region for  values of $S_{T} > 1.5$ which signals towards an increase in transverse activity emanating primarily from underlying events. A similar trend is also seen for $R_{T}$ where the increase is  stronger than  $S_{T}$ and can be attributed to the autocorrelation effect. The study was also performed for different identified particles to see the evolution of their multiplicity with $S_{T}$ (and $R_{T})$. Figures \ref{imeanmultST} and \ref{imeanmultRT} show the variation of mean multiplicity of pions, kaons and protons as a function of $S_{T}$ and $R_{T}$, respectively. The sensitivities for different topological regions is also shown. The trend is similar to that observed for charged multiplicity case.  However, neutral $V_{0}$ particles like $K_{S}^{0}$ and $\Lambda^{0}$, do not exhibit autocorrelation effects as both $S_{T}$ and $R_{T}$ are defined in terms of charged particles.  One can observe that  values for toward region remain consistently higher than the other regions up to $S_{T}$ and $R_{T}$  $\sim$ 2.0 indicating the dominance of $V_{0}$ production by fragmenting jets. The gradual increase for transverse region with increase in $S_{T}$ values indicates that  hadronisation mechanism which is sensitive to the non-perturbative effects affects 
the strangeness and baryon content of the final state.

\subsection{Transverse momentum distributions}
\begin{figure}[!h]
\includegraphics[width=0.5\textwidth, height = 0.35\textwidth ]{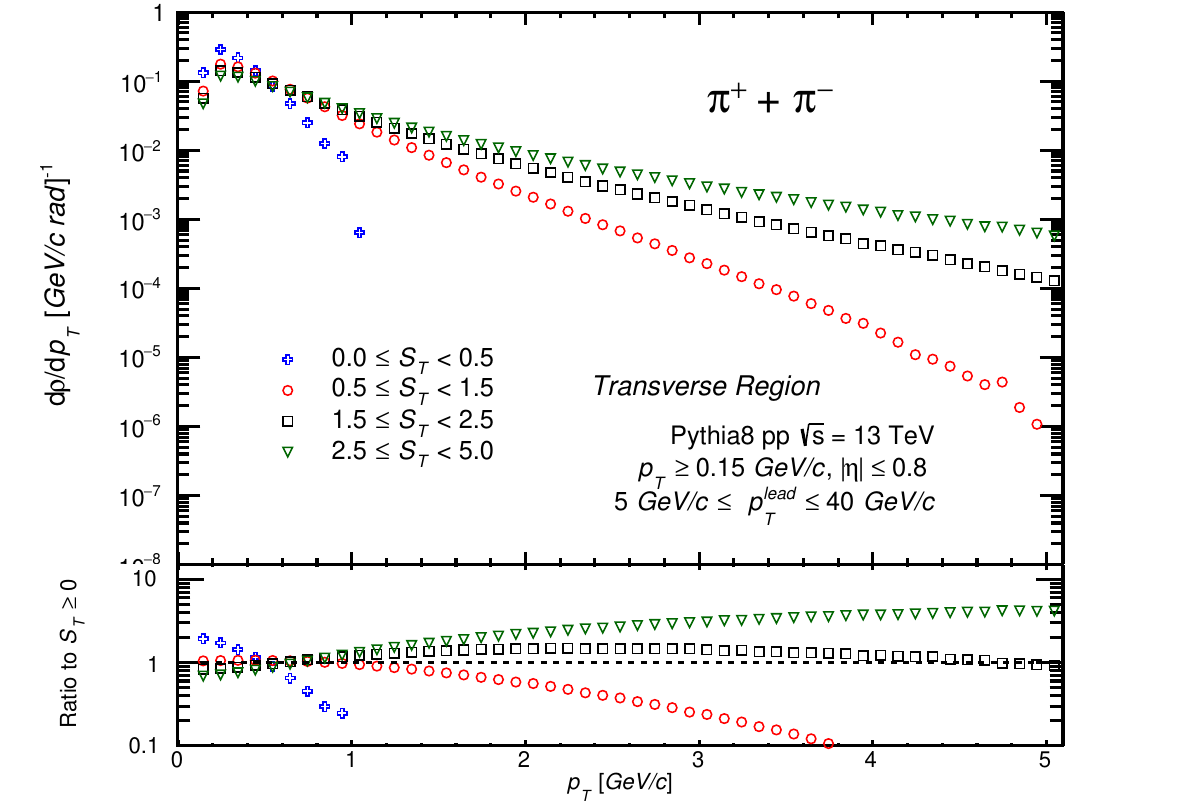}
\includegraphics[width=0.5\textwidth, height = 0.35\textwidth ]{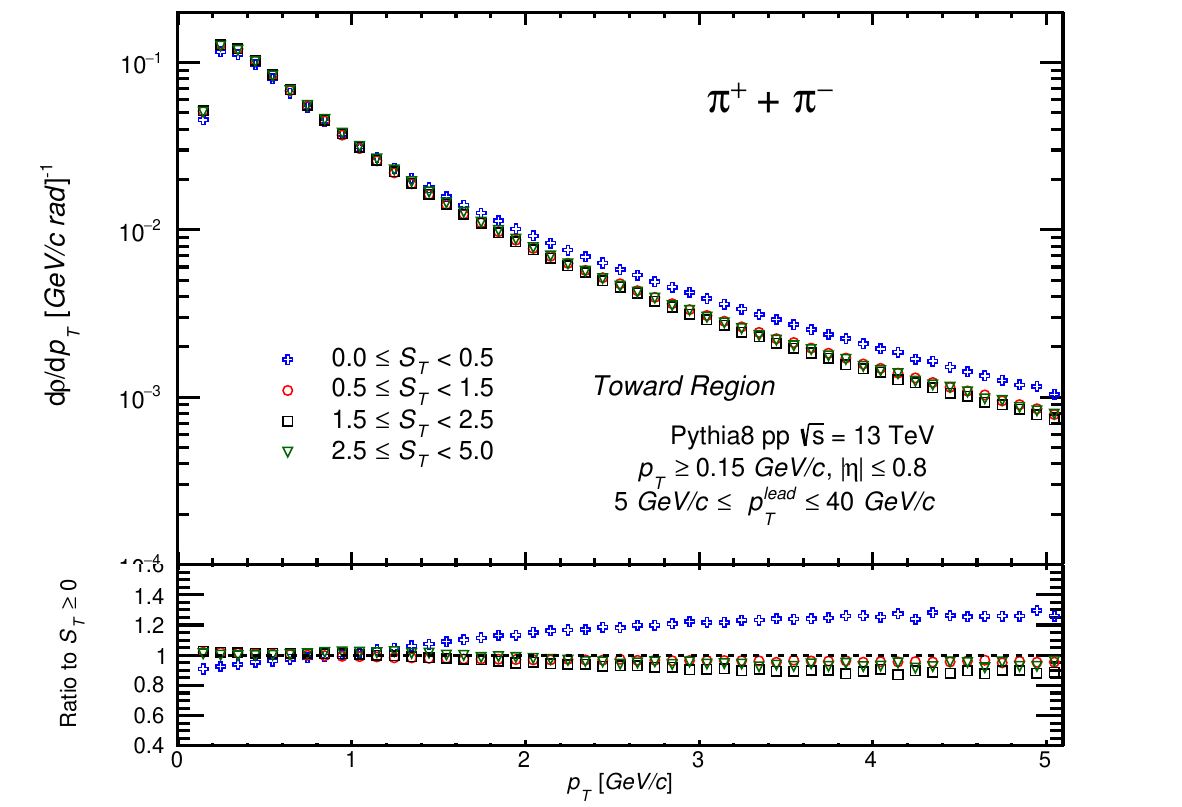}
\includegraphics[width=0.5\textwidth, height = 0.35\textwidth ]{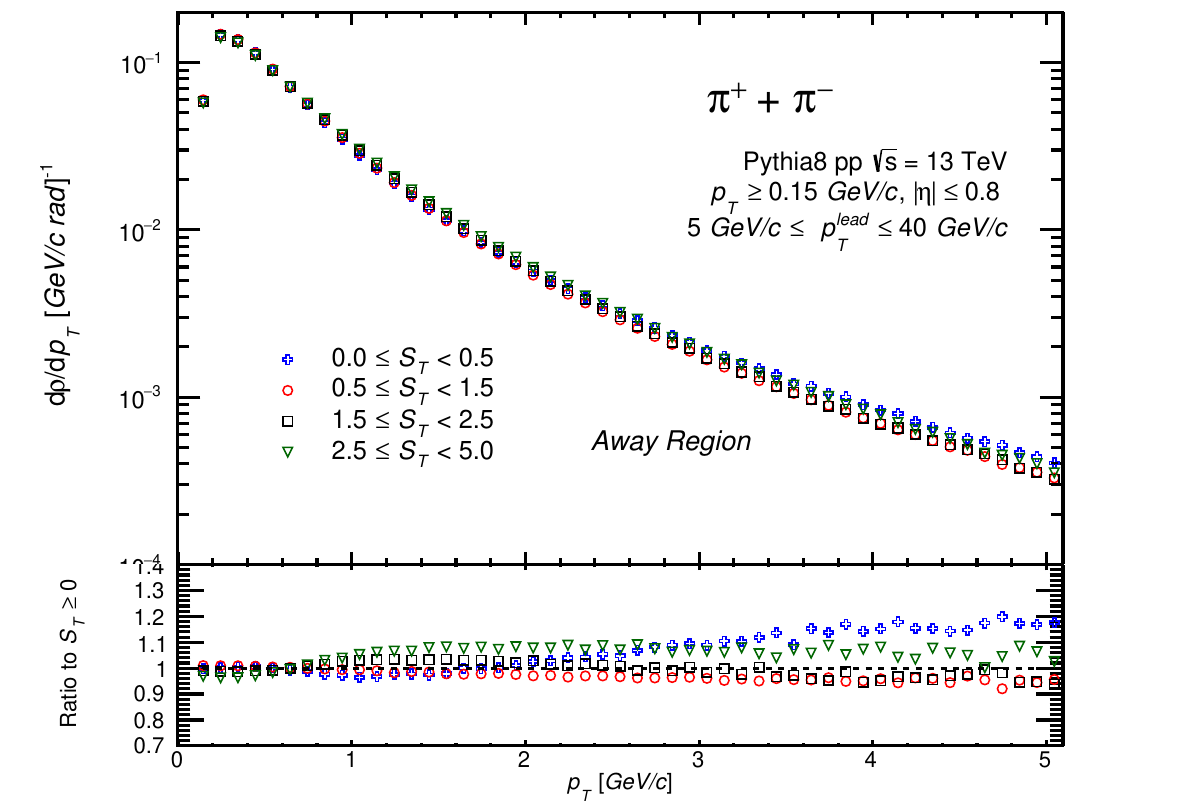}
\caption{The $p_{T}$ spectra of charged pions in transverse, toward and away regions for different $S_{T}$ classes in p$-$p collisions at $\sqrt{s} =$ 13 TeV.}
\label{pionspectraST}
\end{figure}

\begin{figure}[!h]
\includegraphics[width=0.5\textwidth, height = 0.35\textwidth ]{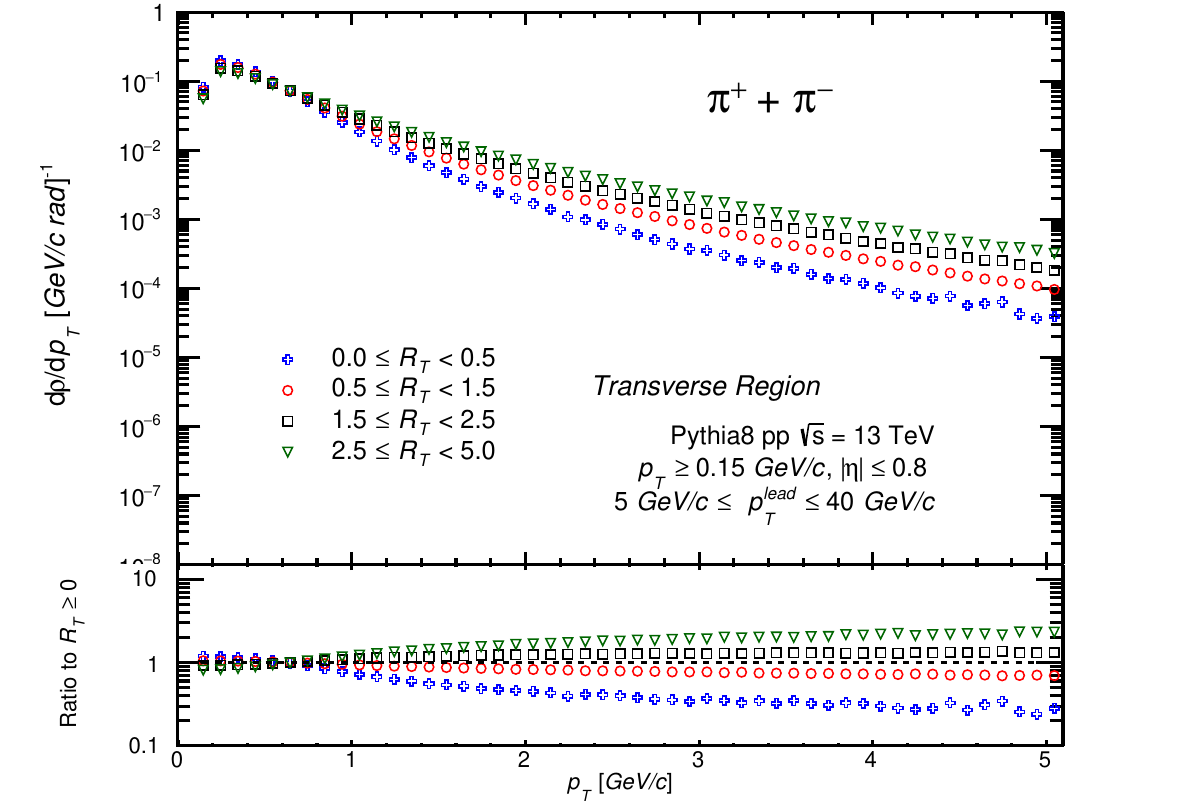} 
\includegraphics[width=0.5\textwidth, height = 0.35\textwidth ]{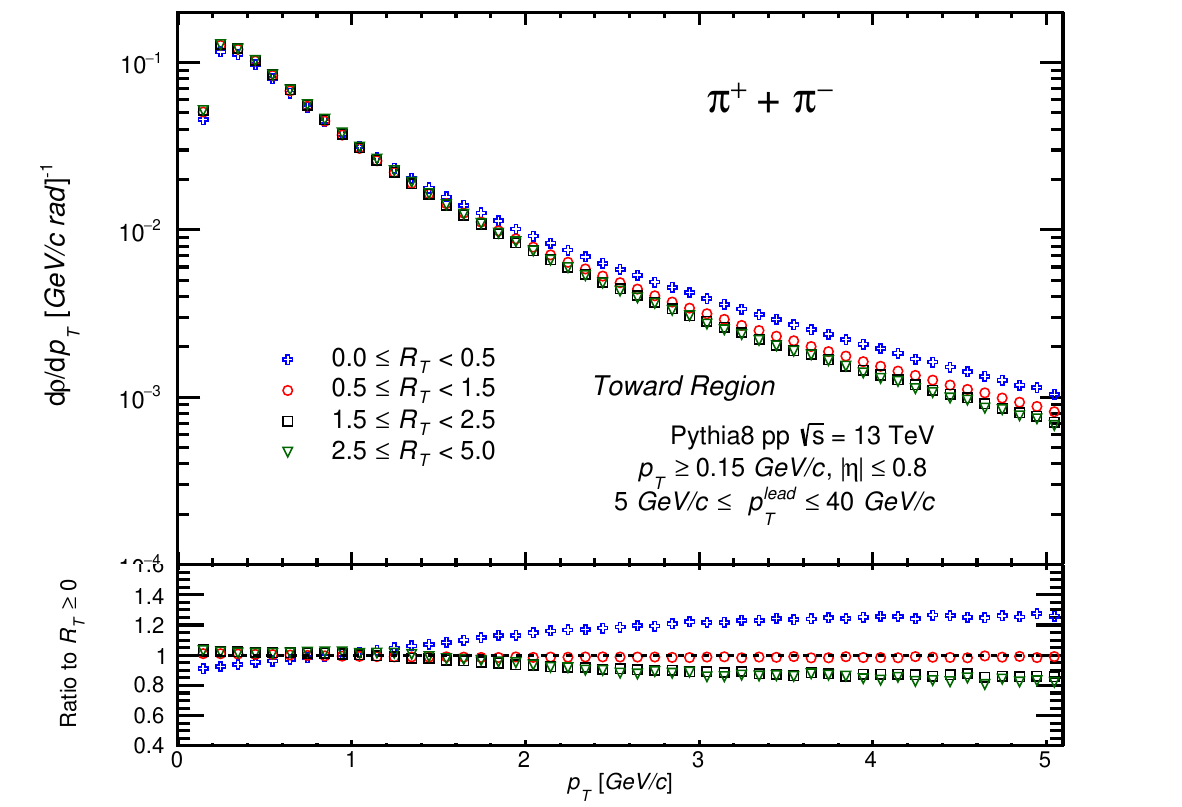}
\includegraphics[width=0.5\textwidth, height = 0.35\textwidth ]{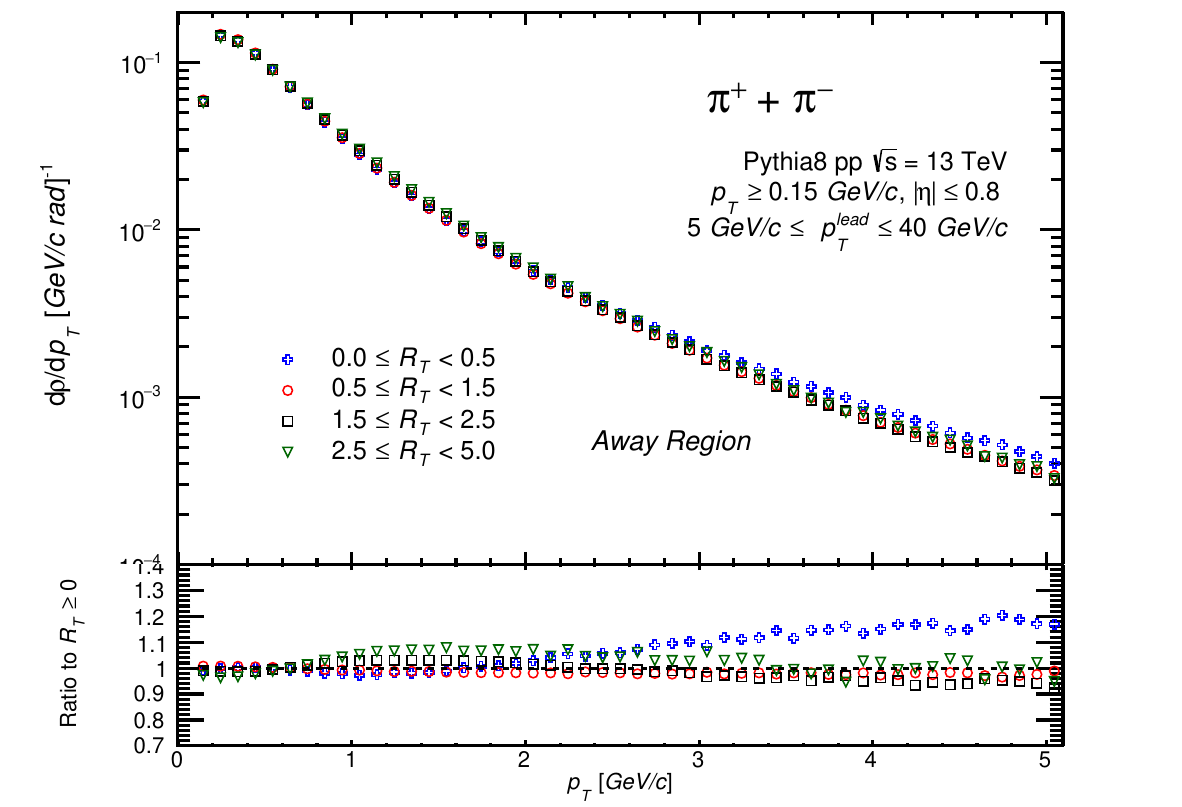}
\caption{The $p_{T}$ spectra of charged pions in transverse, toward and away regions for different $R_{T}$ classes in p$-$p collisions at $\sqrt{s} =$ 13 TeV.}
\label{pionspectraRT}
\end{figure}

\begin{figure}[h!]
\includegraphics[width=0.5\textwidth, height = 0.35\textwidth ]{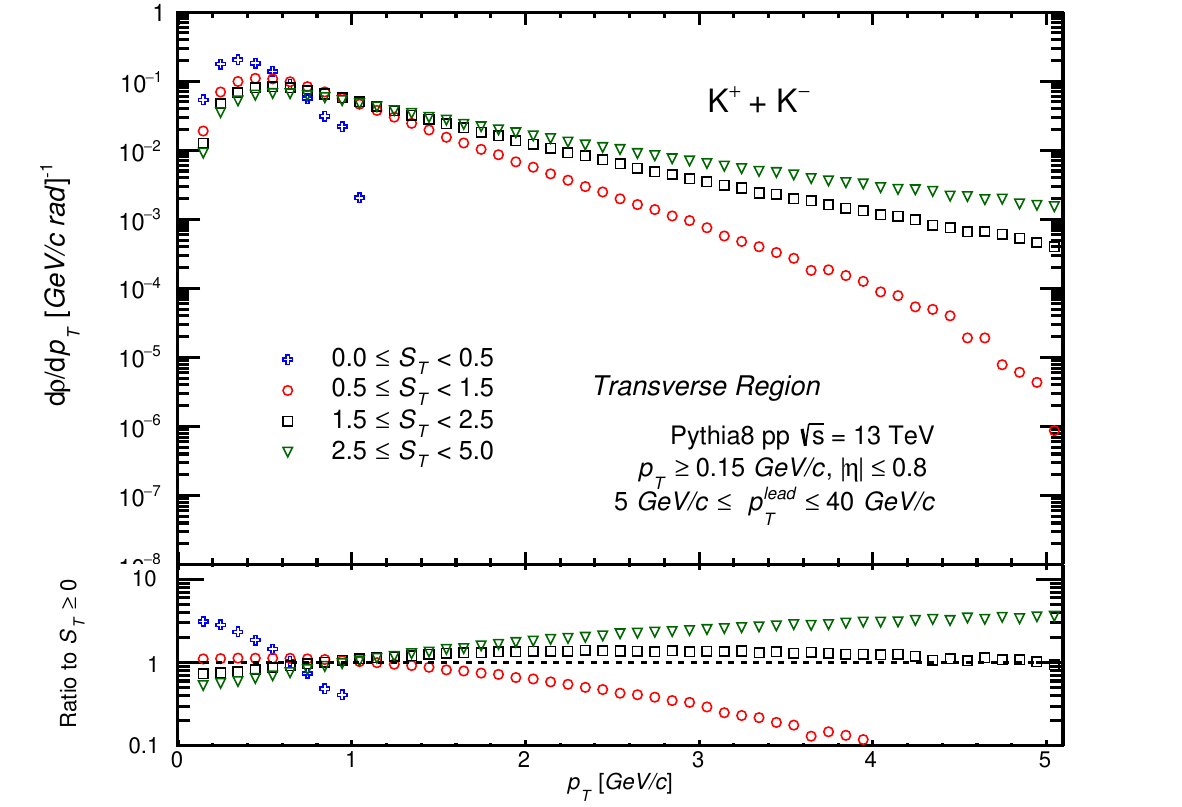} 
\includegraphics[width=0.5\textwidth, height = 0.35\textwidth ]{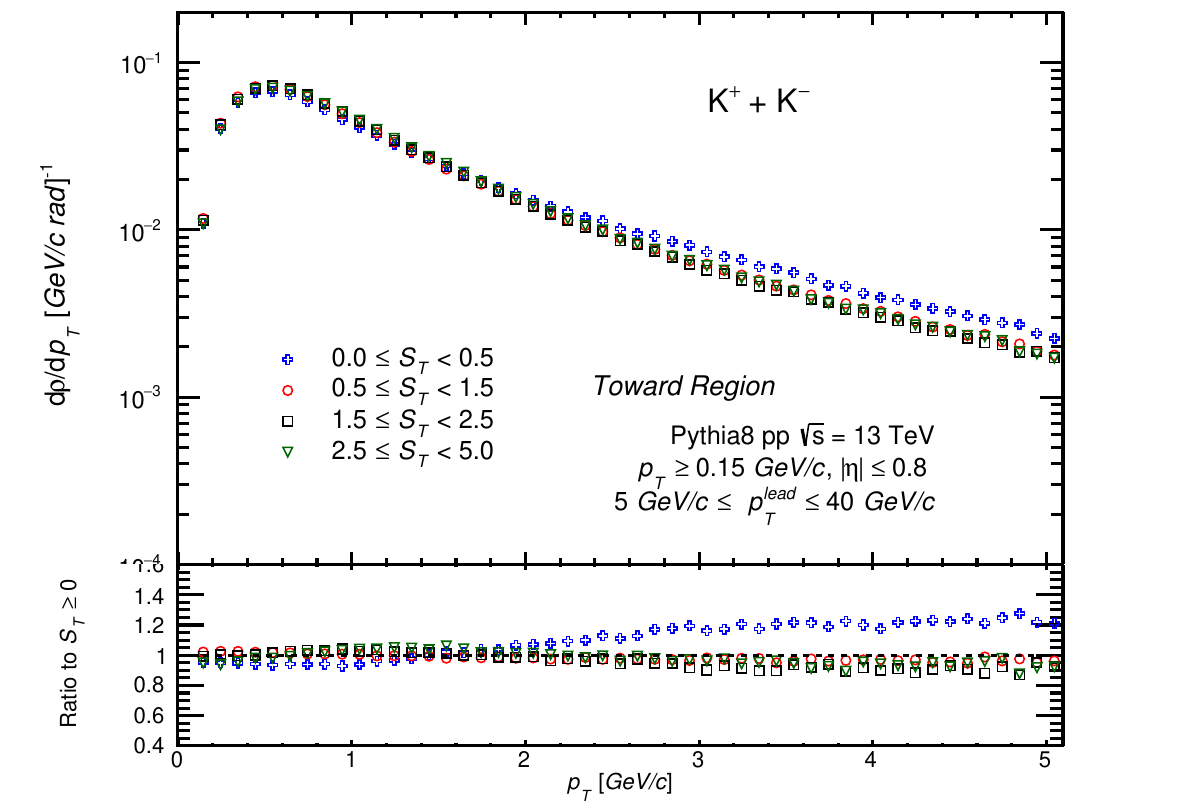}
\includegraphics[width=0.5\textwidth, height = 0.35\textwidth ]{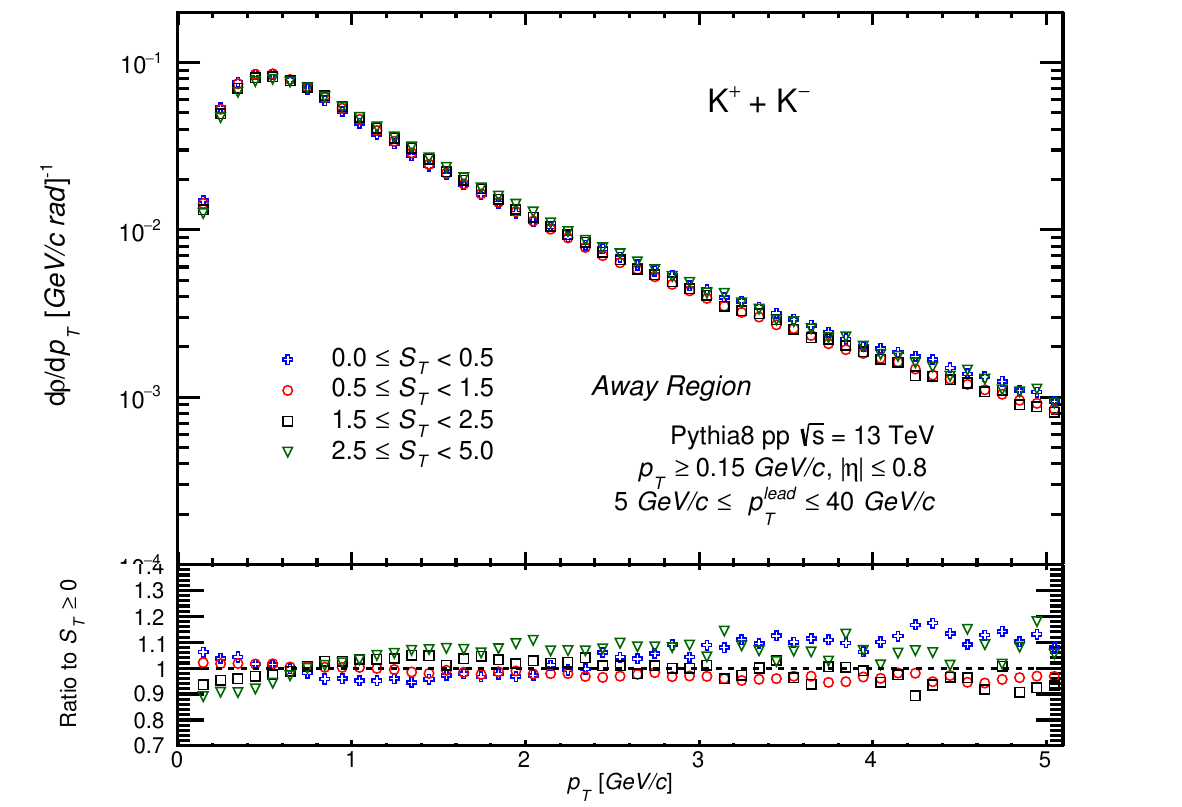}
\caption{The $p_{T}$ spectra of charged kaons in transverse, toward and away regions for different $S_{T}$ classes in p$-$p collisions at $\sqrt{s} =$ 13 TeV.}
\label{kaonspectraST}
\end{figure}

\begin{figure}
\includegraphics[width=0.5\textwidth, height = 0.35\textwidth ]{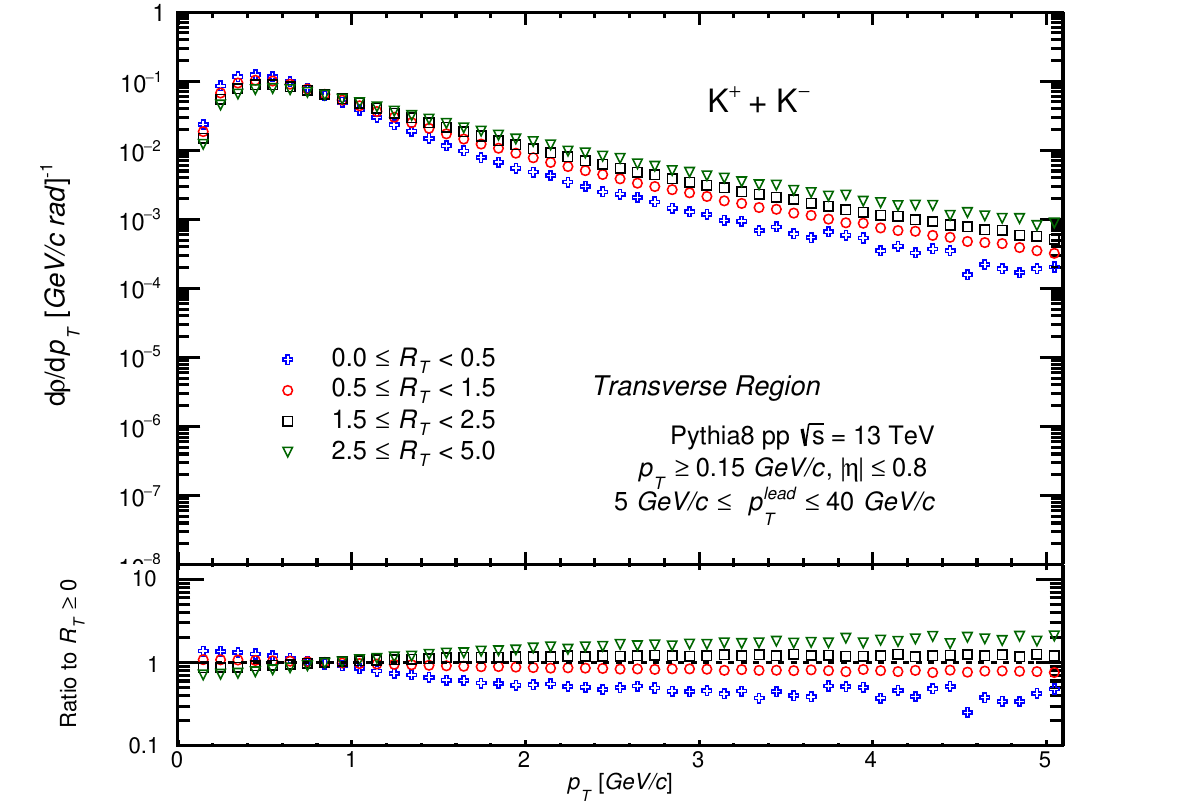} 
\includegraphics[width=0.5\textwidth, height = 0.35\textwidth ]{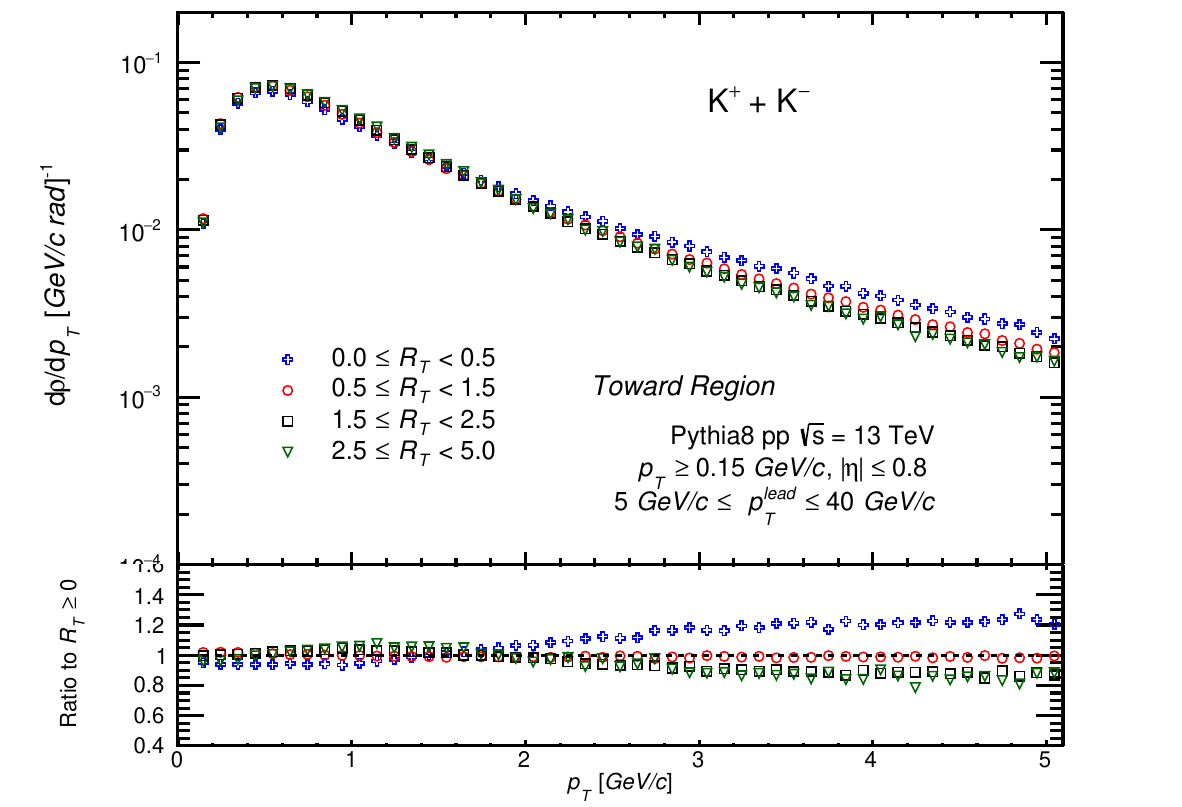}
\includegraphics[width=0.5\textwidth, height = 0.35\textwidth ]{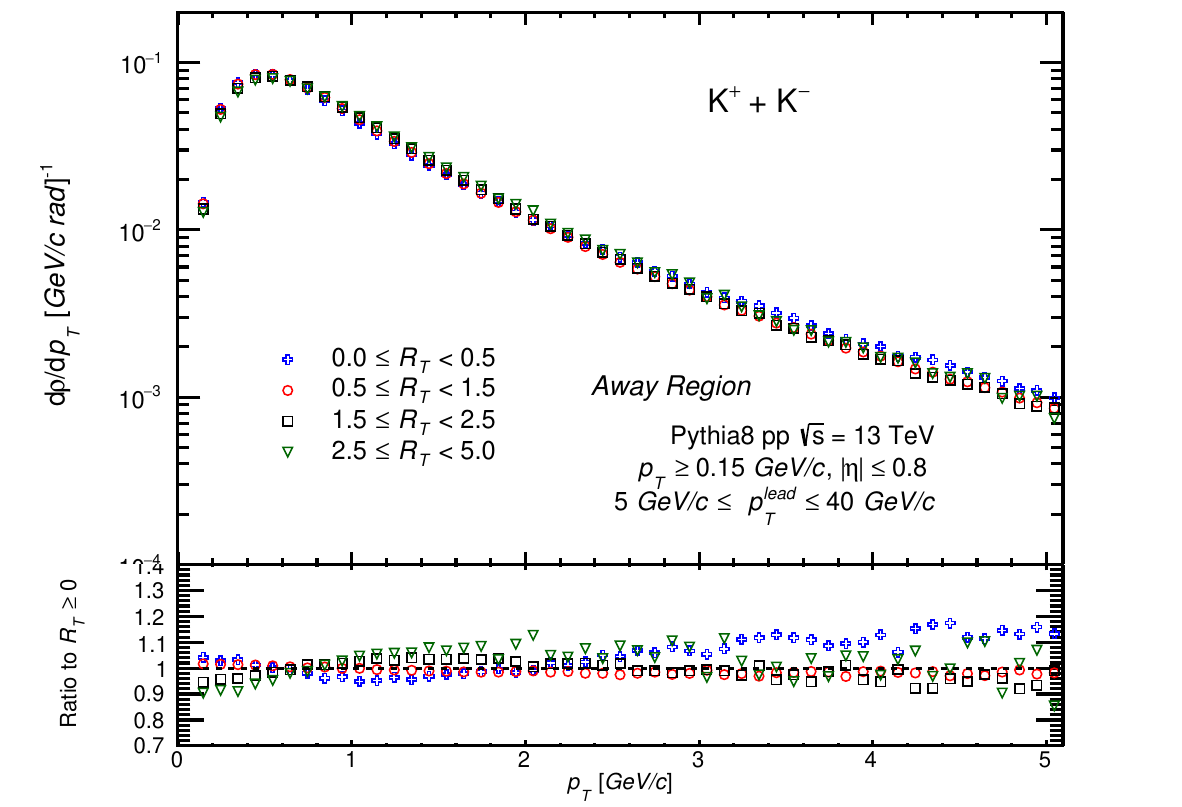}  
\caption{The $p_{T}$ spectra of charged kaons in transverse, toward and away regions for different $R_{T}$ classes in p$-$p collisions at $\sqrt{s} =$ 13 TeV.}
\label{kaonspectraRT}
\end{figure}

\begin{figure}[h!]
\includegraphics[width=0.5\textwidth, height = 0.35\textwidth ]{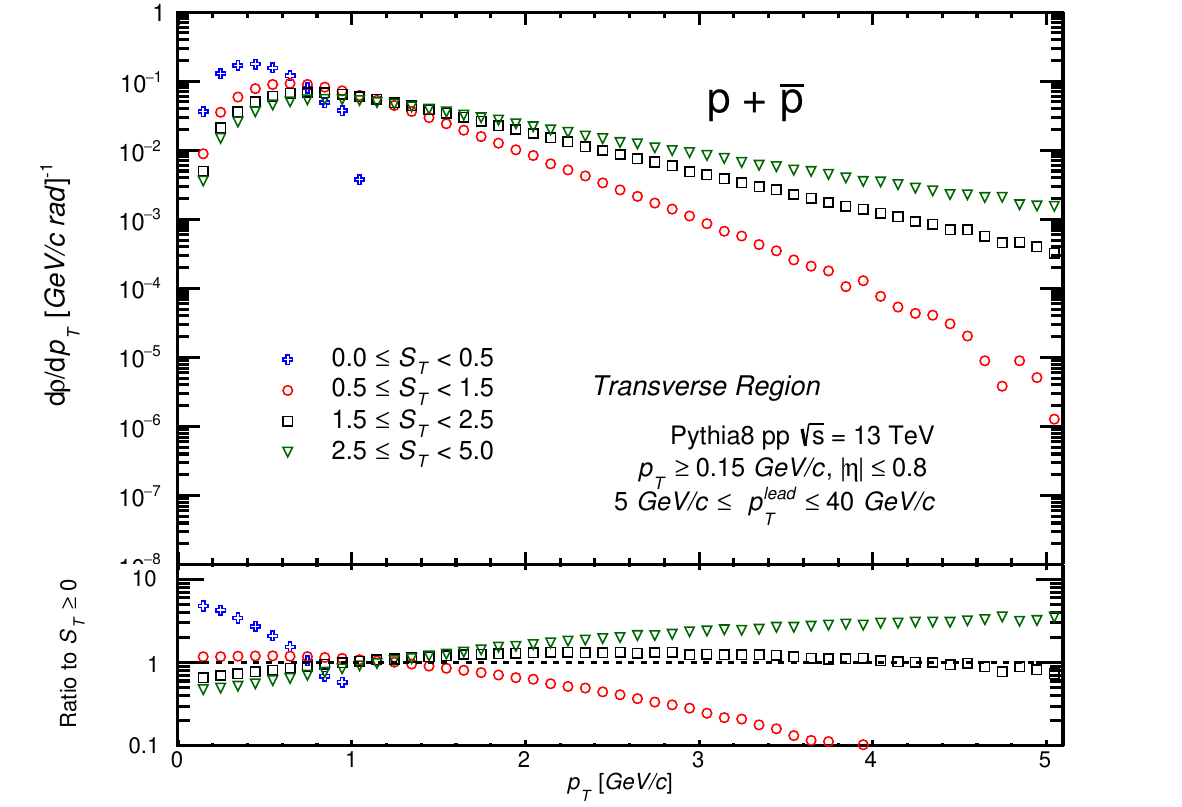}
\includegraphics[width=0.5\textwidth, height = 0.35\textwidth ]{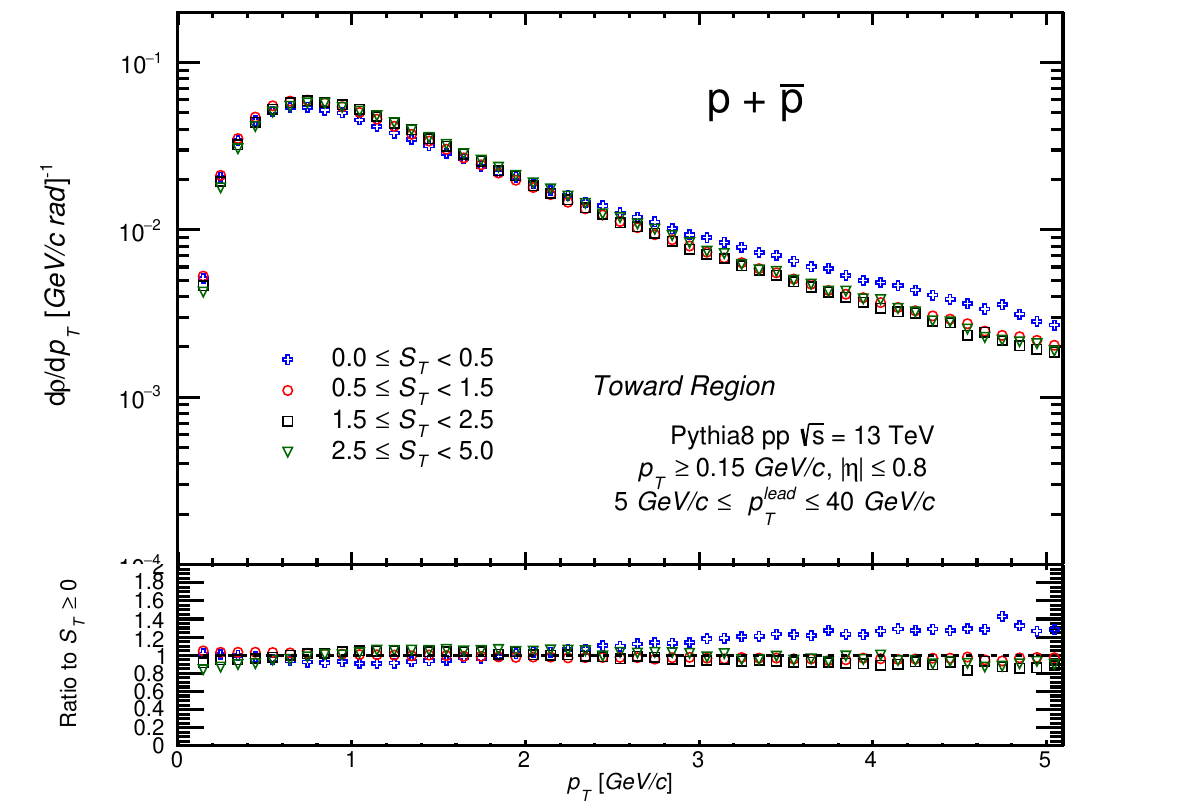} 
\includegraphics[width=0.5\textwidth, height = 0.35\textwidth ]{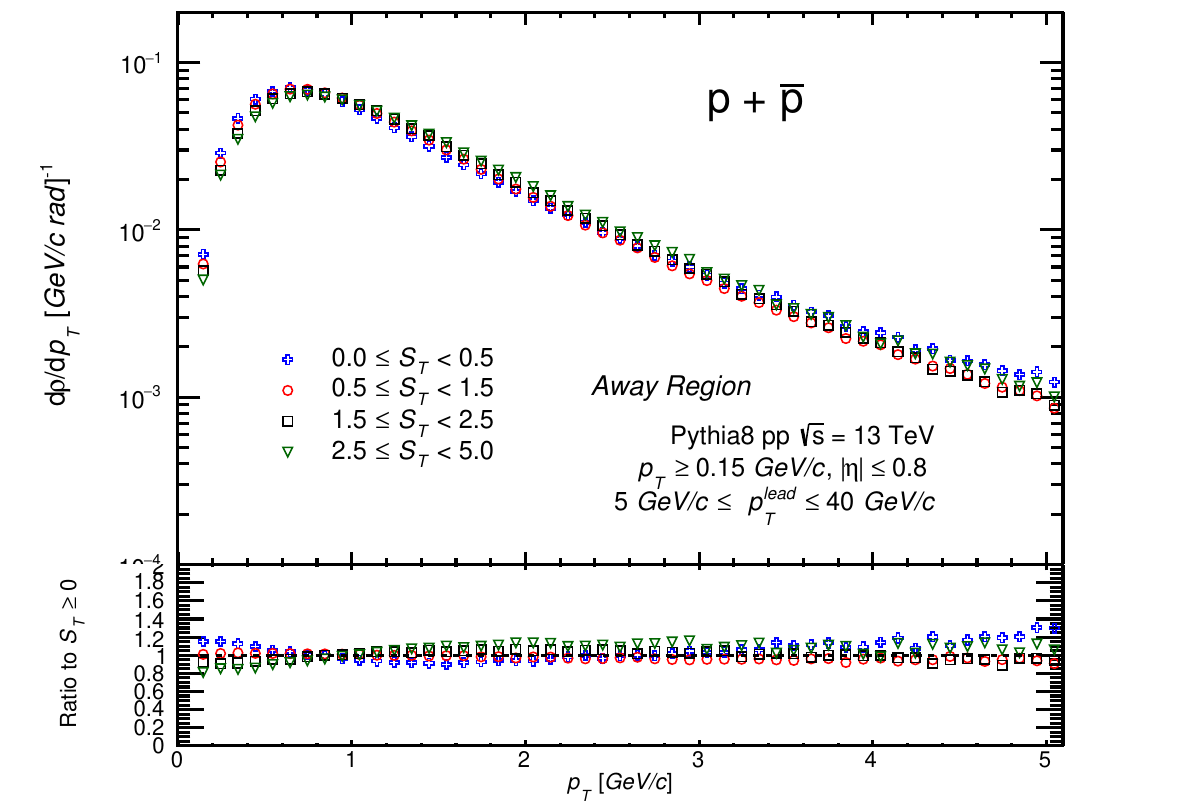} 
\caption{The $p_{T}$ spectra of protons in transverse, toward and away regions for different $S_{T}$ classes in p$-$p collisions at $\sqrt{s} =$ 13 TeV.}
\label{protonspectraRT}
\end{figure}

\begin{figure}
\includegraphics[width=0.5\textwidth, height = 0.35\textwidth ]{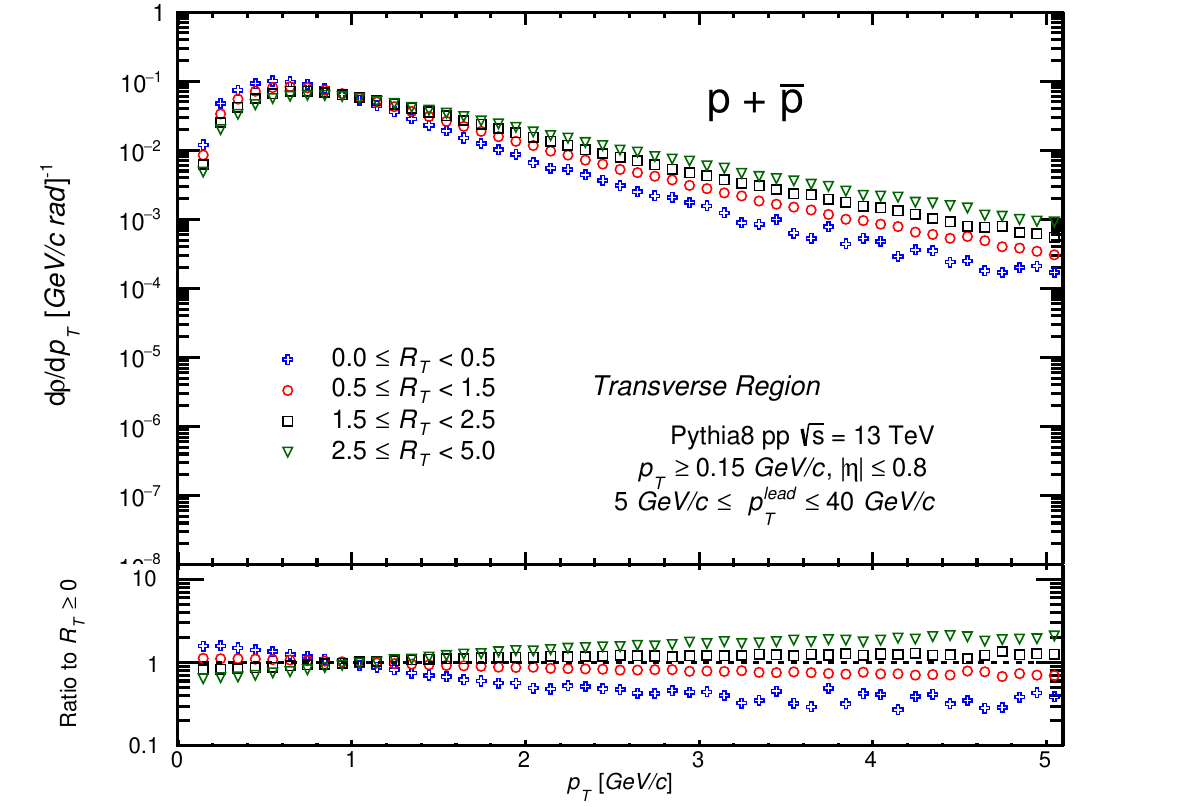}  
\includegraphics[width=0.5\textwidth, height = 0.35\textwidth ]{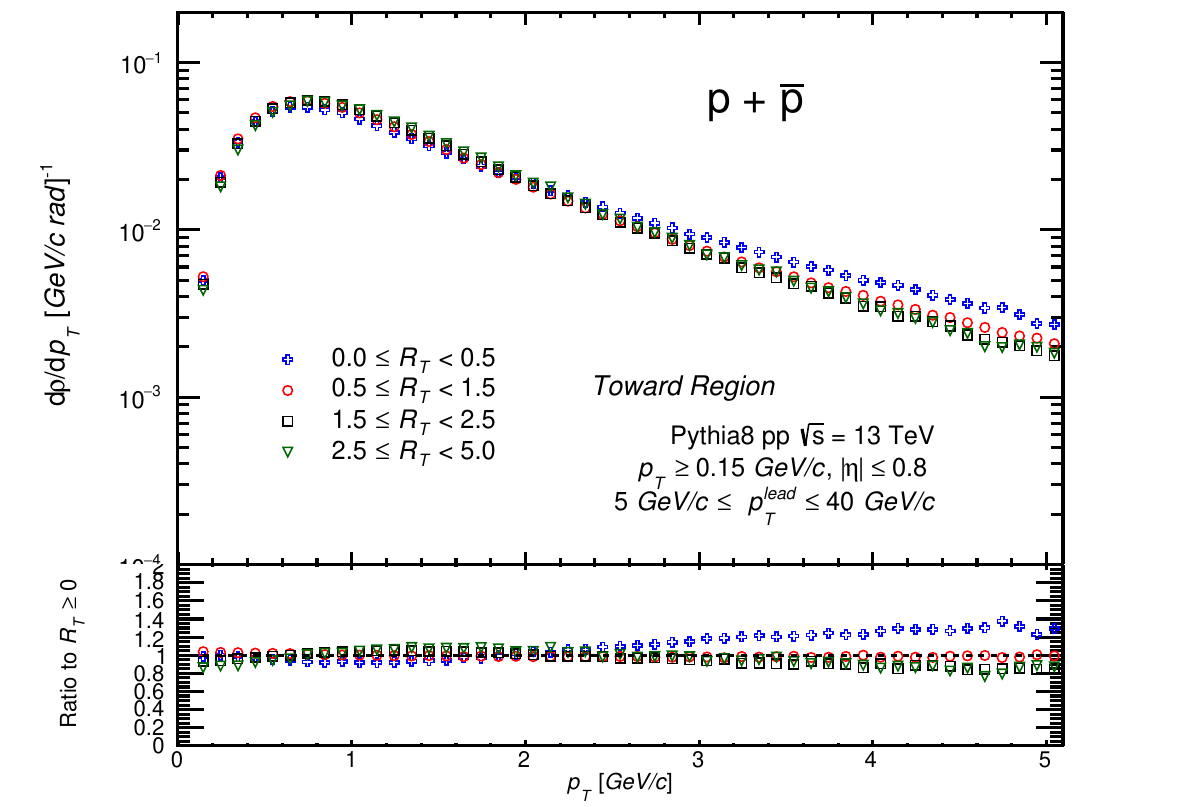}   
\includegraphics[width=0.5\textwidth, height = 0.35\textwidth ]{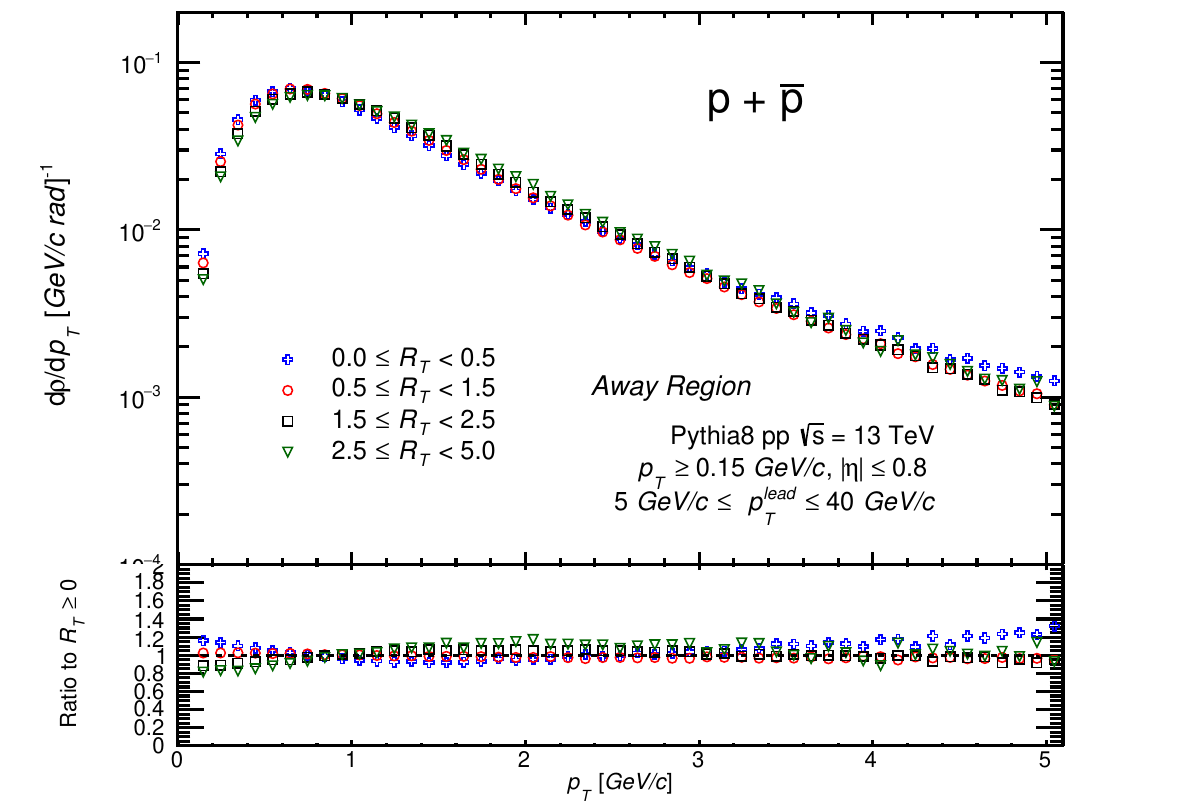}   
\caption{The $p_{T}$ spectra of protons in transverse, toward and away regions for different $R_{T}$ classes in p$-$p collisions at $\sqrt{s} =$ 13 TeV.}	
\label{protonspectraST}
\end{figure}

\begin{figure}[h!]
\includegraphics[width=0.5\textwidth, height = 0.35\textwidth ]{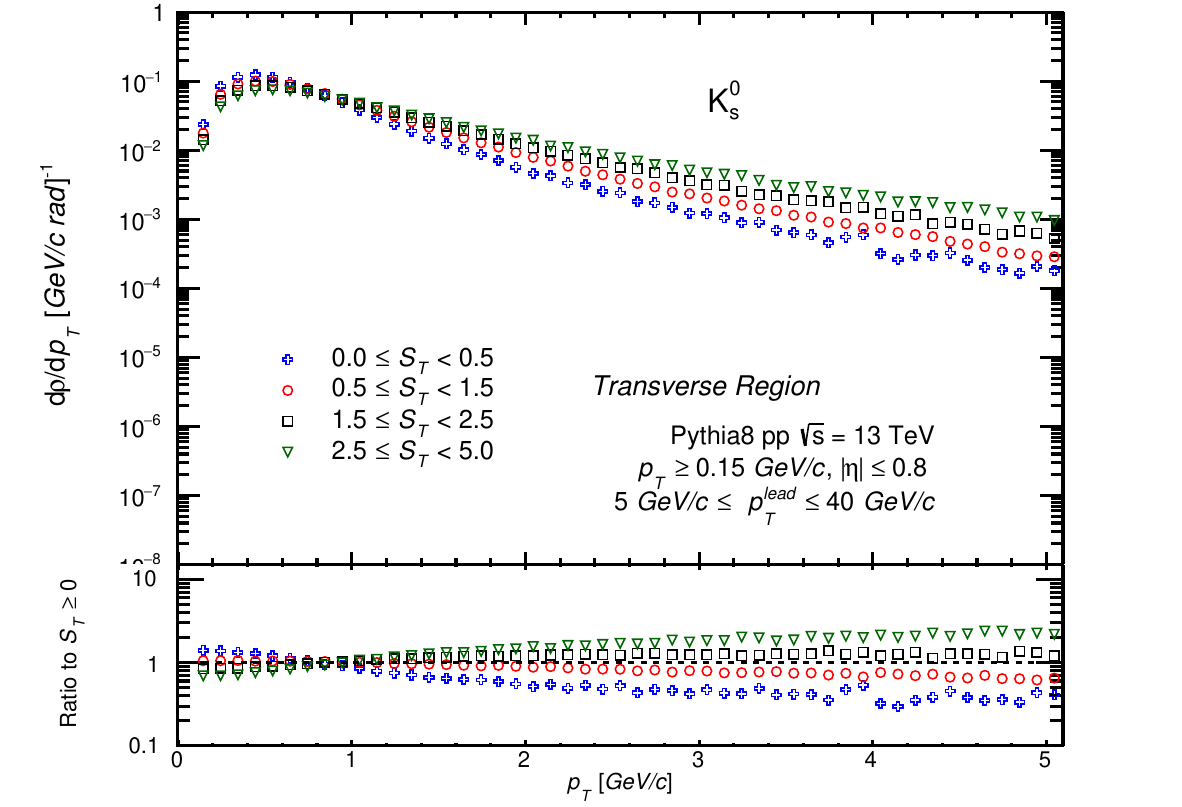} 
\includegraphics[width=0.5\textwidth, height = 0.35\textwidth ]{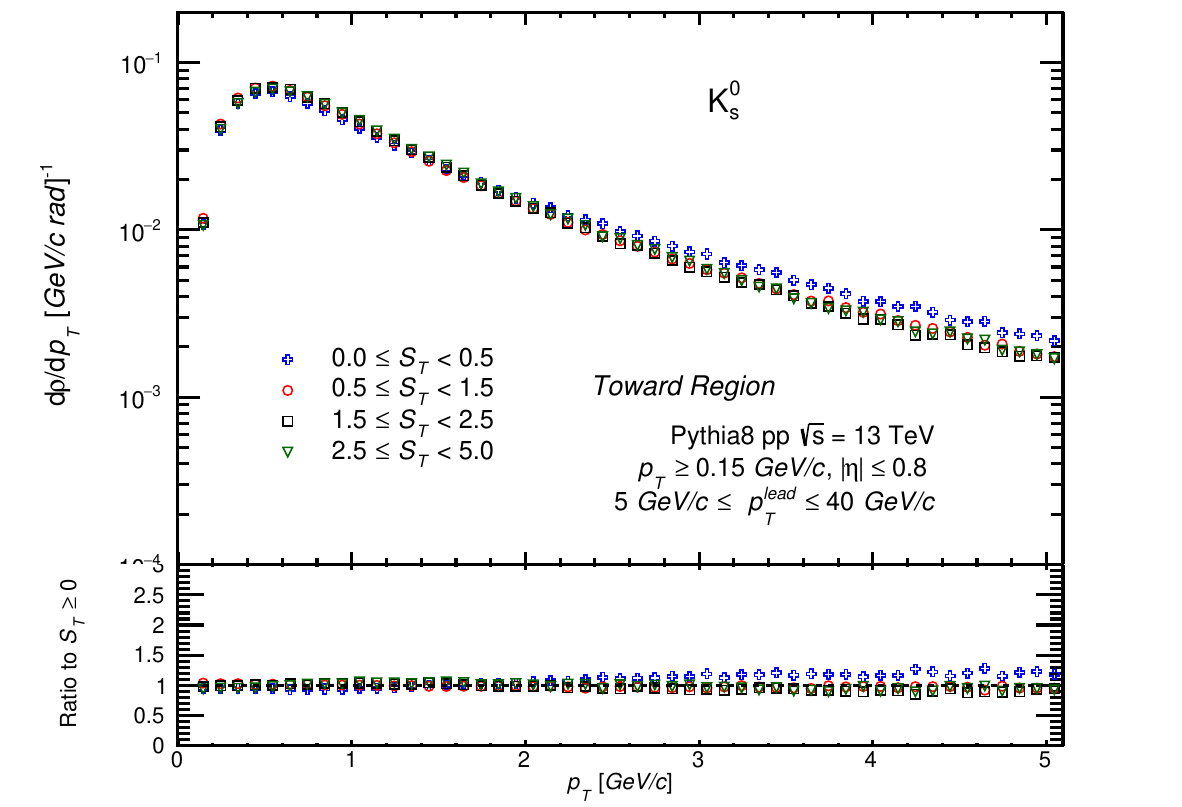} 
\includegraphics[width=0.5\textwidth, height = 0.35\textwidth ]{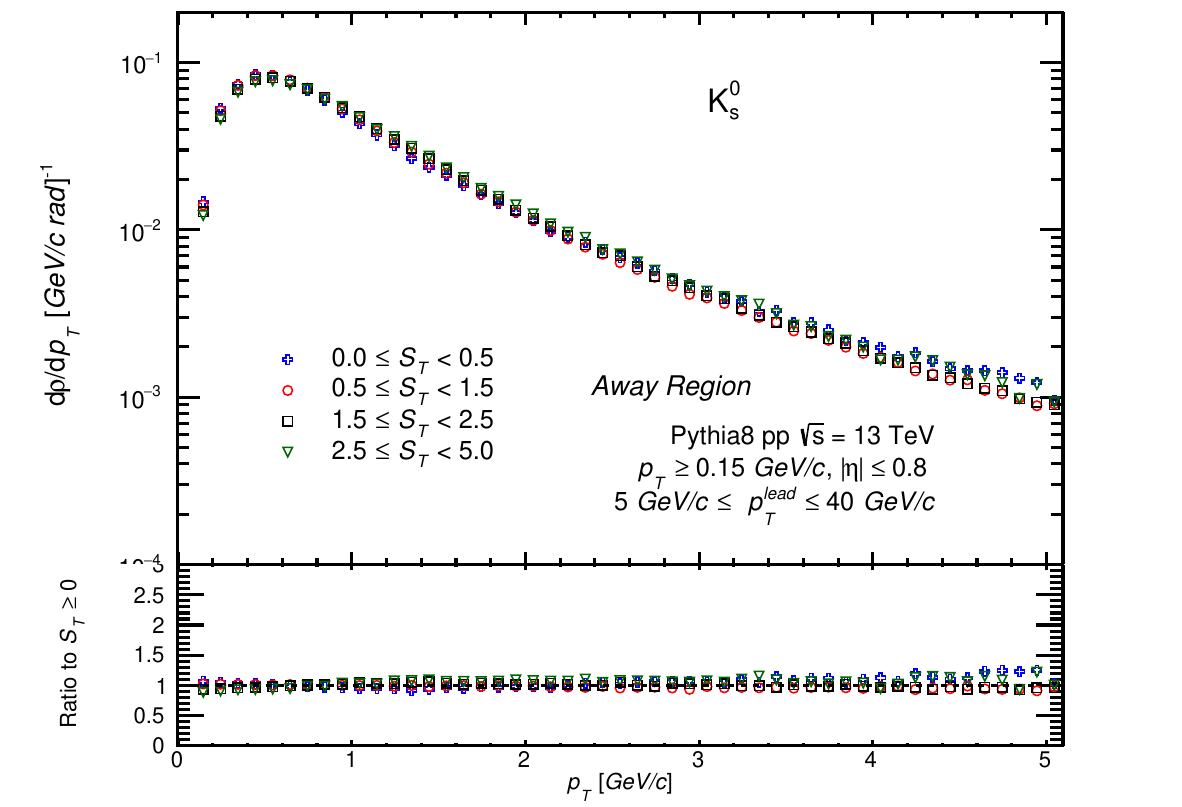} 
\caption{The transverse momentum spectra of $K_{S}^{0}$ particles in transverse, toward and away regions for different $S_{T}$ classes  in p$-$p collisions at $\sqrt{s} =$ 13 TeV.}
\label{kshortspectraRT}
\end{figure}

\begin{figure}
\includegraphics[width=0.5\textwidth, height = 0.35\textwidth ]{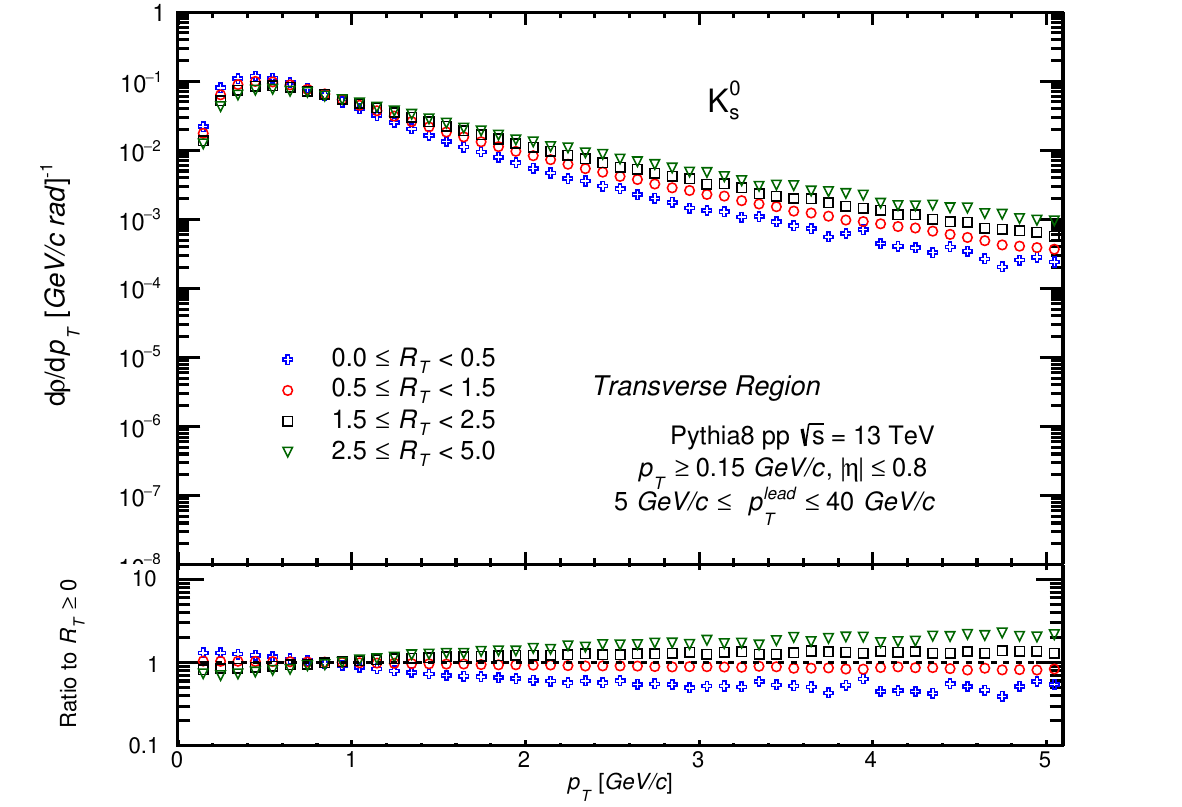}    
\includegraphics[width=0.5\textwidth, height = 0.35\textwidth ]{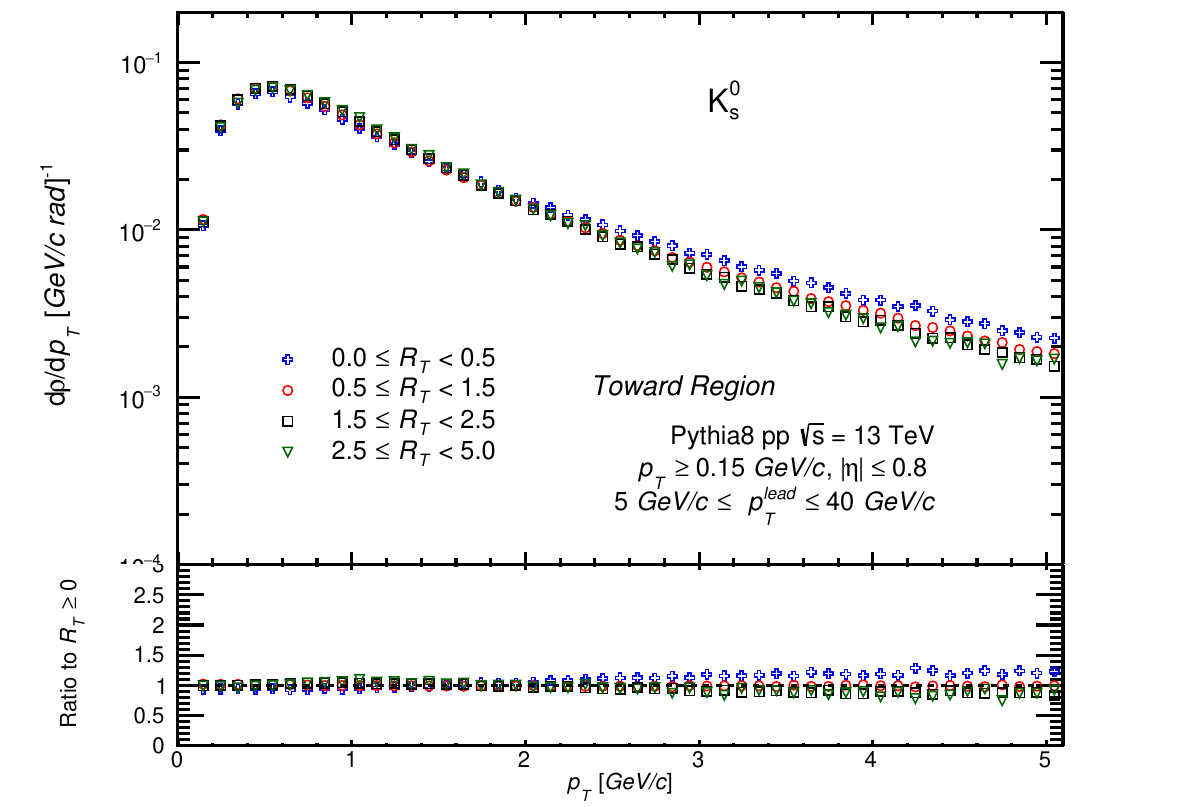}   
\includegraphics[width=0.5\textwidth, height = 0.35\textwidth ]{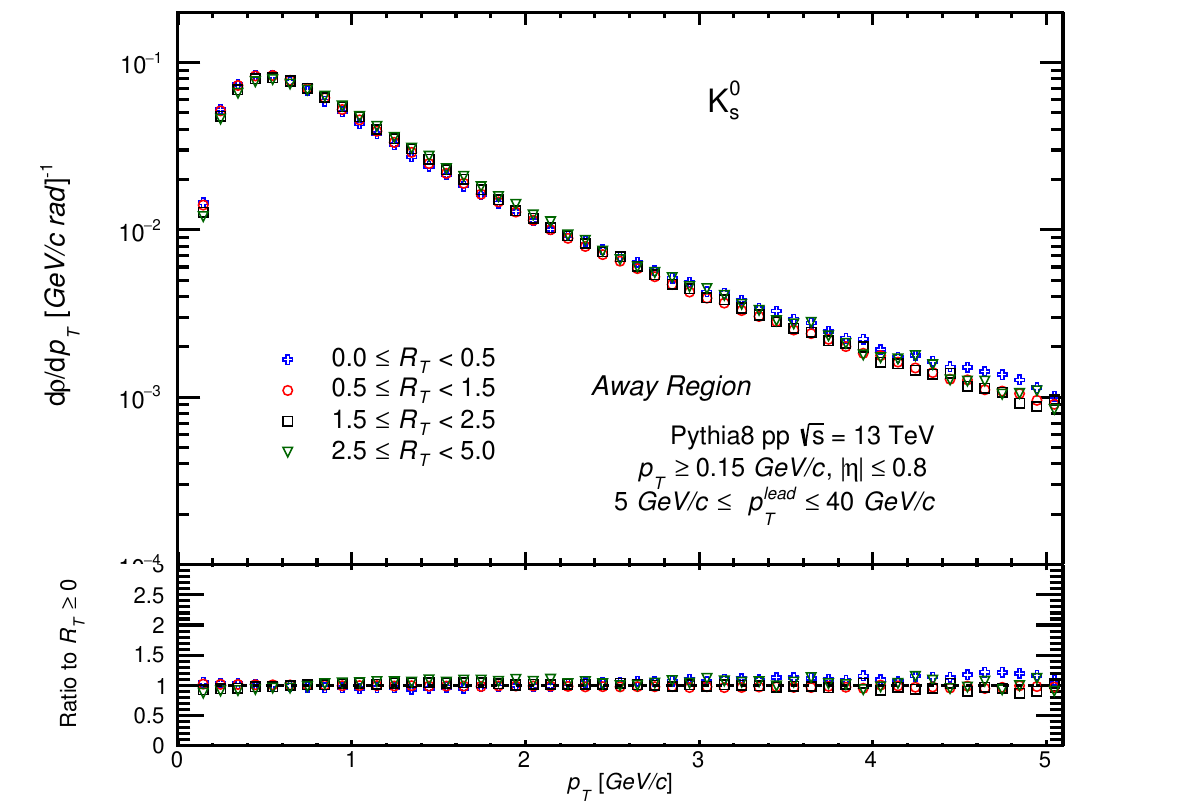}    
\caption{The transverse momentum spectra of $K_{S}^{0}$ particles in transverse, toward and away regions for different $R_{T}$ classes  in p$-$p collisions at $\sqrt{s} =$ 13 TeV.}

\label{kshortspectraST}	
\end{figure}

\begin{figure}[h!]
\includegraphics[width=0.5\textwidth, height = 0.35\textwidth ]{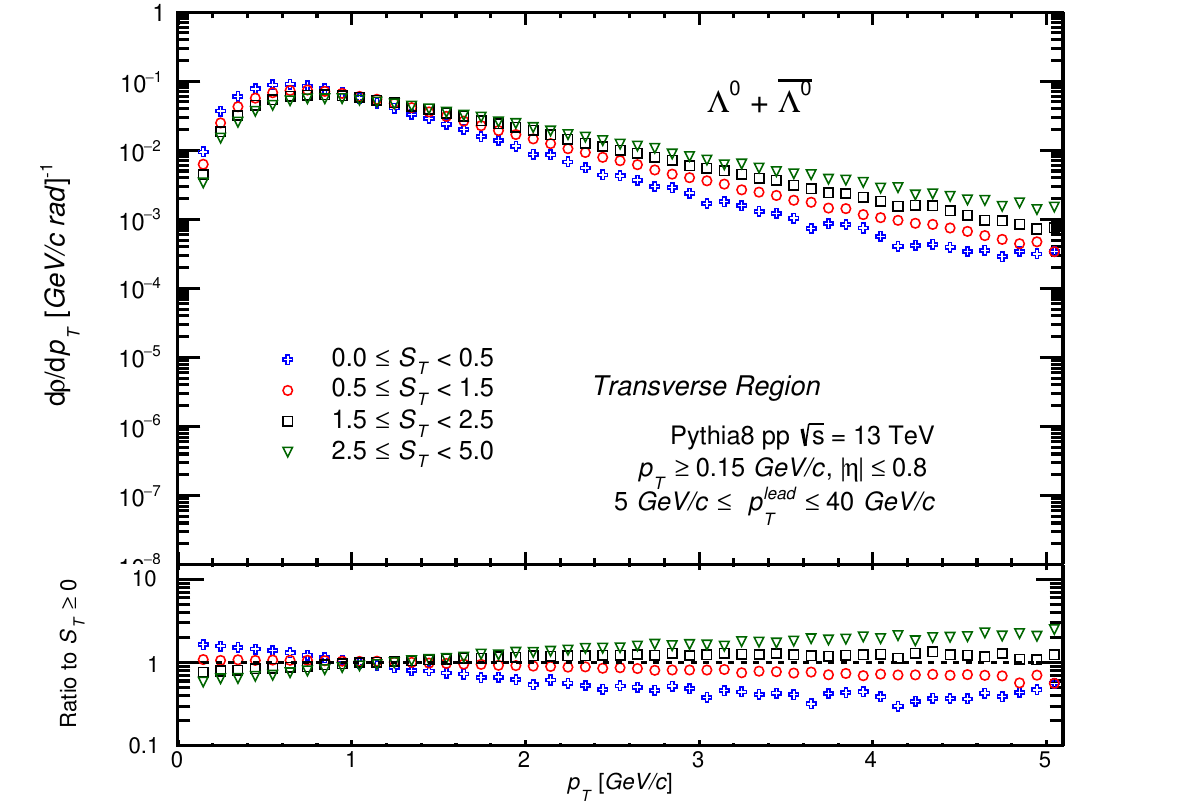}
\includegraphics[width=0.5\textwidth, height = 0.35\textwidth ]{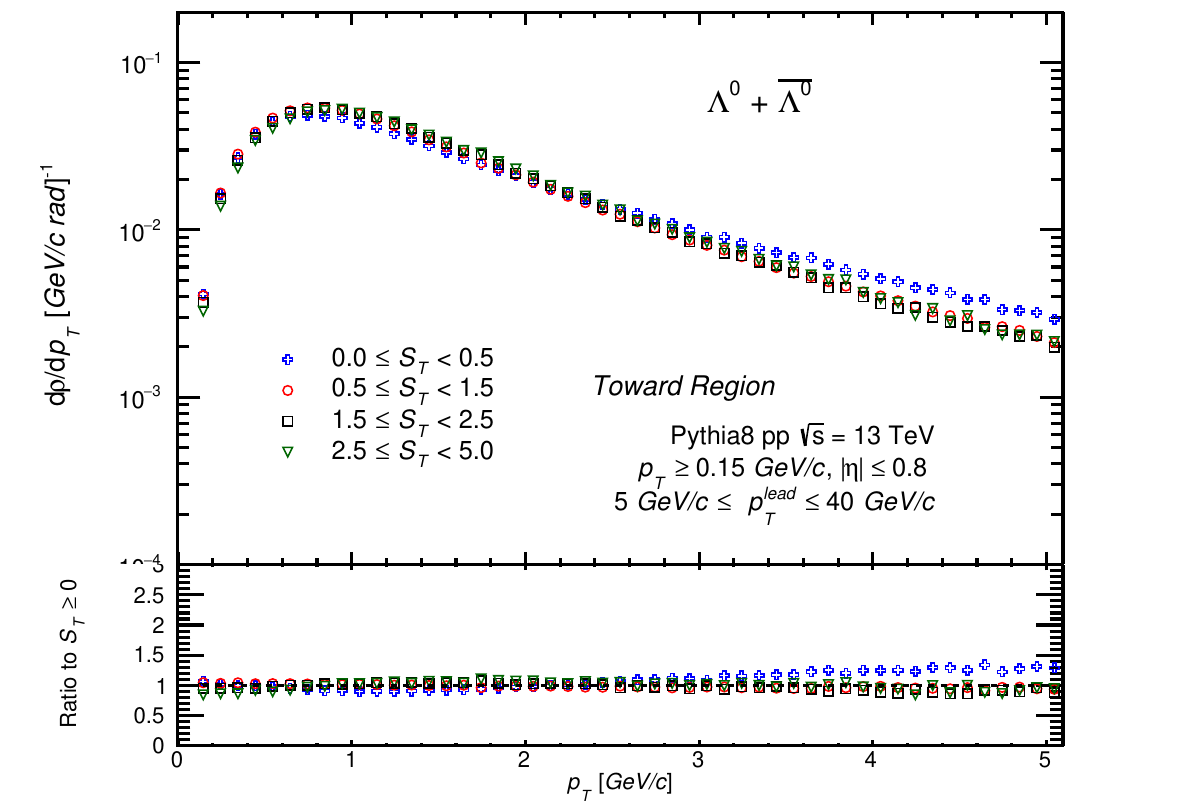}
\includegraphics[width=0.5\textwidth, height = 0.35\textwidth ]{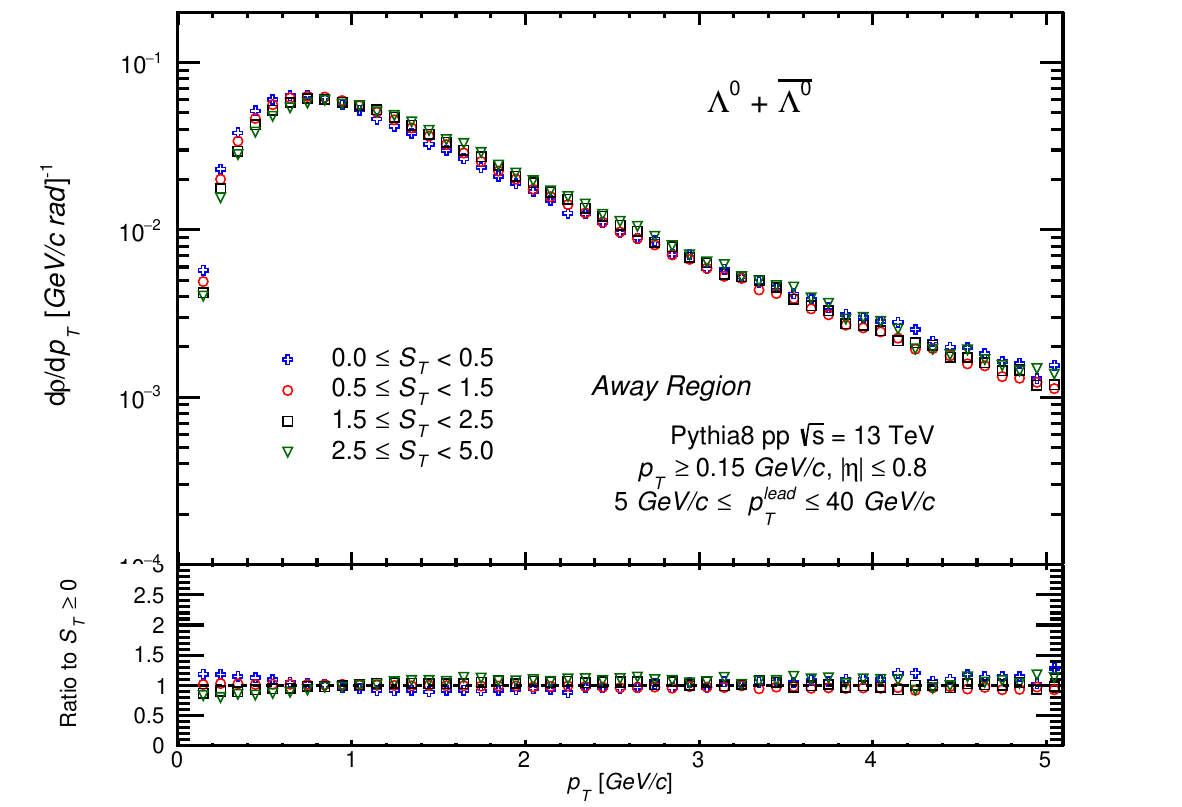}
\caption{The transverse momentum spectra of $\Lambda^{0}$ in transverse, toward and away regions for different $S_{T}$ classes  in p$-$p collisions at $\sqrt{s} =$ 13 TeV.}
\label{lambdaspectraRT}
\end{figure}

\begin{figure}
\includegraphics[width=0.5\textwidth, height = 0.35\textwidth ]{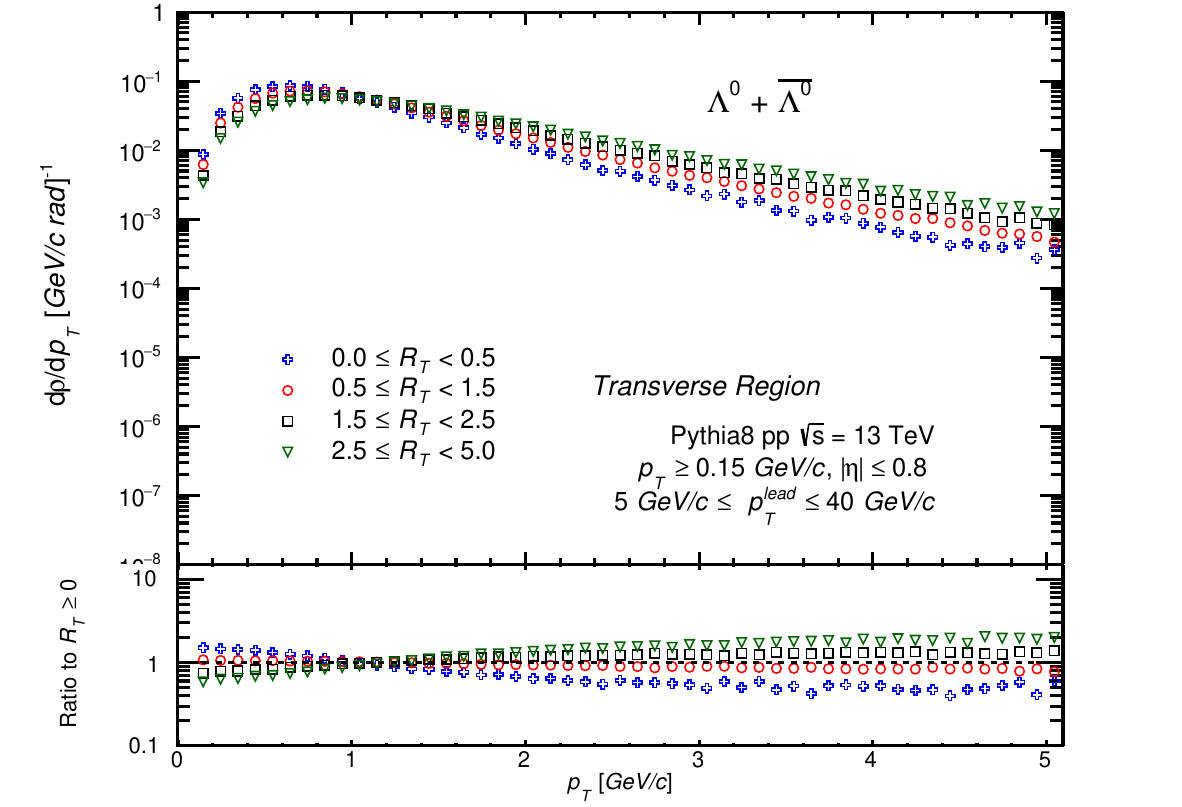} 
\includegraphics[width=0.5\textwidth, height = 0.35\textwidth ]{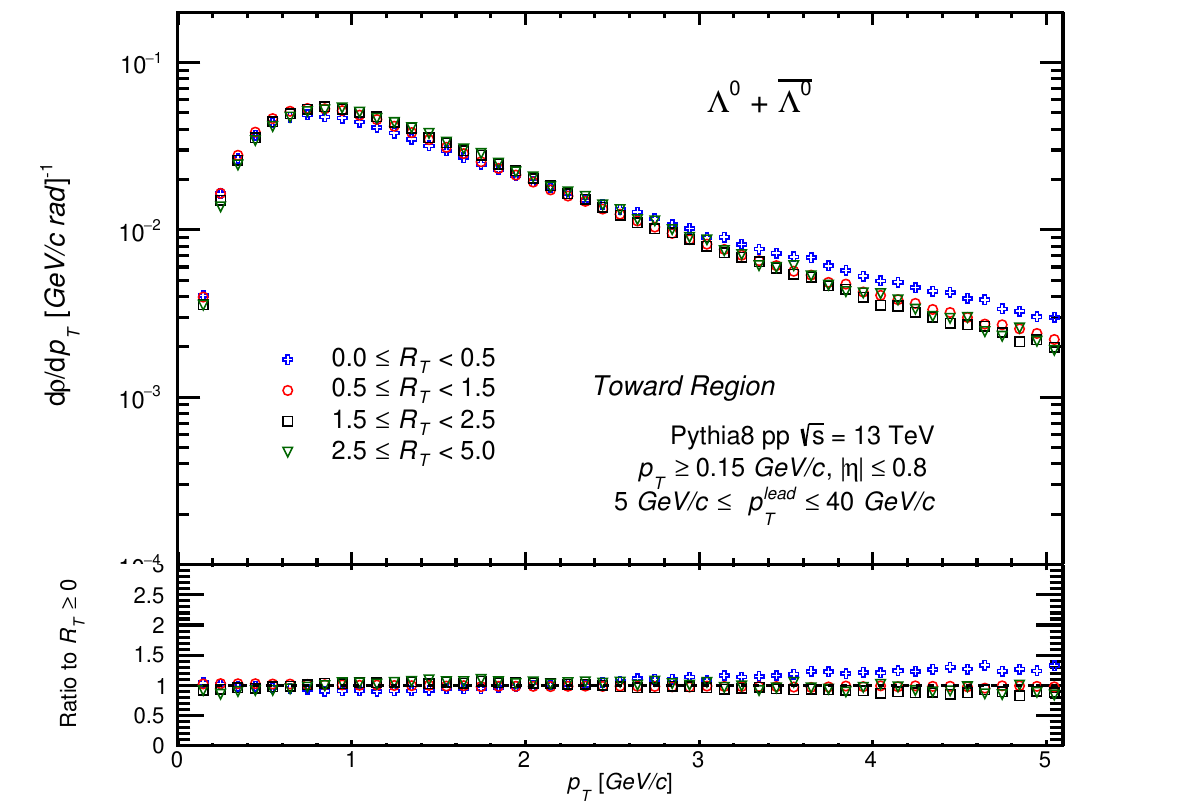}  
\includegraphics[width=0.5\textwidth, height = 0.35\textwidth ]{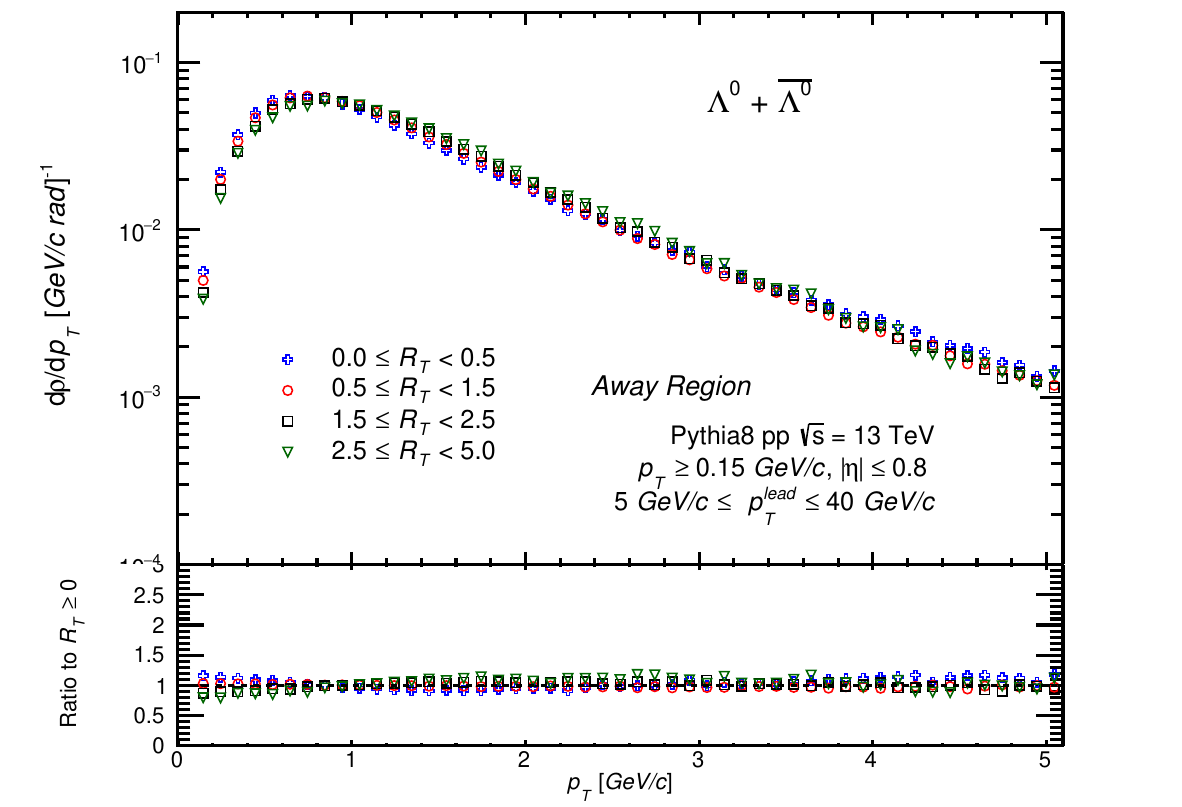}   
\caption{The transverse momentum spectra of $\Lambda^{0}$ in transverse, toward and away regions for different $R_{T}$ classes  in p$-$p collisions at $\sqrt{s} =$ 13 TeV.}
\label{lambdaspectraST}	
\end{figure}

\vspace{-0.3cm}
The evolution of the $p_{T}$ distributions of identified particles in different regions of transverse activity quantified by $R_{T}$ and $S_{T}$ can provide additional information about the particle production mechanism. Figures  \ref{pionspectraST},  \ref{pionspectraRT},  \ref{kaonspectraST}, \ref{kaonspectraRT}, \ref{protonspectraRT} and \ref{protonspectraST} show the  $p_{T}$ spectra of charged pions, kaons and protons in different classes of $S_{T}$ and $R_{T}$.  The same is shown for  $K_{S}^{0}$ and  $\Lambda^{0}$  in Figure \ref{kshortspectraRT}, \ref{kshortspectraST},  \ref{lambdaspectraRT} and \ref{lambdaspectraST}. The $p_{T}$ spectra obtained in  {\bf transverse}, {\bf towards} and {\bf away} regions are  shown in the top, middle and bottom panels, respectively. The corresponding ratios to the $S_{T}$ (and $R_{T}$) integrated spectra is shown in the bottom section of each figure. It can be observed that for all particle species, spectral shapes harden with increasing UE activity i.e. with increasing values of $S_{T}$ and $R_{T}$ in the transverse region and is reminiscent of radial flow effects. However, in the toward and away region, the differentiation of these classes is not pronounced and the spectra is harder for low values of $S_{T}$ and 
$R_{T}$ for toward as. well as away region . This is observed for all the identified particle species studied and can also be seen from ratio to $ S_{T} \geq 0 $ ( and $R_{T} \geq 0 $) integrated spectrum in lower panels of the figures. Additionally, it can be observed that the $p_{T}$ spectra in the transverse region is more strongly differentiated by $S_{T}$ classes than $R_{T}$ classes for charged particles. This effect seems to be a manifestation of the construction of the observable as this differentiation vanished for neutral particles.
 
 



\subsection{Mean $p_{T}$}
\begin{figure}[!h]
\centering
\includegraphics[width=0.5\textwidth, height = 0.35\textwidth ]{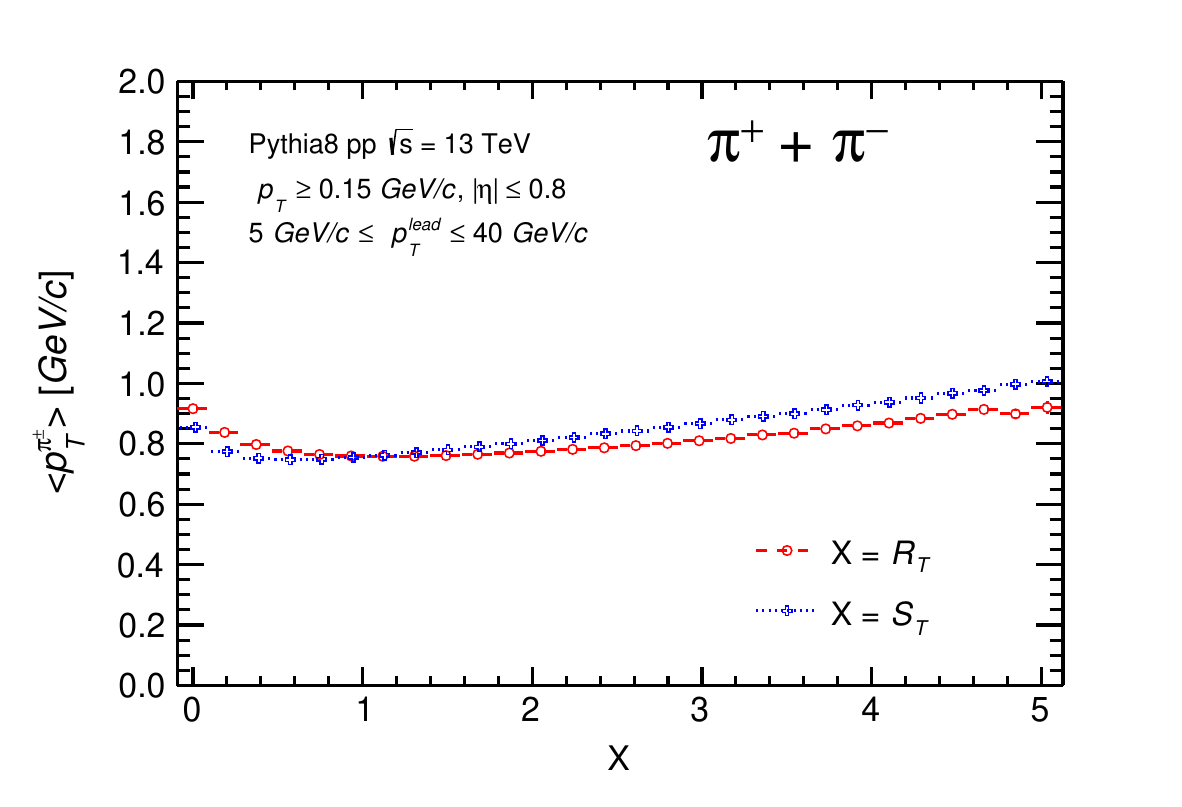}
\includegraphics[width=0.5\textwidth, height = 0.35\textwidth ]{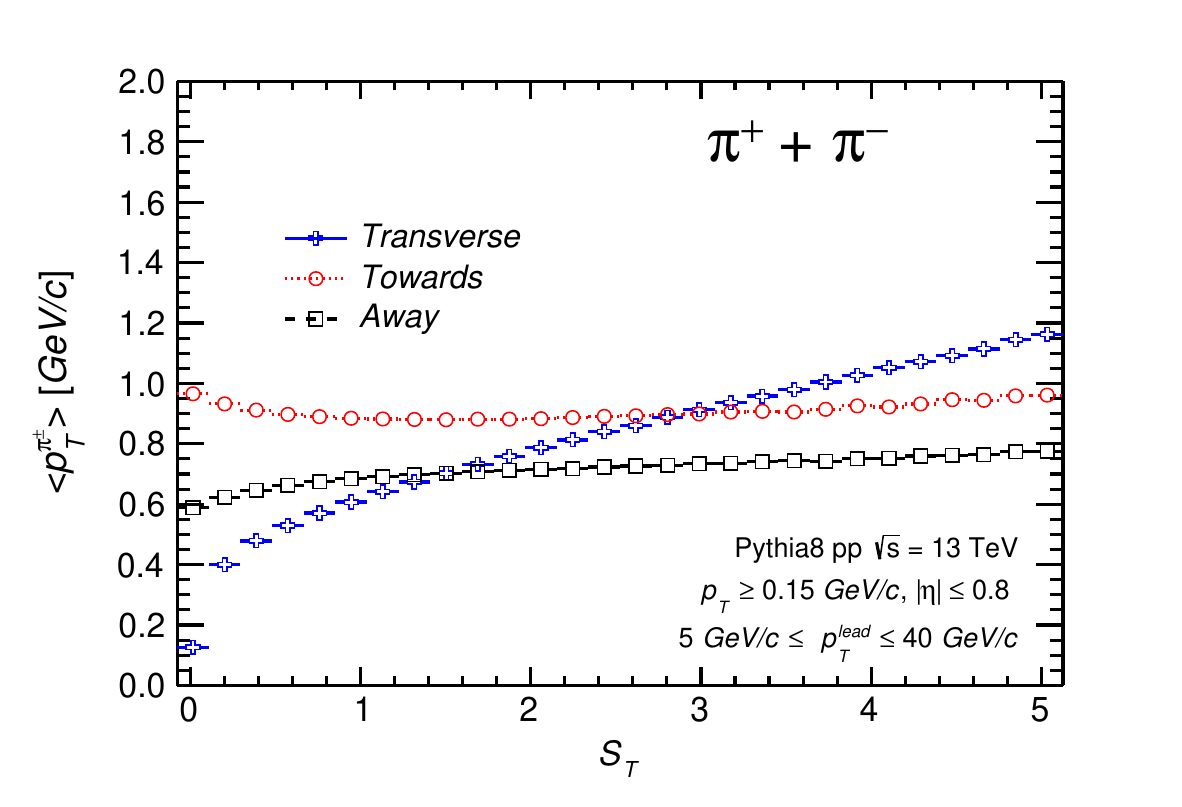}
\includegraphics[width=0.5\textwidth, height = 0.35\textwidth ]{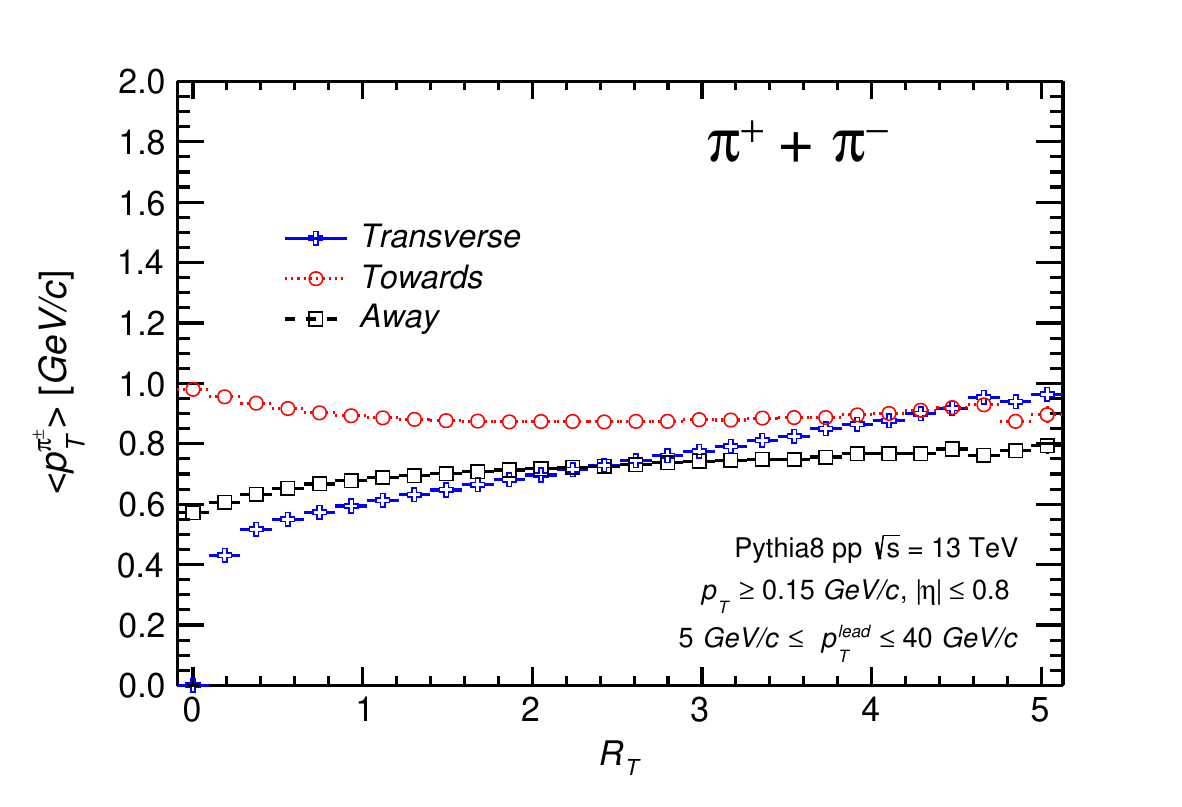}
\caption{ (Upper panel)The variation of  $<p_{T}>$  of pions with respect to $S_{T}$ (and $R_{T}$) with l$p_{T}^{lead} \ge  5$ GeV/c for p$-$p collisions at $\sqrt{s} =$ 13 TeV. The middle and bottom panels show the evolution of $<p_{T}>$  of pions as a function of  $S_{T}$ and  $R_{T}$, respectively. The comparison is shown for  the three topological regions.}
\label{meanptpion}
\end{figure}

\begin{figure}[!h]
\centering
\includegraphics[width=0.5\textwidth, height = 0.35\textwidth ]{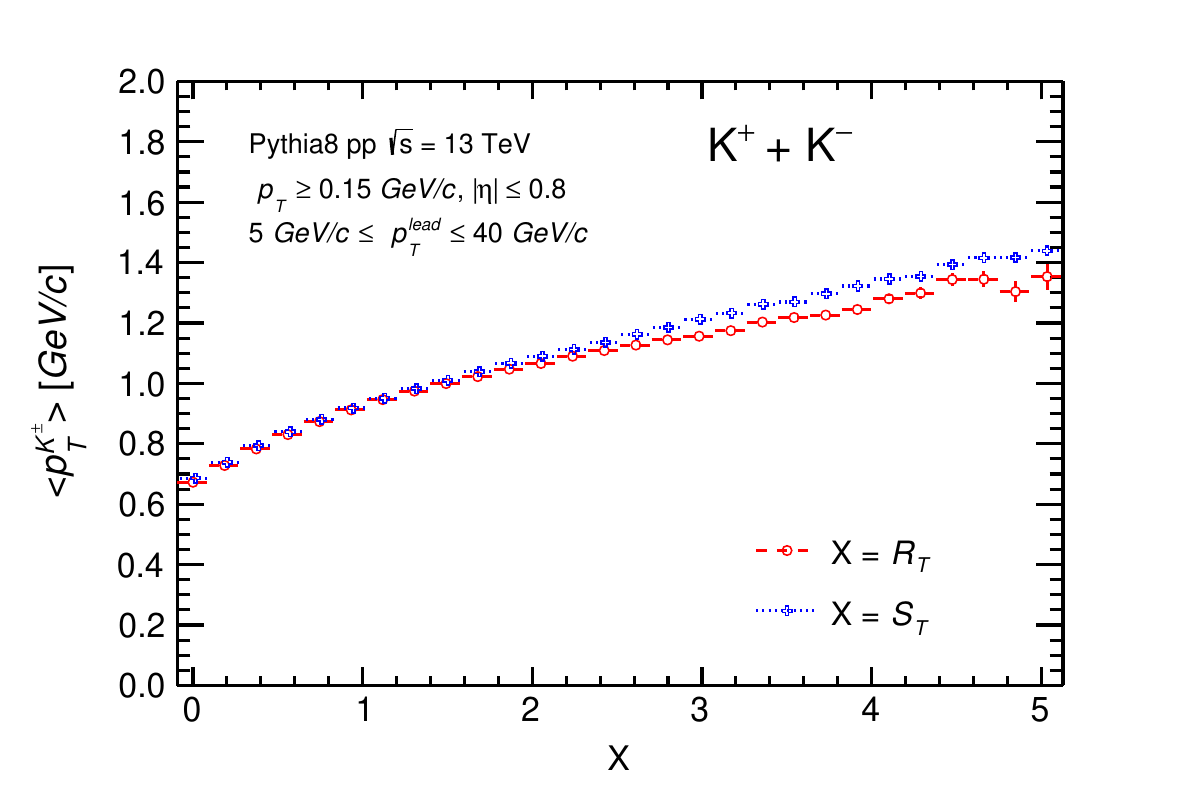}
\includegraphics[width=0.5\textwidth, height = 0.35\textwidth ]{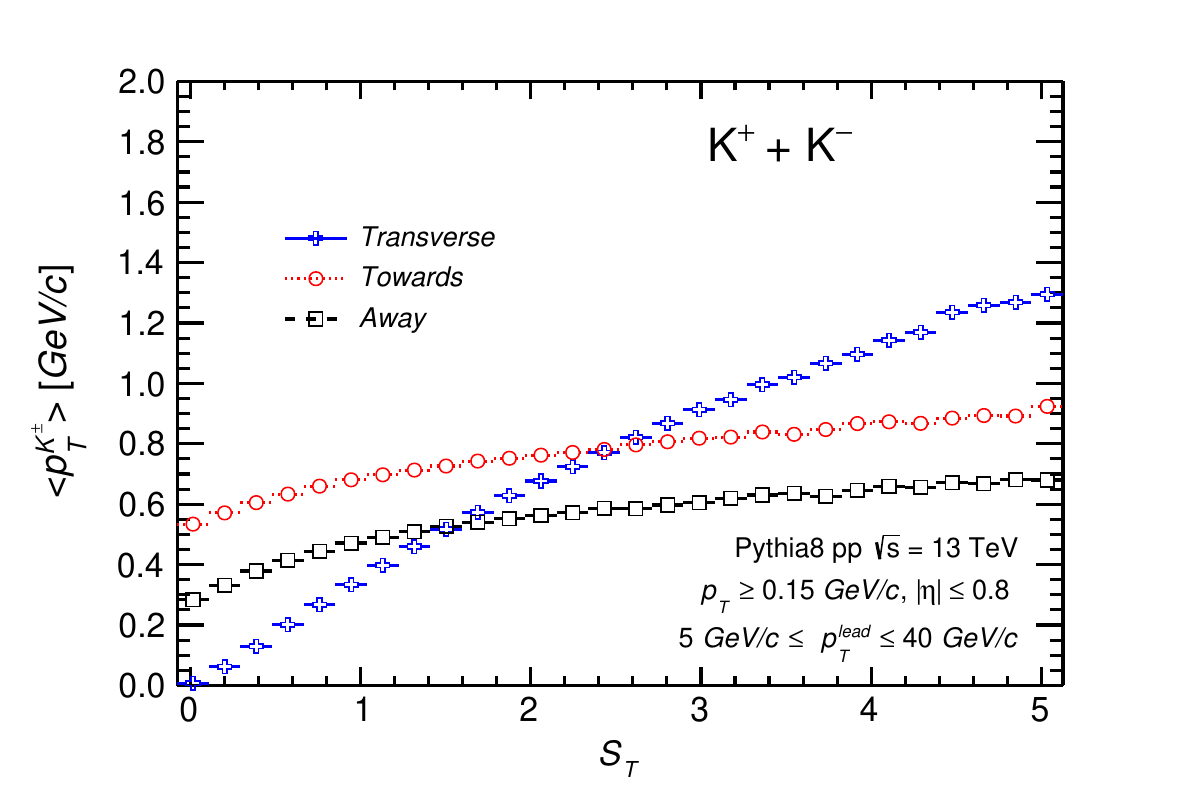}
\includegraphics[width=0.5\textwidth, height = 0.35\textwidth ]{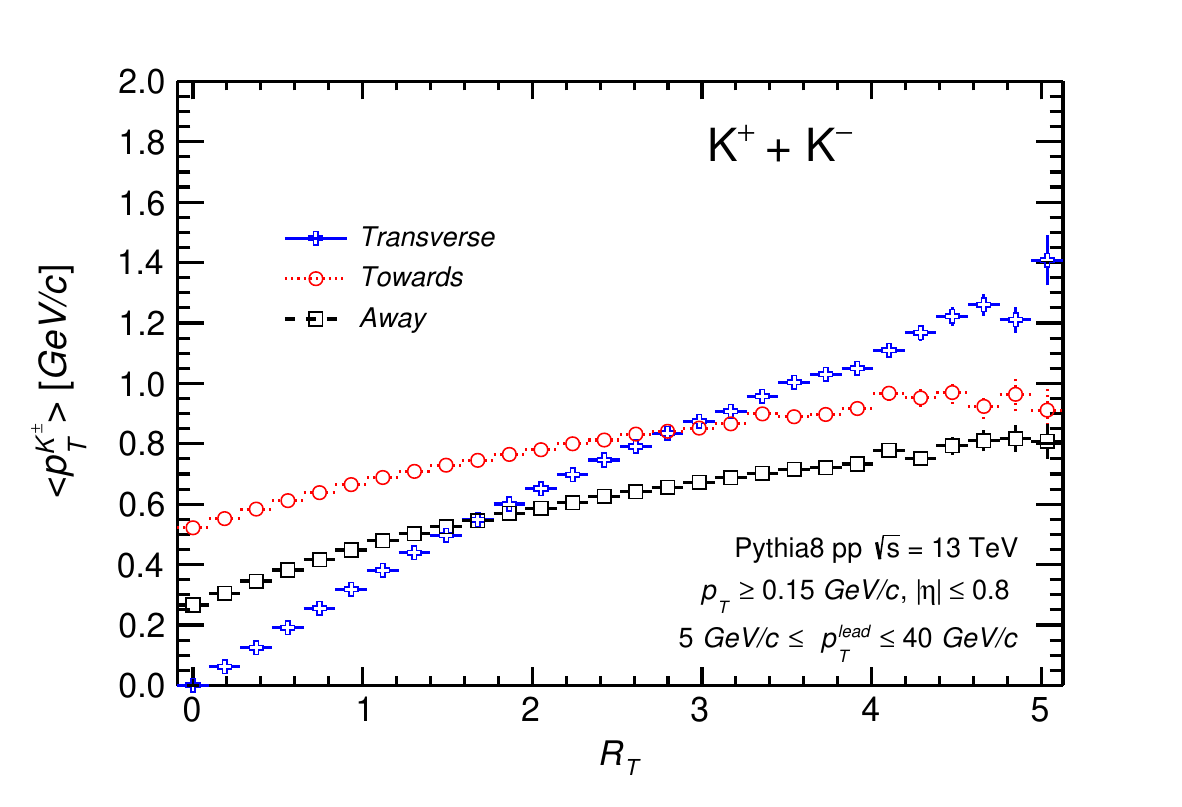}
\caption{(Upper panel)The variation of  $<p_{T}>$  of kaons with respect to $S_{T}$ (and $R_{T}$) with leading $p_{T}^{lead} \ge  5$ GeV/c for p$-$p collisions at $\sqrt{s} =$ 13 TeV. The middle and bottom panels show the evolution of $<p_{T}>$  of kaons as a function of  $S_{T}$ and  $R_{T}$, respectively. The comparison is shown for  the three topological regions.}
\label{meanptkaon}
\end{figure}

\begin{figure}[!h]
\centering
\includegraphics[width=0.5\textwidth, height = 0.35\textwidth ]{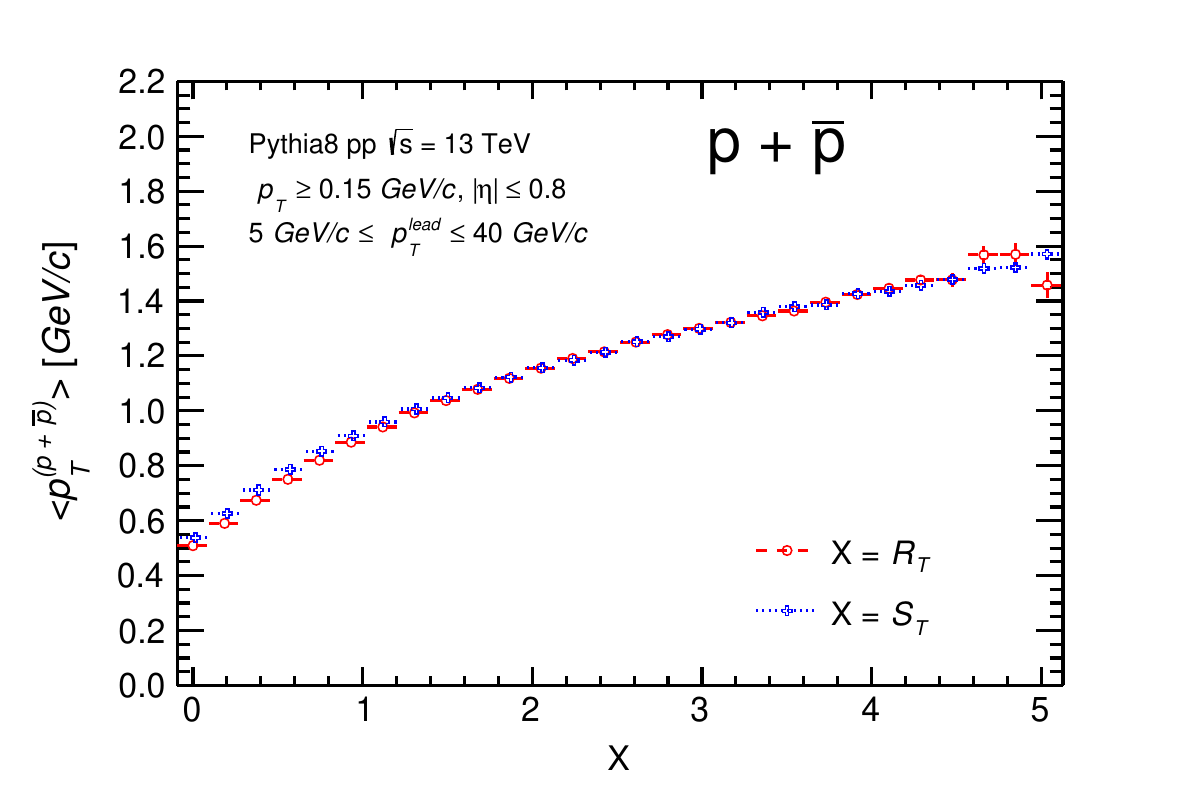}
\includegraphics[width=0.5\textwidth, height = 0.35\textwidth ]{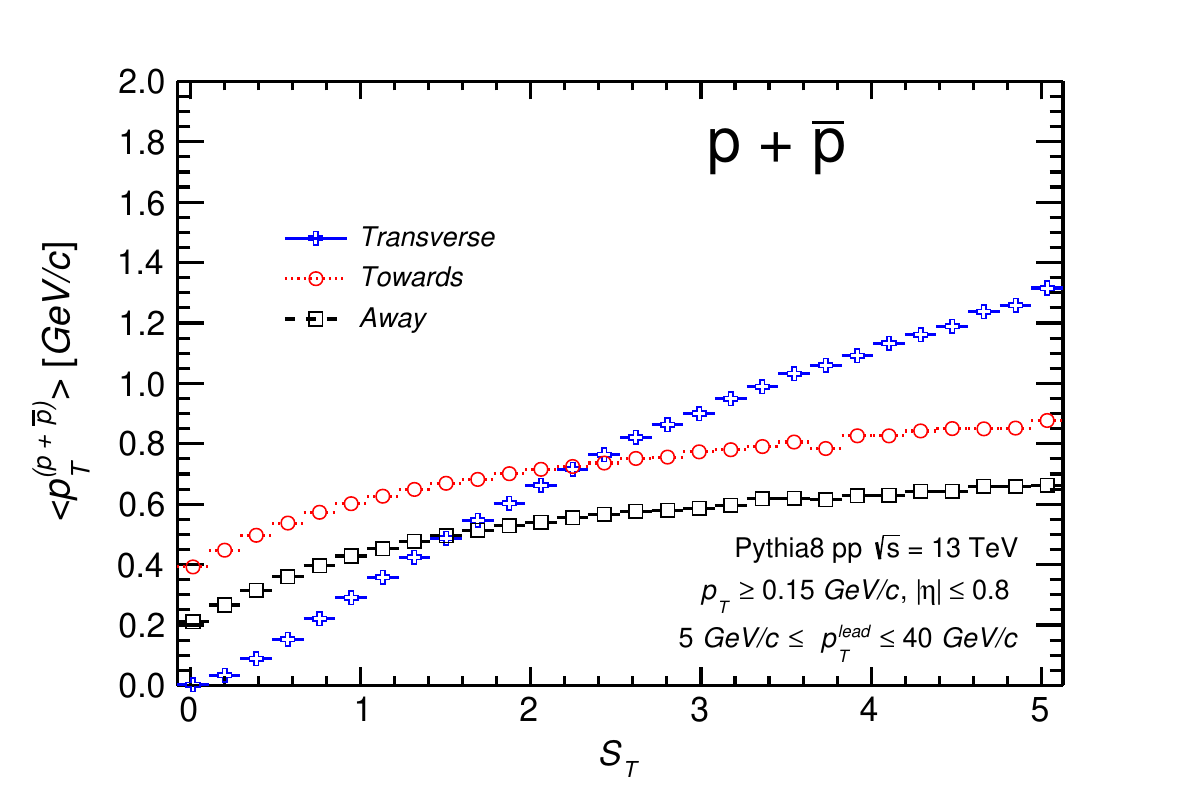}
\includegraphics[width=0.5\textwidth, height = 0.35\textwidth ]{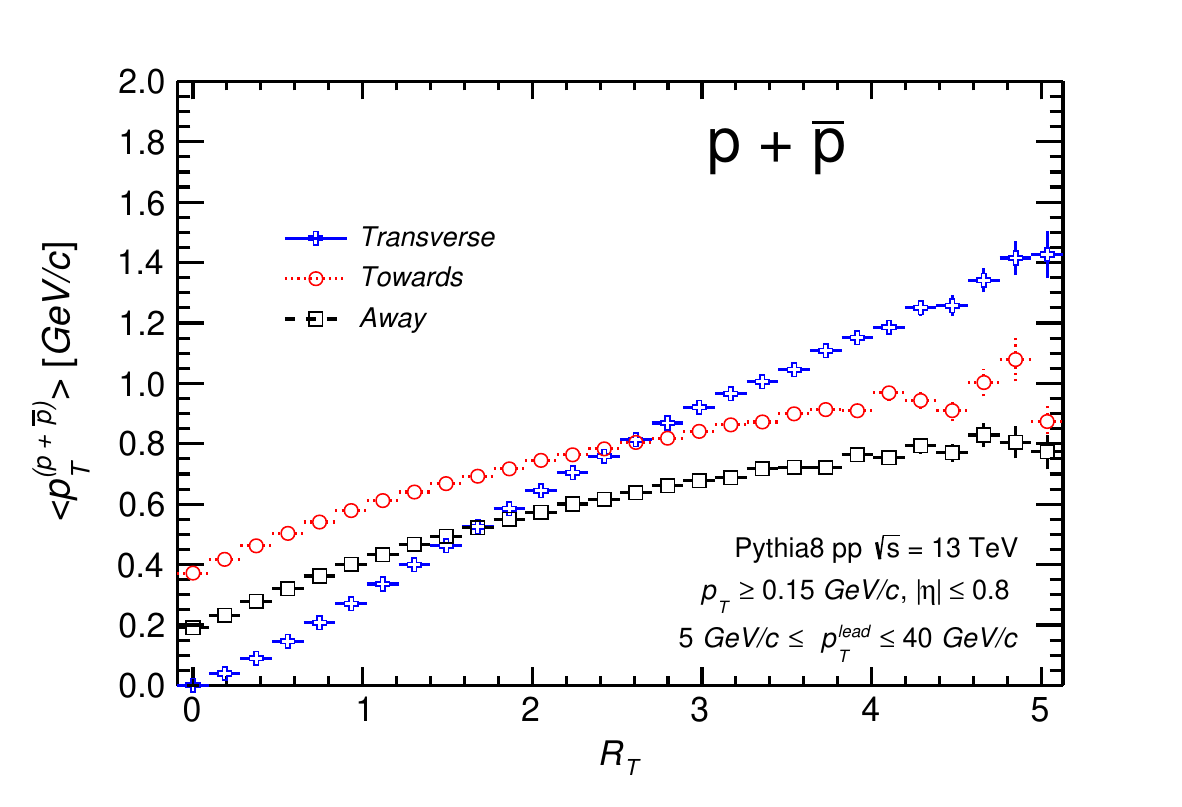}
\caption{(Upper panel)The variation of  $<p_{T}>$  of protons with respect to $S_{T}$ (and $R_{T}$) for $p_{T}^{lead} \ge  5$ GeV/c in p$-$p collisions at $\sqrt{s} =$ 13 TeV. The middle and bottom panels show the evolution of $<p_{T}>$  of protons as a function of  $S_{T}$ and  $R_{T}$, respectively. The comparison is shown for  the three topological regions.}
\label{meanptproton}
\end{figure}

\begin{figure}[!h]
\centering
\includegraphics[width=0.5\textwidth, height = 0.35\textwidth ]{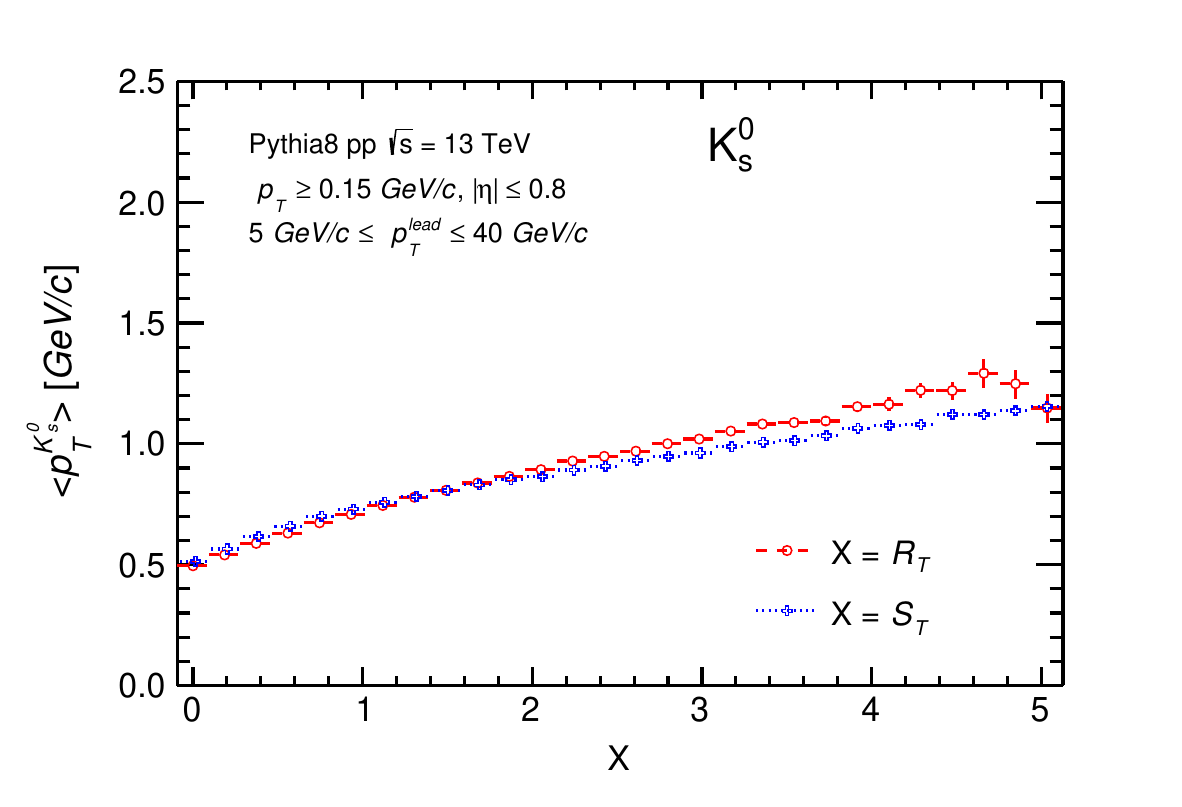}
\includegraphics[width=0.5\textwidth, height = 0.35\textwidth ]{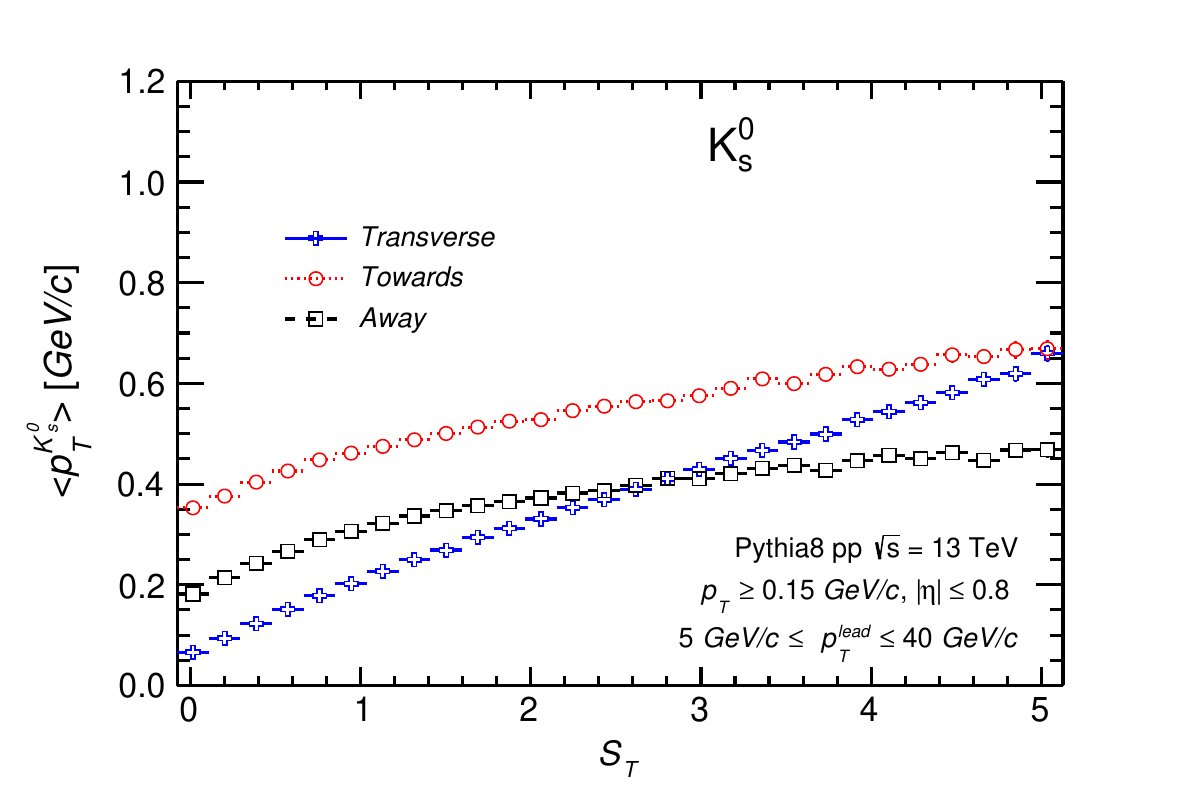}
\includegraphics[width=0.5\textwidth, height = 0.35\textwidth ]{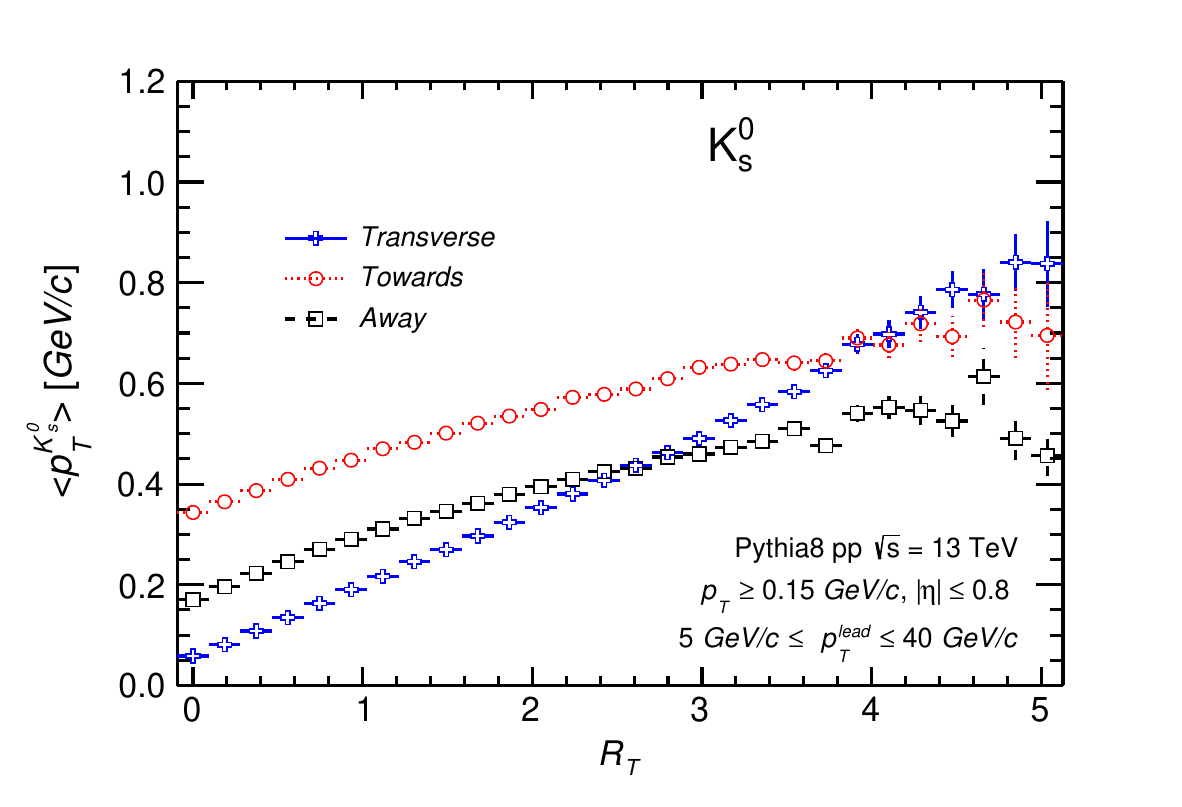}
\caption{(Upper panel)The variation of  $<p_{T}>$  of $K^{0}_{S}$ with respect to $S_{T}$ (and $R_{T}$) for $p_{T}^{lead} \ge  5$ GeV/c in p$-$p collisions at $\sqrt{s} =$ 13 TeV. The middle and bottom panels show the evolution of $<p_{T}>$  of $K^{0}_{S}$  as a function of  $S_{T}$ and  $R_{T}$, respectively. The comparison is shown for  he three topological regions.}
\label{meanptkshort}
\end{figure}

\begin{figure}[!h]
\centering
\includegraphics[width=0.5\textwidth, height = 0.35\textwidth ]{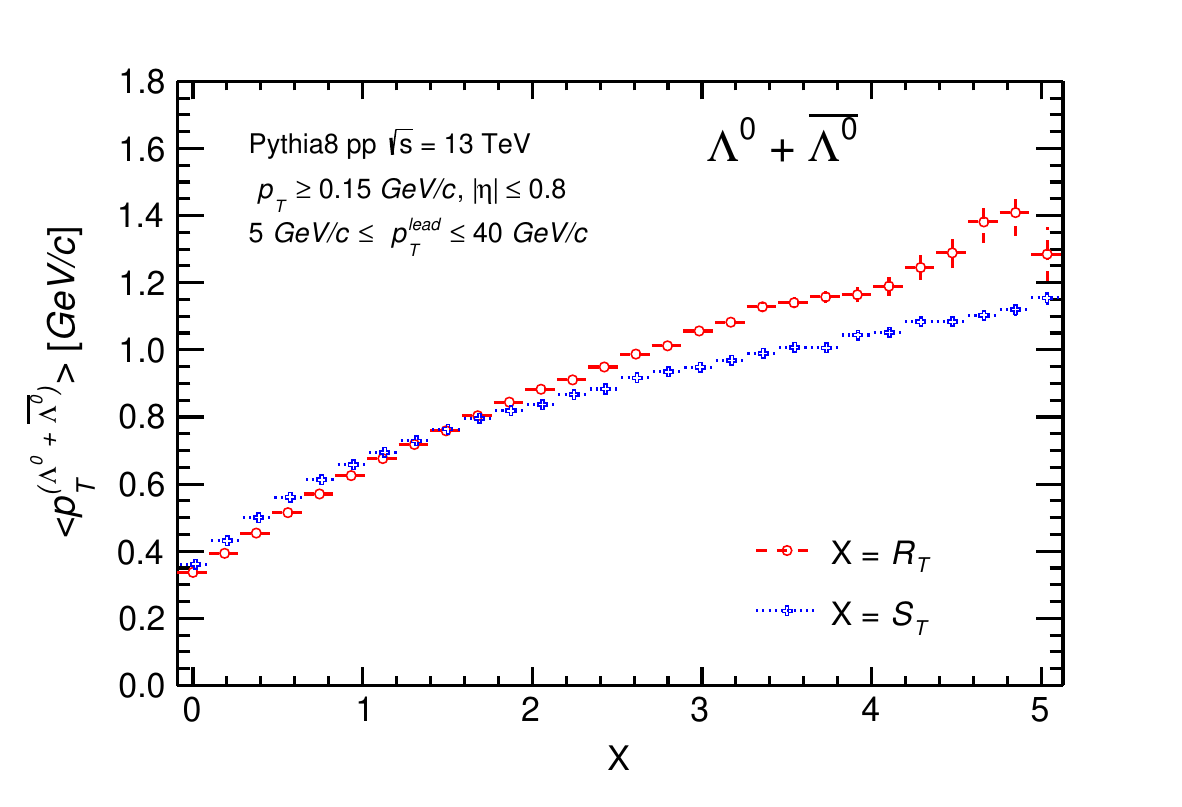}
\includegraphics[width=0.5\textwidth, height = 0.35\textwidth ]{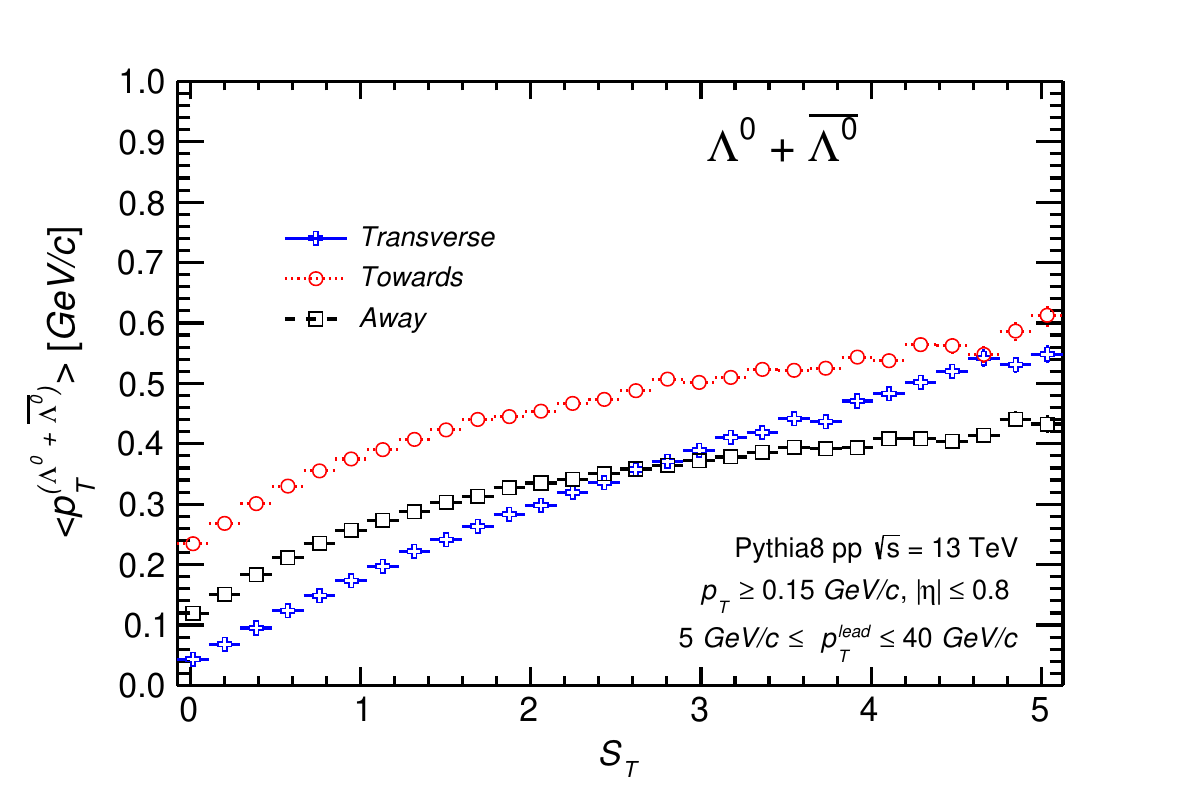}
\includegraphics[width=0.5\textwidth, height = 0.35\textwidth ]{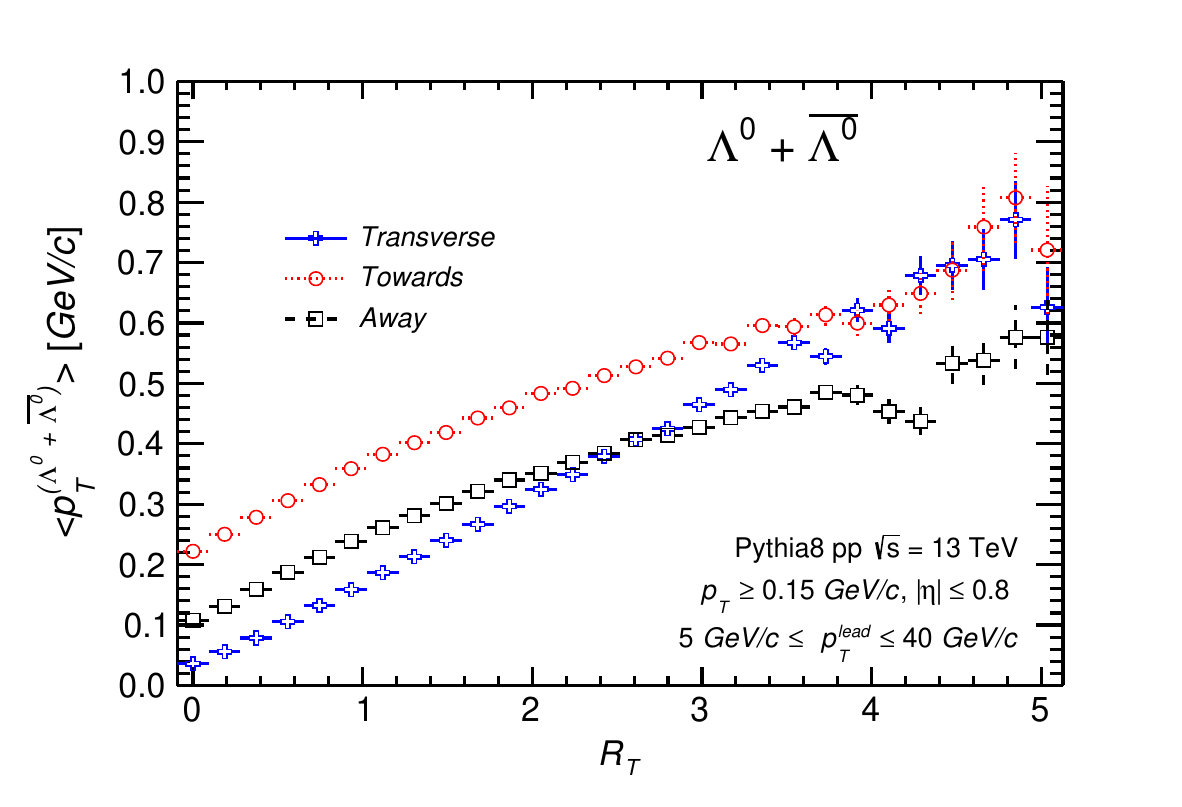}
\caption{(Upper panel)The variation of  $<p_{T}>$  of $\Lambda^{0}$ with respect to $S_{T}$ (and $R_{T}$) for  $p_{T}^{lead} \ge  5$ GeV/c in p$-$p collisions at $\sqrt{s} =$ 13 TeV. The middle and bottom panels show the evolution of $<p_{T}>$  of $\Lambda^{0}$ as function of  $S_{T}$ and  $R_{T}$, respectively. The comparison is shown for the three topological regions.}
\label{meanptlambda}
\end{figure}

The evolution of the $<p_{T}>$  of  pions, kaons ,protons, $K^{0}_{S}$ and  $\Lambda^{0}$ were studied as a function of  $S_{T}$ (and $R_{T}$) for the three topological regions. Figures \ref{meanptpion}, \ref{meanptkaon}, \ref{meanptproton}, \ref{meanptkshort}, and \ref{meanptlambda} show the variation of $<p_{T}>$  with respect to  $S_{T}$ and $R_{T}$. 
One can observe that  $<p_{T}>$ of the toward region is consistently higher for all the identified particles for lower ranges of $S_{T}$ and $R_{T}$. The values for away region is also higher than transverse region for lower ranges. This is expected as these regions are dominated by particle production via jet fragmentation and hence carry the highest $p_{
T}$. However, the trend observed for pions is different than other identified particles. The $<p_{T}>$  of pions are considerably higher for lower ranges showing a slight decrease with $S_{T}$ (and $R_{T}$) and remains more or less uniform throughout.  The values for other particles show a consistent increasing trend with $S_{T}$ (and $R_{T}$). The  values of $<p_{T}>$ of identified charged particles are highest in the transverse region for higher ranges of $S_{T}$ and $R_{T}$ where the  transverse activity due to underlying events dominates. For charged  particles, there is a smooth crossing over of the $<p_{T}>$ in transverse region with the one of away and toward region for $S_{T}$ $\sim$ 1.5  and $R_{T}$ $\sim$ 1.5. The same is not observed for the neutral particles for which the contribution from toward region dominates throughout.  The crossing point for the $V_{0}$ particles with away region is seen at  $S_{T}$ , $R_{T}$ $\sim$ 2.5 but with towards region happen at larger values.  This indicates that the contribution to $<p_{T}>$ is essentially  driven by jet fragmentation for neutral particles  while  the charged particles are affected by underlying events.

\section{Summary}

\vspace{-0cm}
The production of various species of particles $\pi^{\pm}$, $K^{\pm}$, $p + \Bar{p}$, $K_{S}^{0}$ , and $\Lambda^{0}$) in p$-$p collisions at $\sqrt{s} = 13 $ TeV is studied as a function of  transverse activity classifier, $S_{T}$, in the three topological regions using pQCD inspired PYTHIA 8 event generator. The classifier $S_{T}$ was introduced for the first time to investigate its performance with respect to the previously defined $R_{T}$. The transverse activity classifiers $R_{T}$ and $S_{T}$ are used to investigate differentially in the three topological regions where the particle production mechanisms are expected to have contributions emanating from soft processes ({ \bf Transverse} region) and hard processes ({\bf Toward} and  {\bf Away} regions). The evolution of mean multiplicity and mean transverse momentum for identified particles were studied as a function of $S_{T}$. It was observed that the particle production in the  transverse region is dominated by underlying events for higher ranges of $S_{T}$. However, for the  production of $V_{0}$ particles, there was an observed dominance from hard processes. The $p_{T}$  spectra of the considered particle species  were observed to be sensitive to UE activity. Moreover, it was observed that $p_{T}$ spectra of identified charged particles were strongly differentiated by $S_{T}$ classes than $R_{T}$ classes. Therefore, it can be observed that by measuring the production of particles as a function of these transverse activity classifiers, one can study various aspects of underlying events. The results presented in this work are expected to shed more light on the understanding of underlying event activity at LHC energies. The obtained results can act as a baseline for the upcoming experimental measurements at LHC energies and can help constrain the Monte-Carlo models.\clearpage

\section{Acknowledgements}
Sadhana Dash would like to acknowledge and thank SERB Power fellowship, {\bf SPF/2022/000014} for supporting the present work.

\end{document}